\RequirePackage{pdf14}

\documentclass[letterpaper]{article}
\usepackage{array,setspace,mathrsfs,amsfonts,yfonts,dsfont,bbm,colonequals,subcaption}
\usepackage[table]{xcolor}
\usepackage{./aux/jheppub}
\newcommand\be{\beg{equation}}
\newcommand\bea{\beg{eqnarray}}

\newcommand\eea{\end{eqnarray}}
\newcommand\ee{\end{equation}}

\newtheorem{con}{Conjecture}

\newcommand\nn{\nonumber}
\newcommand\ph{\phantom}

\newcommand\eg{{\it e.g.}}
\newcommand\ie{{\it i.e.}}

\newcommand\e{\epsilon}

\newcommand\D{\Delta}

\newcommand\Rb{\mathbb{R}}
\newcommand\Zb{\mathbb{Z}}

\renewcommand\AA{\mathcal{A}}
\newcommand\BB{\mathcal{B}}
\newcommand\CC{\mathcal{C}}
\newcommand\DD{\mathcal{D}}

\newcommand\FF{\mathcal{F}}
\newcommand\GG{\mathcal{G}}

\newcommand\MM{\mathcal{M}}
\newcommand\NN{\mathcal{N}}
\newcommand\OO{\mathcal{O}}

\renewcommand\SS{\mathcal{S}}
\newcommand\TT{\mathcal{T}}

\newcommand\XX{\mathcal{X}}

\newcommand\zb{{\bar z}}

\def\be{\begin{equation}}
\def\ee{\end{equation}}
\def\bea{\begin{eqnarray}}
\def\eea{\end{eqnarray}}

\def\nn{\nonumber}

\def\ph{\phantom}

\newcommand\fverb{\setbox\fverbbox=\hbox\bgroup\verb}
\newcommand\fverbdo{\egroup\medskip\noindent%
			\fbox{\unhbox\fverbbox}\ }
\newcommand\fverbit{\egroup\item[\fbox{\unhbox\fverbbox}]}
\newbox\fverbbox

\def\bea{\begin{eqnarray}}
\def\eea{\end{eqnarray}}
\def\be{\begin{equation}}
\def\ee{\end{equation}}

\def\e{\epsilon}
\def\D{\Delta}

\def\OO{\mathcal{O}}

\def\TT{\mathcal{T}}
\def\NN{\mathcal{N}}
\def\MM{\mathcal{M}}

\def\Tr{\mbox{Tr}}

\def\TT{\mathcal{T}}
\def\SS{\mathcal{S}}

\def\AA{\mathcal{A}}
\def\BB{\mathcal{B}}
\def\CC{\mathcal{C}}
\def\DD{\mathcal{D}}

\def \a{\alpha}
\def \b{\beta}

\def \e{\epsilon}

\def \ph{\phantom}

\def \d{\delta}

\def\sof{\mathfrak{so}}
\def\suf{\mathfrak{su}}
\def\psuf{\mathfrak{psu}}
\def\gf{\mathfrak{g}}
\def\slf{\mathfrak{sl}}
\def\uf{\mathfrak{u}}

\def \hf{\frac{1}{2}}
\def \qt{\frac{1}{4}}
\def \del{\partial}

\newcommand\restr[2]{{
  \left.\kern-\nulldelimiterspace 
  #1 
  \vphantom{\big|} 
  \right|_{#2} 
  }}

\title{More $\NN=4$ superconformal bootstrap}

\author[\!1,2]{Christopher Beem,}
\author[\!3]{Leonardo Rastelli,}
\author[\,4]{and Balt C. van Rees}
 
\emailAdd{christopher.beem@maths.ox.ac.uk}
\emailAdd{leonardo.rastelli@stonybrook.edu}
\emailAdd{balt.van-rees@durham.ac.uk}

\affiliation[1]{Mathematical Institute, University of Oxford, Woodstock Road, Oxford, OX2 6GG, UK}
\affiliation[2]{School of Natural Sciences, Institute for Advanced Study, Einstein Drive, Princeton, NJ 08540, USA}
\affiliation[3]{C.~N.~Yang Institute for Theoretical Physics, Stony Brook University, Stony Brook, NY 11794-3840, USA}
\affiliation[4]{Centre for Particle Theory \& Department of Mathematical Sciences, Durham University, Durham, UK }
\preprint{YITP-SB-16-49}
\bigskip
\abstract{In this long overdue second installment, we continue to develop the conformal bootstrap program for $\NN=4$ superconformal field theories in four dimensions via an analysis of the correlation function of four stress-tensor supermultiplets. We review analytic results for this correlator and make contact with the SCFT/chiral algebra correspondence of \cite{Beem:2013sza}. We demonstrate that the constraints of unitarity and crossing symmetry require the central charge $c$ to be greater than or equal to $3/4$ in any interacting $\NN=4$ SCFT. We apply numerical bootstrap methods to derive upper bounds on scaling dimensions and OPE coefficients for several low-lying, unprotected operators as a function of the central charge. We interpret our bounds in the context of $\NN=4$ super Yang-Mills (SYM) theories, formulating a series of conjectures regarding the embedding of the conformal manifold --- parametrized by the complexified gauge coupling --- into the space of scaling dimensions and OPE coefficients. Our conjectures assign a distinguished role to points on the conformal manifold that are self-dual under a subgroup of the S-duality group. This paper contains a more detailed exposition of a number of results previously reported in \cite{Beem:2013qxa} in addition to new results.
}


\begin{document}
\maketitle


\section{Introduction}
\label{sec:intro}

In this paper we elaborate upon our earlier investigations into the space of $\NN=4$ superconformal field theories (SCFTs) in four dimensions \cite{Beem:2013qxa}. Though it is widely believed that this space comprises only the $\NN=4$ supersymmetric Yang-Mills (SYM) theories, this remains a conjecture at present and there may in principle exist exotic $\NN=4$ SCFTs without a Lagrangian description. By adopting the methods of the conformal bootstrap, we can study $\NN=4$ SCFTs entirely on the basis of their symmetries and general consistency requirements without taking a view on their (non-)Lagrangian nature. For this reason we consider this to be an attractive framework within which to address the classification of these theories.

Within the framework of the conformal bootstrap, a conformal field theory is taken to be characterized by its algebra of local operators, the structure of which is determined entirely by the scaling dimensions and spins of all conformal primary operators along with their OPE coefficients. These \emph{conformal data} are subject to additional constraints that follow from unitarity and from associativity of the operator algebra. For $\NN=4$ SCFTs, there are additional constraints that follow from the requirement of \emph{super}conformal invariance. In this paper we explore the intersection of these various constraints both analytically and numerically.

Whenever possible, we restrict our attention to ``simple'' theories, which cannot be decomposed as tensor products of simpler theories. Local simple theories should have a unique stress tensor, which by $\NN=4$ superconformal symmetry belongs to a supermultiplet whose bottom component is a scalar operator of dimension two that transforms in the ${\bf 20'}$ representation of the $\suf(4)_R$ symmetry. Superconformal Ward identities fix the two- and three-point functions of the stress tensor supermultiplet in terms of a single real number, the central charge $c$ of the theory \cite{Howe:1995aq,Howe:1998zi,Dolan:2004mu}.%
\footnote{The existence of a unique superconformally invariant structure for the stress-tensor three-point function implies that the $a$ and $c$ Weyl anomaly coefficients are necessarily related, and in fact equal (see, \eg, \cite{Dolan:2001tt}).}
This is the most basic parameter of an $\NN =4$ SCFT. 

Some general features of $\NN=4$ SCFTs follow directly from the representation theory of the superconformal algebra $\psuf(2,2|4)$, and these help to frame our investigation. For example, the stress tensor supermultiplet necessarily contains an exactly marginal complex scalar that is an $\suf(4)_R$ singlet and is annihilated by all the Poincar\'e supercharges. Conversely, any such operator must sit in the stress tensor supermultiplet. Consequently, any simple $\NN=4$ SCFT is endowed with a conformal manifold of complex dimension one, \ie, each theory is really a fixed (complex) curve of SCFTs parameterized by an exactly marginal coupling.%
\footnote{Here we are only describing the situation for $\NN=4$ marginal operators. There are additional exactly marginal operators that preserve less than maximal supersymmetry \cite{Leigh:1995ep}, but they will not play a role in our analysis.}
Additionally, interacting $\NN =4$ theories cannot have global symmetries beyond $\suf(4)_R$. This is because in the absence of higher-spin conserved currents (which should always be absent for interacting theories, \emph{cf.} \cite{Maldacena:2011jn}) representation theory dictates that conserved currents can only appear in the stress tensor multiplet, and these are precisely the $\suf(4)_R$ Noether currents. 

The aforementioned properties are readily verified in the Lagrangian $\NN=4$ theories. The only massless free supermultiplet with no fields of spin greater than one is the $\NN=4$ vector multiplet. The vector multiplet contains gauge fields and so must transform in the adjoint representation of a reductive Lie algebra $\gf$. For a simple theory, $\gf$ must be simple or $\mathfrak{u}(1)$ -- in the latter case the theory is free. The Lagrangian is uniquely fixed by $\NN =4$ supersymmetry, with the complexified gauge coupling $\tau \in H$ being the only free parameter. The central charge is easily calculated to be $c_{\mathfrak{g}} = {\rm dim}(\mathfrak{g}) /4$. The set of distinct theories constructed in this way is obtained by modding out the space of pairs $(\gf, \tau)$ by the action of the discrete S-duality group.%
\footnote{This classification ignores finer distinctions that arise upon consideration of extended defects -- such as Wilson and 't Hooft line operators -- and spacetimes with nontrivial topology. For example, line defects can detect the distinction between different gauge groups associated to the same gauge algebra; see \cite{Aharony:2013hda} for a comprehensive discussion. By considering theories at the level of their algebra of local operators, we evade these subtleties for the time being.}%
${}^,$%
\footnote{For simply-laced gauge algebras the S-duality group is simply $(P)SL(2,\mathbb Z)$, which acts on $\tau$ by fractional linear transformations. For more general choices of $\gf$, the duality group is more complicated, and S-duality interchanges $\gf$ with its Langlands dual ${}^L \gf$. See \cite{Dorey:1996hx,Argyres:2006qr} for details.}

From the abstract CFT viewpoint, there is an asymmetry between the two parameters $c$ and $\tau$: while the central charge $c$ is a fundamental conformal datum, easily read off from the stress-tensor OPE, the coupling $\tau$ is not directly observable. However, unprotected scaling dimensions and OPE coefficients vary over the conformal manifold, and one may hope to use one of these quantities (say the scaling dimension of the leading-twist unprotected singlet scalar operator) as a proxy for the coupling in the bootstrap framework. 

The bootstrap program for $\NN=4$ SCFTs is facilitated by the existence of a protected, solvable subsector of the local operator algebra. This is a special case of the structure discovered in \cite{Beem:2013sza}, which exists in any four-dimensional $\NN =2$ SCFT. For a general $\NN = 2$ theory, the protected subsector is isomorphic to a non-supersymmetric chiral algebra, which universally possesses a Virasoro subalgebra of central charge $c_{2d} = -12 c$ descending from the stress tensor multiplet. In the $\NN=4$ case the chiral algebra is in fact supersymmetric, and always contains a small $\NN=4$ super-Virasoro algebra. A proposal for the description of the chiral algebras associated to the $\NN=4$ SYM theories as super $W$-algebras was put forward in \cite{Beem:2013sza}. For $\gf = \suf(2)$, the chiral algebra is conjectured to be precisely the small $\NN=4$ super-Virasoro algebra with Virasoro central charge $c_{2d}=-9$. The super W-algebras associated to SYM theories of higher rank possess additional generators that are in one-to-one correspondence with the generators of the one-half BPS chiral ring. 

The chiral algebra captures an infinite amount of protected conformal data of the four-dimensional SCFT. An important analytic insight that follows from the embedding of the chiral algebra into its four-dimensional parent is the existence of novel unitarity bounds for four-dimensional central charges \cite{Beem:2013qxa, Beem:2013sza, Liendo:2015ofa, Lemos:2015orc}. In the $\NN=4$ case, one finds that $c$ must be greater or equal to $3/4$ in any \emph{interacting} $\NN=4$ SCFT --- a result announced in \cite{Beem:2013qxa} and discussed in detail below. Indeed, for $c < 3/4$ the constraints of unitarity and crossing symmetry require the introduction of supermultiplets containing higher-spin conserved currents. The bound is saturated by the interacting $\NN =4$ SYM theory with minimal central charge, namely the theory with $\suf(2)$ gauge algebra. The existence of exotic theories with smaller central charge is rigorously ruled out. We hope that similar analytic considerations, possibly combined with suitable assumptions about the chiral ring, will show that the allowed central charges coincide with the discrete set realized in the $\NN=4$ super Yang-Mills theories. In the present work we take $c$ to be a continuous parameter. We will mostly be concerned with the range $c \geqslant 3/4$ of unitarity interacting theories, but we will find it instructive to also consider positive values smaller than $3/4$ in order to investigate the expected approach to free field theory as $c \to1/4$.

To proceed beyond the protected subsector, we turn to the numerical bootstrap methods pioneered in \cite{Rattazzi:2008pe}. In recent years the ideas of \cite{Rattazzi:2008pe} and their generalizations and extensions have been used with great success to obtain constraints on the spectrum and OPE coefficients in a variety of conformal theories, both with and without supersymmetry. Most striking are perhaps the accurate determination of the critical exponents in the three-dimensional Ising CFT \cite{ElShowk:2012ht,El-Showk:2014dwa}, and in particular the exclusion of all but a small island in the space of scaling dimensions for the two relevant operators of the theory \cite{Kos:2014bka, Simmons-Duffin:2015qma}. 

In light of these kinds of results, we envision the goal of the numerical $\NN=4$ bootstrap program as follows. Let $\TT_c$ denote the space of $\NN=4$ superconformal field theories with central charge $c$, and $\SS$ the infinite-dimensional space of conformal data that are compatible with unitarity bounds and Ward identities imposed by $\NN=4$ superconformal symmetry. There then exists a map $\rho: \TT_c \to \SS$ defined by identifying a given theory with its conformal data. The map $\rho$ is not surjective, because the associativity constraints of the operator algebra are not satisfied everywhere on $\SS$. In this language, our goal is to delineate the image $\rho(\TT_c) \subset \SS$. For a given allowed value of $c$, we expect $\rho(\TT_c)$ to be a real-two-dimensional submanifold of $\SS$, corresponding to an embedding of the conformal manifold into $\SS$. A construction of this embedding would implicitly supply a nonperturbative solution of the $\NN=4$ SCFTs, at least for the low-lying spectrum and OPE coefficients accessible by the numerical bootstrap.

This program is in its early stages. The present paper is devoted to the analysis of constraints imposed on $\rho(\TT_c)$ by consistency of a single correlation function, the four-point functions of stress tensor multiplets. In fact, since four-point functions of half-BPS operators have a unique structure in superspace \cite{Dolan:2004mu, Korchemsky:2015ssa}, there is no loss of generality in restricting our attention to the four-point function of the superconformal primary operator in this multiplet --- namely the ${\bf 20'}$ operator. This is a much simpler object with which to work. We derive upper bounds on scaling dimensions and OPE coefficients of several low-lying unprotected operators in the theory, as a function of the central charge $c$.\footnote{Upper bounds on OPE coefficients in the ${\bf 20'}$ four-point function were first derived using numerical bootstrap methods in \cite{Alday:2014qfa}.} Our bounds constrain the image $\rho(\TT_c)$, which must lie entirely within the allowed subregions of $\SS$ that are carved out numerically. (See Fig.~\ref{fig:embedding} for a sketch.) We formulate a series of conjectures about $\rho$ which assign a distinguished role to the self-dual points at $\tau = i$ or $\tau = \exp(i \pi/3)$ on the conformal manifolds of the $A$ or $D$ series of $\NN=4$ SYM theories. We emphasize that these conjectures pertain to the non-perturbative behavior of the non-planar gauge theories at finite coupling. This is a regime beyond the usual reach of either ordinary perturbation theory or integrability. 

It would be very interesting to confirm (or refute) our results by other methods. For example, our conjectures are generally compatible with the S-duality improved resummations of perturbative results performed in \cite{Beem:2013hha, Alday:2013bha,Chowdhury:2016hny}, but it would be better if such resummation methods could be improved and put on firmer conceptual footing, perhaps with the help of resurgence theory \cite{Dorigoni:2014hea,Dunne:2015eaa}. We also look forward to progress in the lattice formulation of $\NN=4$ super Yang-Mills theory \cite{Catterall:2005fd,Catterall:2009it,Schaich:2016jus}, which may offer an independent approach to extracting the same data that we are studying here with the bootstrap. At the same time, there is a great deal of room to improve the bootstrap results using established methods. One can, for example, study different correlation functions and different observables; initial results in this direction were reported in \cite{Alday:2013opa,Alday:2014qfa}, and the groundwork for further numerical analysis is provided by the solution of the superconformal Ward identities for half-BPS correlation functions presented in \cite{Doobary:2015gia,Bissi:2015qoa}.\footnote{Similar results were obtained in \cite{Liendo:2016ymz} for $\NN=4$ theories with defects.} In future work we hope to move this program forward through the simultaneous analysis of multiple correlators in the style of \cite{Kos:2014bka}.

The rest of the paper is organized as follows. We begin in the next section with a self-contained review of the structure of the ${\bf 20'}$ four-point function, tailored to the numerical analysis that we have undertaken. We then provide in Section \ref{sec:methods} a short review of the numerical bootstrap approach pioneered in \cite{Rattazzi:2008pe}. In Section \ref{sec:firstresults} we present in detail the first batch of numerical results (a preview of which appeared in \cite{Beem:2013qxa}) and speculate on the nonperturbative behavior of scaling dimensions in $\NN=4$ super Yang-Mills theories. In particular, we conjecture to have obtained scaling dimensions in som`'e of these theories for a self-dual value of the coupling. In Section \ref{sec:secondresults} we present new results involving OPE coefficients and scaling dimensions of other operators in the theory. These results potentially connect the isolated points of Section \ref{sec:firstresults} through a line on the conformal manifold to the free field theory.

\section{The \texorpdfstring{${\bf 20'}$}{20'} four-point function}
\label{sec:fourptfunction}

Our primary object of study is the four-point function of ${\bf 20'}$ operators. This operator is the superconformal primary of the stress tensor super-multiplet, so it is universally present in any local $\NN=4$ SCFT. This correlator has been studied extensively in $\NN=4$ super Yang-Mills theories, both at weak coupling (see \cite{Bourjaily:2015bpz} for a state-of-the-art computation, and references therein for further results) and at strong coupling in the planar limit using holography \cite{D'Hoker:1999pj}. It also sits at one of the three vertices of the Wilson loop/correlator/amplitude triangle of observables in planar $\NN=4$ super Yang-Mills \cite{Eden:2010zz,Alday:2010zy}. We will make contact with some of these results below, though our main results are nonperturbative and apply to the non-planar theories.

Superconformal invariance strongly constrains the structure of this correlation function and imposes selection rules that determine the superconformal blocks in terms of which the correlator will be decomposed. These constraints were solved in \cite{Dolan:2001tt} and we discuss them in detail below. We will take this opportunity to re-phrase them in light of the discovery of a protected subsector of the OPE algebra that is isomorphic to a chiral algebra \cite{Beem:2013sza}. The chiral algebra controls certain meromorphic functions that enter the solution of the superconformal Ward identities. The heart of this section is the detailed derivation of several results announced in \cite{Beem:2013qxa}, notably the unitarity bound $c\geqslant 3/4$ that is valid for any interacting $\NN=4$ SCFT and the bootstrap equation that controls the unprotected operator dimensions and OPE coefficients appearing in the ${\bf 20'}$ four-point function. Representation theory of the $\NN=4$ superconformal algebra plays a crucial role in this analysis, so we begin with a short review of shortening conditions for $\NN =4$ superconformal multiplets.

\subsection{Representations of the \texorpdfstring{$\NN=4$}{N=4} superconformal algebra}
\label{subsec:superconformal_reps}

The four-dimensional $\NN=4$ superconformal algebra is $\psuf(2,2|4)$ and its unitary irreducible highest-weight representations are described in detail in \cite{Dolan:2002zh}. These representations are traditionally labeled by the quantum numbers of their superconformal primary operator -- the scaling dimension $\Delta$, Lorentz spins $(j, \bar j)$, and $\suf(4)$ Dynkin labels $[d_1,d_2,d_3]$. Because we are studying the four-point function of scalar operators, we will mostly be concerned with symmetric traceless Lorentz tensors with $j = \bar j$ a half-integer. For such representations we will frequently define the spin $\ell \colonequals j + \bar j = 2j \in \Zb$.

If a superconformal primary is \emph{not} annihilated by any special combination of supercharges then its representation is a \emph{long} multiplet and is denoted $\AA^\D_{[d_1,d_2,d_3](j,\bar j)}$. The superconformal descendants in a long multiplet all have positive norm if and only if
\begin{equation}\label{eq:deltalongbound}
\Delta \geqslant \Delta_\AA \colonequals \max\left(2 + 2 j + \hf(3 d_1 + 2 d_2 + d_3), 2 + 2\bar j + \hf(d_1 + 2 d_2 + 3 d_3)\right)~.
\end{equation}
Alternatively, certain combinations of supercharges may annihilate a superconformal primary, in which case the multiplet is called \emph{short} or \emph{semishort}. There is one basic shortening and one basic semishortening condition (which takes an exceptional form for primaries with $j=0$), which are respectively given by
\begin{align}
b: &  & Q^i_\alpha |~[d_1,d_2,d_3]~(j,\bar j)~\rangle_{\text{h.w.}} &= 0 & \alpha = +,-\nn\\
c: & & \left(Q^i_- - \frac{1}{2 j + 1}J_{-}^{\ph{-}+} Q^i_+\right) |~[d_1,d_2,d_3]~(j,\bar j)~\rangle_{\text{h.w.}} &= 0~& j>0~,\,~~\\
   & & \varepsilon^{\alpha\beta}Q^i_\alpha Q^i_{\beta} |~[d_1,d_2,d_3]~(0,\bar j)~\rangle_{\text{h.w.}} &= 0~& j=0~.~~~\nn
\end{align}
The labels `$b$' and `$c$' follow the notation introduced in \cite{Dolan:2002zh}. For a given representation, these shortening conditions may hold for a range of values of $i =1, \dots, s$, in which case they are denoted $b^{\frac{s}{4}}$ and $c^{\frac{s}{4}}$. Similar conditions can be imposed for the action of the supercharges $\widetilde{Q}_{i\alpha}$ on the highest weight state, with $i = 4, \dots 4-\bar s$. We denote these by $\bar b^{\frac{\bar s}{4}}$ and $\bar c^{\frac{\bar s}{4}}$. A representation can obey both barred and unbarred shortening conditions, but not in every combination. The shortening conditions on the two sides are subject to compatibility conditions --- the complete list of possible representations are those displayed in Table~\ref{tab:irrepslist}, along with conjugate representations obtained by interchanging $j \leftrightarrow \bar j$ and $d_1 \leftrightarrow d_3$. As in \cite{Dolan:2002zh}, we denote representations by $\XX^{\frac{s}{4},\frac{\bar s}{4}}_{[d_1,d_2,d_3](j,\bar j)}$ with $\XX$ a letter from the first column of Table~\ref{tab:irrepslist} encoding the different shortening conditions.\footnote{A similar table appeared recently in the comprehensive work \cite{Cordova:2016emh}. Although their organization is slightly different, the resulting set of multiplets is the same.}

\begin{table}
\begin{tabular}{>{}r<{} >{}l<{} >{}l<{} >{}l<{} >{}l<{} >{}l<{} l} 
& type & $(s, \bar s)$ & $[d_1,d_2,d_3]$ & $(j,\bar j)$ & $\Delta$ & comments\\
\hline
$\AA$ & $(0,0)$ 	 & $(0,0)$ & $[d_1,d_2,d_3]$ & $(j,\bar j)$ & $\geqslant \Delta_{\AA}$ 		 & generic long\\
$b$   & $(b,0)$ 	 & $(1,0)$ & $[\underline{d_1},d_2,d_3]$ & $(0,\bar j)$ & $\hf(3 d_1 + 2 d_2 + d_3)$ 	 & $d_1 - d_3 > 2 + 2 \bar j$\\
$c$   & $(c,0)$ 	 & $(1,0)$ & $[\underline{d_1},d_2,d_3]$ & $(j,\bar j)$ & $2 + 2j + \hf (3d_1+2d_2+d_3)$ & $d_1 - d_3 > 2 (\bar j - j)$\\
      &       	 	 & $(2,0)$ & $[0,\underline{d_2},d_3]$ 	 & $(j,\bar j)$ & $2 + 2j + \hf (2 d_2 + d_3)$   & $ - d_3 > 2 (\bar j - j)$\\
      &       	 	 & $(3,0)$ & $[0,0,\underline{d_3}]$ 	 & $(j,\bar j)$ & $2 + 2j + \hf d_3$			 & $ - d_3 > 2 (\bar j - j)$\\
      &       	 	 & $(4,0)$ & $[0,0,0]$ 		 & $(j,\bar j)$ & $2 + 2j$ 						 & $j > \bar j$\\
$\BB$ & $(b,\bar b)$ & $(1,1)$ & $[\underline{d_1},d_2,\underline{d_1}]$ & $(0,0)$ 	  	& $d_2 + 2 d_1$ 				 & 1/4 BPS \\
 	  &  			 & $(2,2)$ & $[0,\underline{d_2},0]$	 & $(0,0)$	  	& $d_2$ 						 & 1/2 BPS\\
 	  &  			 & $(4,4)$ & $[0,0,0]$		 & $(0,0)$ 	  	& $0$ 							 & identity\\
$\CC$ & $(c,\bar c)$ & $(1,1)$ & $[d_1,d_2,\underline{d_3}]$ & $(j,\bar j)$ & $2 + j + \bar j +d_1+d_2+d_3$  & $d_1 - d_3 = 2(\bar j - j)$\\
 	  & 			 & $(1,2)$ & $[\underline{d_1},\underline{d_2},0]$ 	 & $(j,\bar j)$ & $2 + j + \bar j + d_1 + d_2$ 	 & $d_1 = 2 (\bar j - j)$\\
 	  & 			 & $(1,3)$ & $[\underline{d_1},0,0]$ 	 & $(j,\bar j)$ & $2 + j + \bar j + d_1$ 		 & $d_1 = 2 (\bar j - j)$\\
 	  & 			 & $(2,2)$ & $[0,\underline{d_2},0]$ 	 & $(j,j)$ 	  	& $2+ 2j + d_2$ 				 &\\
 	  & 			 & $(4,4)$ & $[0,0,0]$ 		 & $(j,j)$ 	  	& $2 + 2 j$ 					 & higher-spin currents\\
$\DD$ & $(b,\bar c)$ & $(1,1)$ & $[d_1,d_2,\underline{d_3}]$ & $(0,\bar j)$ & $1 + \bar j + d_1 + d_2 + d_3$ & $d_1 - d_3 = 2 + 2 \bar j$\\
 	  & 			 & $(1,2)$ & $[d_1,\underline{d_2},0]$ 	 & $(0,\bar j)$ & $1 + \bar j + d_1 + d_2$  	 & $d_1 = 2 + 2 \bar j$\\
 	  & 			 & $(1,3)$ & $[\underline{d_1},0,0]$ 	 & $(0,\bar j)$ & $1 + \bar j + d_1$ 			 & $d_1 = 2 + 2 \bar j$\\
\end{tabular}
\caption{\label{tab:irrepslist}The unitary irreducible representations of $\mathfrak{psu}(2,2|4)$. Underlined Dynkin labels must be nonzero. We omitted the CPT conjugate multiplets which can be obtained by interchanging $j \leftrightarrow \bar j$ and $d_1 \leftrightarrow d_3$.}
\end{table}

Let us point out a couple of important entries in the table. The maximally semishort multiplets are denoted by $\CC^{1,1}_{[0,0,0](j,j)}$. These semishort multiplets contain higher-spin conserved currents, which are the hallmark of free CFTs \cite{Maldacena:2011jn}. We will consequently demand that such multiplets are absent in our analysis of interacting $\NN=4$ SCFTs.
The maximally short multiplets, also known as half-BPS multiplets, are denoted by $\BB^{\hf,\hf}_{[0,p,0](0,0)}$. In particular, the stress tensor supermultiplet is such a representation with $p=2$. The superconformal primary is a dimension two scalar that is the lowest weight of the $\bf{20'}$ representation of $\suf(4)$ with Dynkin labels $[0,2,0]$. This is the two-index symmetric traceless representation of $\sof(6)$, so we will denote this operator schematically as $\OO_{\bf{20'}}^{IJ}$ with $I$, $J$ fundamental indices of $\sof(6)$. In $\NN=4$ SYM theories this operator is simply $\Tr(\Phi^{\{I} \Phi^{J\}})$ with $\Phi^I$ the elementary scalar fields of the theory. The superconformal descendants of $\OO_{\bf{20'}}^{IJ}$ include the supercurrents, $R$-symmetry currents, and the stress tensor.

\subsection{The \texorpdfstring{$20^\prime$}{20'} four-point function}
\label{subsec:20prime_four_point}

The four-point function of stress tensor multiplets can be analyzed in superspace and can be shown to allow only a single superconformally invariant structure \cite{Dolan:2004mu}. From this follows the remarkable fact that four-point functions of arbitrary combinations of superconformal descendants in this multiplet are determined completely in terms of the four-point function of the ``top-component'' scalars $\OO_{\bf{20'}}^{IJ}$. Without loss of information we can therefore restrict our analysis to the latter four-point function. From a bootstrap perspective, it is a huge simplification to be able to analyze the (complicated) four-point function of stress tensors in terms of a scalar four-point function. This is one benefit of applying bootstrap methods to superconformal field theories.

It is be convenient to contract the $\sof(6)$ indices on $\OO_{\bf{20'}}^{IJ}$ with a six-dimensional complex ``polarization vector'' $w_I$:
\begin{equation}
\OO(w,x) := w_I w_J \OO_{\bf{20'}}^{IJ} (x)~.
\end{equation}
We may demand that $w_I w^I = 0$ because $\delta_{IJ} \OO_{\bf{20'}}^{IJ} = 0$. The four-point function in question can then be written as follows,
\begin{equation}
\langle \OO(w_1,x_1)\OO(w_2,x_2)\OO(w_3,x_3)\OO(w_4,x_4)\rangle = \frac{1}{x_{12}^{4} x_{34}^4} \FF(w_{ij}; u,v)~,
\end{equation}
where the form of the right-hand side is dictated by conformal invariance. Here $u$ and $v$ denote the conformal cross ratios
\begin{equation}
\label{crossratios}
u \equiv  z \bar z \colonequals \frac{x_{12}^2 x_{34}^2}{x_{13}^2 x_{24}^2}~,\qquad \qquad v \equiv (1-z)(1-\bar z) \colonequals \frac{x_{14}^2 x_{23}^2}{x_{13}^2 x_{24}^2}\,.
\end{equation}
Global $\suf(4)$ invariance further implies that $\FF(w_{ij};u,v)$ is a function only of $w_{ij} \colonequals  w_i \cdot w_j$. Because by construction the correlator is a homogeneous polynomial of degree four in the $w_i$ with each vector appearing exactly twice, we can write
\begin{multline}
\FF(w_{ij};u,v) = w_{12}^2 w_{34}^2 a_1 (u,v) + w_{13}^2 w_{24}^2 u^2 a_2 (u,v) + w_{14}^2 w_{23}^2 u^2 v^{-2} a_3(u,v)\\
 + w_{12} w_{23} w_{34} w_{41} u v^{-1}c_1(u,v) + w_{13} w_{23} w_{24} w_{41} u^2 v^{-1} c_2(u,v) + w_{12} w_{24} w_{34} w_{31} u c_3(u,v)~.
\end{multline}
This leaves six functions of the conformal cross-ratios, where extra factors of $u$ and $v$ have been introduced to make the six functions $a_i(u,v)$ and $c_i(u,v)$ match those defined in \cite{Dolan:2001tt}.

The full four-point function is invariant under permutations of the four external operators. At the level of the six functions defined above, invariance upon interchanging the first and the second operator requires that
\begin{equation}
\label{eq:crossing12}
\begin{split}
&a_1(u,v) = a_1(u/v, 1/v) \qquad a_2(u,v) = a_3(u/v, 1/v)~,\\
&c_1(u,v) = c_3(u/v,1/v) \qquad c_2(u,v) = c_2(u/v,1/v)~,
\end{split}
\end{equation}
while interchanging the first and the third operator requires that
\begin{equation}
\label{eq:crossing13}
\begin{split}
&a_1(u,v) = a_3(v,u) \qquad a_2(u,v) = a_2(v,u)~,\\
&c_1(u,v) = c_1(v,u) \qquad c_2(u,v) = c_3(v,u)~.
\end{split}
\end{equation}
Additional permutations do not lead to any further constraints.

We have to this point imposed all of the constraints following from conformal symmetry, $\suf(4)$ invariance, and crossing symmetry on the structure of the four-point function. The constraints of \emph{super}conformal invariance impose further conditions which greatly simplify our numerical analysis. These constraints were analyzed in detail in \cite{Dolan:2001tt} and in the next few paragraphs we summarize their findings.

It is shown in \cite{Dolan:2001tt} that the six functions $a_i(u,v)$ and $c_i(u,v)$ can all be expressed in terms of a \emph{single} two-variable function $\GG(u,v)$, plus \emph{three} (meromorphic) single-variable functions $f_i(z)$. Explicitly, they find that
\begin{equation}
\label{suconfwardidsdo}
\begin{split}
c_1(z,\bar z) - \frac{1-z}{z} a_1(z,\bar z) - \frac{z}{1-z} a_3(z,\bar z) &\equalscolon f_1(z)~,\\
c_2(z,\bar z) + (1-z) a_2(z,\bar z) +\frac{1}{1-z} a_3(z,\bar z) & \equalscolon f_2(z)~,\\
c_3(z,\bar z) + z a_2(z,\bar z) + \frac{1}{z} a_1(z,\bar z) & \equalscolon f_3(z)~,
\end{split}
\end{equation}
and similarly with $z \leftrightarrow \bar z$. The only independent combination of the $a_i$ and $c_i$ not fixed by \eqref{eq:crossing13} and \eqref{suconfwardidsdo} in terms of the single variable functions is then given by
\begin{equation}
\label{eq:GGdefn}
a_2(z,\bar z) + \frac{1}{v^2}a_3(z,\bar z) + \frac{1}{v} c_2(z,\bar z) \equalscolon \GG(u,v)~.
\end{equation}
The $a_i$ and $c_i$ are all then completely determined in terms of the $f_i$ and $\GG$, subject to \eqref{eq:crossing12}.

\subsubsection{Bootstrapping the meromorphic functions}
\label{subsubsec:mero_bootstrap}

It was noted in \cite{Dolan:2001tt} that the meromorphic functions are protected by supersymmetry and can therefore be fixed by their free-field values in super Yang-Mills theories. Their appearance was understood to be connected to a much larger physical picture in \cite{Beem:2013sza}, where it was shown that they themselves are correlators of an auxiliary chiral algebra. The existence of this chiral algebra follows generally from $\NN=2$ superconformal invariance. It will consequently be useful to reconsider our four-point function in terms of the $\NN = 2$ superconformal algebra $\suf(2,2|2)$.

From this perspective, the $\suf(4)_R$ symmetry splits into the $\suf(2)_R\times \uf(1)_r$ of the $\NN = 2$ superconformal algebra, along with an additional $\suf(2)_F$ flavor symmetry. In the $\sof(6)$ language this amounts to the natural split $\sof(6) \to \sof(4) \times \sof(2)$, with $\sof(4) \cong \suf(2)_R \times \suf(2)_F$. The $\bf{6}$ of $\sof(6)$ decomposes as the $\bf4$ of $\sof(4)$ and a $\bf2$ of $\sof(2)$, and the first of these can be interpreted as the $(\bf2,\bf2)_0$ of $\suf(2)_R \times \suf(2)_F \times \uf(1)$. The $\bf{20'}$ decomposes according to
\begin{equation}
\bf{20'} \to (\bf3,\bf3)_0 + (\bf 2,\bf2)_{+1} +  (\bf 2,\bf2)_{-1} + (\bf 1,\bf1 )_{+2}  +(\bf 1, \bf1 )_{-2} +  (\bf 1,\bf1 )_{0}~.
\end{equation}
The relevant term for our purposes is the $(\bf3,\bf3)_0$. The operator in this representation is the superconformal primary of an $\NN =2$ $\suf(2)_F$ flavor current multiplet. In the language of \cite{Beem:2013sza} it is a \emph{Schur} operator, and after the appropriate superconformal twist it becomes an affine current $j^A(z)$, with $A$ an adjoint index of $\suf(2)_F$. Its OPE in the chiral algebra is standard,
\begin{equation}
j^A(z) j^B(0) \sim \frac{k_{2d}\delta^{AB}}{z^2} + \frac{i f^{AB}_{\phantom{AB}C} j^C(0)}{z}~,
\end{equation}
where the level $k_{2d} = - \hf k_{4d}$, and by $\NN=4$ supersymmetry the four-dimensional level $k_{4d} = 4 a = 4 c$ with $a=c$ the central charges of the theory. The singular terms in the OPE completely fix the current four-point function,
\begin{multline}
\label{4ptcurrents}
z_{12}^2 z_{34}^2 \langle j^A(z_1) j^B(z_2) j^C(z_3) j^D(z_4) \rangle =\\   \d^{AB} \d^{CD} + z^2 \d^{AC} \d^{BD} + \frac{z^2}{(1-z)^2} \d^{AD} \d^{BC} - \frac{z}{k_{2d}} f^{ACE}f^{BDE} - \frac{z}{k_{2d}(z-1)} f^{ADE} f^{BCE}~,
\end{multline}
where $z \colonequals z_{12} z_{34}/z_{13} z_{24}$ the holomorphic cross-ratio, and we have rescaled the currents $j^A(z)$ so that they have unit norm.

In terms of the polarization vectors $w^I$ we must recover this four-point function by first replacing $w^I \to w^\mu$ for $I \in \{1,2,3,4\}$ and setting $w^5$ and $w^6$ to zero. The $w^\mu$ can then be split as $w^\mu = \sigma^\mu_{\a \dot \a} v^\a \tilde v^{\dot \a}$ where we adopt the convention that dotted (undotted) indices are for $\suf(2)_R$ ($\suf(2)_F$). Altogether this amounts to
\begin{equation} \label{wtov}
w_{ij} \to (v_i \cdot v_j) (\tilde v_i \cdot \tilde v_j)~,
\end{equation}
where $v_i \cdot v_j = \e_{\a \b} v_i^\a v_j^\b$. The superconformal twist of \cite{Beem:2013sza} then proceeds as follows. We restrict all four $x_i$ to lie in a two-plane, where we introduce complex coordinates $(z_i,\bar z_i)$. We then set $\tilde v_i = (1,\bar z_i)$. The resulting correlator must be \emph{meromorphic} in its positions --- \ie, the dependence on the $\bar z_i$ drops out --- and in fact should match the affine current four-point function above. 

For the correlation function at hand this is easily verified. In complex coordinates, and after the substition \eqref{wtov}, the correlator becomes
\begin{equation}
\frac{1}{|z_{12}|^4 |z_{34}|^4} \FF(v_{ij}, \tilde v_{ij}; z,\bar z)~.
\end{equation}
The superconformal twist is implemented by setting $\tilde v_{ij} \to \bar z_{ij}$. If we also recall that two-dimensional vectors obey the following identity
\begin{equation}
(v_1 \cdot v_2) (v_2 \cdot v_3) (v_3 \cdot v_4) (v_4 \cdot v_1) = \hf \left( (v_1 \cdot v_2)^2 (v_3 \cdot v_4)^2 + (v_1 \cdot v_4)^2 (v_2 \cdot v_3)^2 - (v_1 \cdot v_3)^2 (v_2 \cdot v_4)^2 \right)~,
\end{equation}
then some simple algebraic manipulations lead to the following expression for the four-point function,
\begin{equation} \label{eq:twisted4ptfn}
\begin{split}
\frac{1}{|z_{12}|^4 |z_{34}|^4} \FF(v_{ij}, \bar z_{ij}; z,\bar z) = &\frac{1}{z_{12}^2 z_{34}^2} \left(\frac{z}{2(z-1)}\right)\\
\times \Big( 
&v_{12}^2 v_{34}^2 \left(\phantom{-} c_1(z,\zb)+z c_2(z,\zb) - (1-z)c_3(z,\zb) -\tfrac{2(1-z)}{z}a_1(z,\zb)\right)\\ 
&\!\!\!\!\!\!\!+ 	v_{13}^2 v_{24}^2 \Big(		    -  c_1(z,\zb)-z c_2(z,\zb) - (1-z)c_3(z,\zb) -{2z(1-z)}a_2(z,\zb)\Big)\\
&\!\!\!\!\!\!\!+ 	v_{14}^2 v_{23}^2 \left(\phantom{-} c_1(z,\zb)-z c_2(z,\zb) + (1-z)c_3(z,\zb) -\tfrac{2z}{1-z}a_3(z,\zb)\right)\Big)~.
\end{split}
\end{equation}
We observe that after some reorganization, demanding meromorphicity of this correlator precisely reproduces the superconformal ward identities of \eqref{suconfwardidsdo}. We can compare this equation with Eqn.~\eqref{4ptcurrents} and, using also that $f^{ACE}f^{BDE} = 2 (\d^{AB} \d^{CD} - \d^{AD} \d^{BC})$ for $\suf(2)$ and the conventions for the meromorphic functions introduced in \eqref{suconfwardidsdo}, we can find that
\begin{equation}
\label{eq:fi_solutions}
\begin{split}
f_1(z) &= -\frac{1}{z}+\frac{1}{z-1}+2-\frac{2}{k_{2d}}~,\\
f_2(z) &= -z+\frac{1}{1-z}+1-\frac{2}{k_{2d}}~,\\
f_3(z) &= z+\frac{1}{z}-\frac{2}{k_{2d}}~.
\end{split}
\end{equation}
This agrees with the results of \cite{Dolan:2001tt}, where the variables $c$ and $a$ in that paper are related to ours according to $c_{\rm DO}=-2/k_{2d}$ and $a_{\rm DO} = 1$. (Note also that $k_{\rm DO} = f_1(z) + f_2(z) + f_3(z) = 3(1 -\frac{2}{k_{2d}})$.) We emphasize that the derivation here does not use any assumption about the existence of a weak-coupling limit described by super Yang-Mills theory in order to arrive at an explicit expression for the meromorphic functions, so this is a completely general input for the bootstrap program for $\NN=4$ SCFTs.

In what follows we will consider the conformal block decomposition of the four-point function according to the operator product expansion in the channel where operators at $x_1$ and $x_2$ are fused. The orthonormal $\suf(2)_F$ polarization vector structures for the singlet, triplet, and quintuplet in that channel are given by
\begin{equation}
P^{\bf1}(v_{ij}) = \frac{1}{3} v_{12}^2 v_{34}^2,\quad P^{\bf3}(v_{ij}) = \frac{1}{2} ( v_{14}^2 v_{23}^2 - v_{13}^2 v_{24}^2 ), \quad P^{\bf5}(v_{ij}) = \frac{1}{2}( v_{13}^2 v_{24}^2 + v_{14}^2 v_{23}^2) - \frac{1}{3} v_{12}^2 v_{34}^2\,.
\end{equation}
The contributions of the singlet, triplet, and quintuplet operators in the chiral algebra are then captured in the following basis for the meromorphic functions,
\begin{equation}
\begin{split}
\hat h_{\bf 1}(z) &\colonequals \frac{z}{2 (z-1)} \left( 3 f_1(z) +  z f_2(z) - 3 (1-z) f_3(z) \right)~,\\
\hat h_{\bf 3}(z) &\colonequals \frac{z}{2 (z-1)} \left( 2 f_1(z)  + 2 (1-z) f_3(z) \right)~,\\
\hat h_{\bf 5}(z) &\colonequals \frac{z}{2 (z-1)} \left(- 2 z f_2(z) \right)~.
\end{split}
\end{equation}
In terms of the functions introduced in \cite{Dolan:2001tt} we have $\hat h_{\bf 3}(z) \equiv - \tilde f(z)$ and $\hat h_{\bf 5}(z) \equiv \tilde f_2 (z)$. 

Notice that the $\NN=4$ superconformal Ward identities dictate that $\hat h_1(z) = 3 k - 3 \left(\frac{1}{2}-\frac{1}{z}\right) \hat h_3(z)+ \left(\frac{3(z-1)}{z^2}-\frac{1}{2}\right) \hat h_5(z)$ where $k$ is a constant defined in \cite{Dolan:2001tt}. This identity should hold at the level of the individual superconformal blocks, and from the expressions below one may indeed verify that $\hat h_1(z)$ is a finite sum over blocks if $\hat h_3(z)$ and $\hat h_5(z)$ are.

\subsection{Superconformal block decomposition}
\label{subsec:block_decomposition}

It is a direct consequence of (super)conformal invariance that any four-point function can be decomposed into (super)conformal blocks that correspond to the exchange of all operators belonging to an irreducible representation of the (super)conformal algebra. Here we will consider the decomposition into superconformal blocks that arises by fusing together the operators at $x_1$ and $x_2$.

The set of superconformal representations that can appear in the OPE of two $\bf{20'}$ operators is restricted by superconformal invariance. Schematically we have \cite{Eden:2001ec,Nirschl:2004pa}
\begin{equation}
\begin{split}
\BB^{\hf,\hf}_{[0,2,0](0,0)} \times \BB^{\hf,\hf}_{[0,2,0](0,0)} \to
\mathbf 1 &+  \BB^{\hf,\hf}_{[0,2,0](0,0)} + \BB^{\hf,\hf}_{[0,4,0](0,0)} + \mathcal B^{\qt,\qt}_{[2,0,2](0,0)} \\ 
&+ \sum_{\ell = 0}^{\infty} \mathcal C^{1,1}_{[0,0,0](j,j)} + \sum_{\ell=0}^{\infty}  \mathcal C^{\hf,\hf}_{[0,2,0](j,j)} + \sum_{\ell = 0}^{\infty} \ \mathcal C^{\qt,\qt}_{[1,0,1](j,j)} + \sum_{\Delta,\ell} \AA^{\Delta}_{[0,0,0](j,j)}\,.
\end{split}
\end{equation}
We will see below that Bose symmetry dictates that all the spins $\ell=2j$ appearing in this equation are even.

Let us make a few preliminary comments concerning this OPE. First of all, the stress tensor multiplet itself appears, with a coefficient that is fixed to be $2/c$ if the operators are unit normalized (see below). The presence of another half-BPS multiplet of type $\BB^{\hf,\hf}_{[0,4,0](0,0)}$ is in agreement with expectations from the half-BPS chiral ring. The other short multiplets are quarter BPS and semishort. They are not of specific significance to us, with the exception of the multiplets of type $\CC^{1,1}_{[0,0,0](j,j)}$: these contain higher-spin currents and should not appear in an interacting theory \cite{Maldacena:2011jn}. Finally, the only long multiplets that can appear are of type $\AA^{\Delta}_{[0,0,0](j,j)}$, with a superconformal primary operator that is an $R$ symmetry singlet and has even Lorentz spin.\footnote{Simple examples of such operators in the $\NN=4$ SYM theories are the Konishi operator $\text{Tr}(\Phi^I\Phi_I)$ and double trace operators of the form $\OO_{\bf{20'}}^{IJ} \partial^{2n}\OO_{\bf{20'}}^{IJ}$.} Eqn.~\eqref{eq:deltalongbound} reduces to
\begin{equation}
\Delta \geqslant 2 + \ell~,
\end{equation}
for these particular multiplets. 

In the remainder of this section we will derive the superconformal blocks for all the multiplets listed above, and show that the chiral algebra essentially determines all of the OPE coefficients for short and semishort multiplets.

\subsubsection{Superconformal blocks}
\label{subsubsec:superblocks}

As a first step towards the superconformal block decomposition let us consider the global $\suf(4)_R$ symmetry. In the tensor product of two $\bf{20'}$ representations we find
\begin{equation}
\begin{split}
(\bf{20'}\otimes\bf{20'})_s &= \bf1 \oplus \bf{20'} \oplus \bf{84} \oplus \bf{105}~, \\
(\bf{20'}\otimes\bf{20'})_a &= \bf{15} \oplus \bf{175}\,.
\end{split}
\end{equation}
 
Using explicit projection tensors for these irreducible representations one may determine the precise combinations of the $a_i(u,v)$ and $c_i(u,v)$ that encode the contributions of operators in a given representation. In particular, we have
\begin{alignat}{2}
&\bf1 &:\qquad  &20 a_1 + u^2 a_2 + \frac{u^2}{v^2} a_3 + \frac{10}{3}\left(\frac{u}{v} c_1 + u c_3 \right) + \frac{u^2}{3v} c_2~, \nonumber \\
&\bf{20'} &:\qquad & u^2 a_2 + \frac{u^2}{v^2} a_3 + \frac{5}{3}\left( \frac{u}{v} c_1 + u c_3\right) + \frac{u^2}{6v} c_2~,\nonumber \\
&\bf{84'} &:\qquad &u^2 a_2 + \frac{u^2}{v^2} a_3 - \frac{u^2}{2v} c_2~, \label{eq:rsychannels}\\
&\bf{105'} &:\qquad &u^2 a_2 + \frac{u^2}{v^2} a_3 + \frac{u^2}{v} c_2~,\nonumber \\
&\bf{15'} &:\qquad &u^2 a_2 - \frac{u^2}{v^2} a_3 - 2 \left( \frac{u}{v} c_1 - u c_3\right)~,\nonumber \\
&\bf{175} &:\qquad &u^2 a_2 - \frac{u^2}{v^2} a_3~.\nonumber
\end{alignat}
These equations may be rewritten in terms of the functions $f_i(z)$ and $\GG(u,v)$. The expressions are not particularly illuminating, with the exception of
\begin{alignat*}{2}
&\bf{105'} &:\qquad &u^2 \GG(u,v)~,
\end{alignat*}
which follows directly from \eqref{eq:GGdefn}.

In general the superconformal block decomposition of four-point functions is fairly involved because of the different superconformal multiplets that can appear in the OPE. Correspondingly, one finds in the literature a variety of methods for computing the superconformal blocks, with no clear winner amongst them. Instead each method is more or less suitable depending on the spacetime dimension, number of supercharges, and possible shortening conditions of the external and the exchanged operators. For our particular four-point function the authors of \cite{Dolan:2001tt} found a remarkable shortcut that we will explain in this subsection.

First define the following \emph{atomic} functions,
\begin{equation}
\begin{split}
\label{eq:atomic_blocks}
h^{(at)}(\beta,z) &\colonequals (-\tfrac{1}{2} z)^{\beta-1} z F[\beta,z]\\
\GG^{(at)}_{\Delta,\ell}(z,\bar z) &\colonequals u^{-2} G_{\Delta + 4}^{(\ell)}(z,\bar z) \\&\colonequals u^{-2} \left( \frac{u^{\hf(\Delta + 4 - \ell)}}{z - \bar z} \left( (-\tfrac12 z)^\ell z F[\tfrac12 (\Delta + 4+ \ell),z] F[\tfrac12(\Delta + 2 - \ell),\zb] - z \leftrightarrow \zb \right) \right)
\end{split}
\end{equation}
with
\begin{equation}
\label{eq:hypergeometric_atomic}
F[\alpha,z]\colonequals {}_2 F_1 (\alpha,\alpha,2\alpha,z)~.
\end{equation}
Here $G_\Delta^{(\ell)}(z,\bar z)$ is an ordinary four-dimensional conformal block corresponding to an conformal primary with scaling dimension $\Delta$ and spin $\ell$ being exchanged in a four-point function of identical scalars. Similarly $z^{\beta}F[\beta,z]$ is an ordinary $\slf(2)$ block, which accounts for the exchange of an operator and all of its holomorphic derivatives.

Next we use the fact that every unitary irreducible representation of the superconformal algebra contains a finite number of representations of the underlying (bosonic) conformal algebra. Correspondingly, each superconformal block decomposes into a finite sum of ordinary conformal blocks in the different $R$-symmetry channels. Furthermore, from the structure of the superconformal twist described above one sees that a superconformal multiplet contributes at most a finite number of $\slf(2)$ blocks to the $\hat{h}_{\bf R}(z)$. It follows that for each superconformal block, the contributions to $h_i(z)$ and $\GG(z,\bar z)$ are finite sums of the functions $h^{(at)}(\beta,z)$ and $\GG^{(at)}_{\Delta,\ell}(z,\bar z)$, respectively. (This explains our nomenclature: in a sense, these atomic functions are the minimal contributions in the superconformal block decomposition.)

The complete analysis of the superconformal blocks has been carried out in this way in \cite{Dolan:2001tt}. Most remarkably, in almost all cases the superconformal block contributes a \emph{single} one of the $\GG^{(at)}_{\Delta,\ell}(z,\bar z)$ atomic blocks to $\GG(u,v)$. In Table~\ref{tab:superblocks} we display the details of the superconformal blocks and their decomposition into atomic functions for all of the relevant supermultiplets. It was verified explicitly in \cite{Dolan:2001tt} that each of these superconformal blocks can be written as a finite sum of ordinary conformal blocks in each of the six $R$-symmetry channels listed in \eqref{eq:rsychannels}. 

\begin{table}
\begin{center}
\renewcommand{\arraystretch}{1.5}
\begin{tabular}{>{}l<{} >{}l<{}  l >{}l<{} >{}l<{}}
$\text{type}$   & $\GG(z,\bar z)$  & $\hat h_1(z)$ & $\hat h_3(z)$ & $\hat h_5(z)$\\
\hline
\hline
$\bf 1$  & $0$  & $3$ & $0$ & $0$\\
\hline
$\mathcal A^{\Delta}_{[0,0,0](j,j)}$ & $\GG^{(at)}_{\Delta,\ell}(z,\bar z)$  & $0$ & $0$ & $0$ \\
\hline
$\mathcal C^{1,1}_{[0,0,0](j,j)}$ & $0$  & $\begin{array}{l}\frac{3}{2} h^{(at)}(\ell + 2,z) \\ \,\, + \frac{3}{2} d_\ell h^{(at)}(\ell + 4,z)\end{array}$ & $- h^{(at)}(\ell+3,z)$  & $0$ \\
\hline
$\mathcal B^{\hf,\hf}_{[0,2,0](0,0)}$ & $0$ & $\frac{1}{2}h^{(at)}(2,z)$ & $-h^{(at)}(1,z)$ & $0$ \\
\hline
$\mathcal C^{\qt,\qt}_{[1,0,1](j,j)}$ & $\GG^{(at)}_{\ell + 4,\ell + 2}(z,\bar z)$ & 
$\begin{array}{l}6 h^{(at)}(\ell + 4,z) \\ \,\, + 6 d_{\ell + 2} h^{(at)}(\ell + 6,z)\end{array}$ & $4 h^{(at)}(\ell+5,z)$ & $0$\\
\hline
$\mathcal B^{\qt,\qt}_{[2,0,2](0,0)}$ & $\GG^{(at)}_{(2,0)}(z,\bar z)$ & $\begin{array}{l}-6 h^{(at)}(2,z)\\ \, \,  - 6 d_0 h^{(at)}(4,z) \end{array}$ & $4h^{(at)}(3,z)$ & $0$\\
\hline
$\mathcal C^{\hf,\hf}_{[0,2,0](j,j)}$ & $\GG^{(at)}_{(\ell+2,\ell+2)}(z,\bar z)$ & $- \frac{1}{2} h^{(at)}(\ell + 4,z)$ & $\begin{array}{l} h^{(at)}(\ell+3,z) \\ \,\,  + d_{\ell + 1} h^{(at)}(\ell+5,z) \end{array}$ & $-2 h^{(at)}(\ell + 4,z)$\\
\hline
$\mathcal B^{\hf,\hf}_{[0,4,0](0,0)}$ & $\GG^{(at)}_{(0,0)}(z,\bar z)$ & $0$ & $d_{-1} h^{(at)}(3,z)$ & $-2h^{(at)}(2,z)$
\end{tabular}
\renewcommand{\arraystretch}{1}
\end{center}
\caption{\label{tab:superblocks}Superconformal blocks. In this table $d_\ell \colonequals (\ell+3)^2/((2\ell + 5)(2\ell + 7))$ and $j = \ell/2$ is a nonnegative integer. Notice that there is a typo in Eqn.~(8.31) in \cite{Dolan:2001tt}, but the correct expression can be read off from Eqn.~(8.21) in that paper.}
\end{table}

\subsubsection{OPE coefficients}
The $\hat{h}_{\bf R}(z)$ are fixed by the chiral algebra and correspondingly we know their exact decomposition into atomic parts. They are given as follows:
\begin{equation}
\begin{split}
\hat h_1(z) &= 3 + \sum_{n=0}^\infty \left( - \frac{4^{n+1} (2 n+1) (2 n+2)! (2 n+1)!}{(4 n+2)!}+ \frac{(2n)!(2n+1)! 4^{n+1}}{c(4n+1)!} \right) h^{(at)}(2n + 2,z)~, \\
\hat h_3(z) &= \sum_{n=0}^\infty \left(\frac{4^n (2 n+1)! (2 n)!}{(4 n-1)!} - \frac{4^{n+1} ((2 n)!)^2}{2 c (4 n)!} \right) h^{(at)}(2 n+1,z)~,\\
\hat h_5(z) &= \sum_{n=0}^{\infty} \left( - \frac{4^{n+1} (2 n+1) (2 n+2)! (2 n+1)!}{(4 n+2)!}- \frac{(2n)!(2n+1)! 4^{n+1}}{2 c(4n+1)!} \right) h^{(at)}(2n + 2,z)~,
\end{split}
\end{equation}
where we made the replacement $k_{2d} = -2c$. From Table~\ref{tab:superblocks}, we see that the OPE coefficients (squared) for the superconformal blocks for almost all multiplets that participate in the chiral algebra are fixed by the chiral algebra correlator. The list of OPE coefficients after imposing chiral algebra constraints is displayed in Table~\ref{tab:superopecoeffs}.

We observe that there are two families of undetermined coefficients. The first series are the $a_{\Delta,\ell}$ for the long multiplets, which have no direct connection to the chiral algebra. The second series are the coefficients $\alpha_\ell$ of the $\CC^{1,1}_{[0,0,0](j,j)}$ multiplets, which are also not fixed by the chiral algebra. This ambiguity arises because the $\hat{h}_{\rm R}(z)$ do not change under the simultaneous addition of these superconformal blocks and a superconformal block of type $\BB^{\qt,\qt}_{[2,0,2](0,0)}$ or $\CC^{\qt,\qt}_{[1,0,1](j,j)}$ (with relative coefficient 4), as is evident by inspection of Table~\ref{tab:superblocks}. Underlying this pattern of undetermined OPE coefficients are two recombination rules for the $\NN=4$ superconformal algebra \cite{Dolan:2002zh},
\begin{equation}
\begin{split}
\AA^{2}_{[0,0,0](0,0)} &\simeq \CC^{1,1}_{[0,0,0](0,0)} \oplus \BB^{\qt,\qt}_{[2,0,2](0,0)} \oplus \DD^{\qt,\frac{3}{4}}_{[2,0,0](0,0)}\oplus \DD^{\frac{3}{4},\qt}_{[0,0,2](0,0)}~,\\
\AA^{2 + 2j}_{[0,0,0](j,j)} &\simeq \CC^{1,1}_{[0,0,0](j,j)} \oplus \CC^{\qt,\qt}_{[1,0,1](j - \hf,j - \hf)} \oplus \CC^{\qt,\frac{3}{4}}_{[1,0,0](j - \hf,j)}\oplus \CC^{\frac{3}{4},\qt}_{[0,0,1](j,j - \hf)}~.
\end{split}
\end{equation}
These rules imply that the long multiplets at threshold decompose into four short multiplets, out of which only the first two multiplets make an appearance in the four-point function in question. For this reason the exchange of this set of operators can be captured in two ways: either by the blocks of two short multiplets, or by the block of a single long multiplet at threshold. Since a single long multiplet cannot contribute to the chiral algebra part of the correlator, the contributions of short multiplets that are indistinguishable from a long multiplet must cancel from the chiral algebra.

Fortunately in the case at hand it is straightforward to fix the $\alpha_\ell$. The $\CC^{1,1}_{[0,0,0](j,j)}$ multiplets contain higher-spin conserved currents for all $\ell$, and we do not expect these to be present in an interacting theory.\footnote{In the free SYM theories the higher-spin current multiplet of type $\CC^{1,1}_{[0,0,0](0,0)}$ is the Konishi multiplet. This is lifted from the spectrum of (semi-)short multiplets at nonzero coupling.} For this reason we are tempted to set the $\alpha_\ell$ to zero. A close inspection of Table~\ref{tab:superopecoeffs} then reveals a surprise: if $c < 3/4$ then the coefficient of the $\BB^{\qt,\qt}_{[2,0,2](0,0)}$ multiplet would become \emph{negative} for $\alpha_0 = 0$, in violation of unitarity. Therefore, in those cases $\alpha_0$ must be nonzero and we conclude that \emph{all unitary $\NN = 4$ theories with $c < 3/4$ necessarily contain higher-spin conserved currents}. We may rephrase this result as a unitarity bound: $c \geqslant 3/4$ for any  interacting $\NN=4$ theory. The threshold value $c = 3/4$ is precisely the smallest possible value in the list of $\NN=4$ SYM theories, corresponding to the choice of $\suf(2)$ gauge algebra. Exotic $\NN =4$ theories with $c < 3/4$ are ruled out.

\begin{table}
\begin{center}
\renewcommand{\arraystretch}{1.8}
\begin{tabular}{>{}l<{} >{}l<{}}
$\text{type} $  & $\text{coefficient}$\\
\hline
\hline
$\bf 1  $& $1$\\
\hline
$\mathcal A^{\Delta}_{[0,0,0](j,j)} $& $a_{\Delta,\ell}$\\
\hline
$\mathcal C^{1,1}_{[0,0,0](j,j)} $& $\alpha_\ell$\\
\hline
$\mathcal B^{\hf,\hf}_{[0,2,0](0,0)}$  &$ \frac{2}{c} $\\
\hline
$\mathcal C^{\qt,\qt}_{[1,0,1](j,j)}$ &$\frac{2^{\ell+1}(\ell+3)!(\ell+4)!}{(2 \ell+7)!}\left((\ell+3) (\ell+6)-\frac{3}{c}\right) + \qt \alpha_{\ell + 2}$\\
\hline
$\mathcal B^{\qt,\qt}_{[2,0,2](0,0)}$ &$  \frac{2}{3} - \frac{1}{2c} + \frac{1}{4} \alpha_0 $\\
\hline
$\mathcal C^{\hf,\hf}_{[0,2,0](j,j)}$ &$\frac{2^{\ell +2} (\ell +2)! (\ell +3)!}{(2 \ell +5)!}\left((\ell+3)(\ell+4)+ \frac{1}{c}\right) $\\
\hline
$\mathcal B^{\hf,\hf}_{[0,4,0](0,0)}$ &$ 2 + \frac{1}{c}$
\end{tabular}
\renewcommand{\arraystretch}{1}
\end{center}
\caption{\label{tab:superopecoeffs}Coefficients of the superconformal blocks. The normalization of the superconformal blocks follows from Table~\ref{tab:superblocks}. As before $j = \ell/2$ is a nonnegative integer. The coefficients of the $\mathcal C^{\qt,\qt}_{[1,0,1](j,j)}$ (with $\alpha_{\ell + 2} = 0$) and of the $\mathcal C^{\hf,\hf}_{[0,2,0](j,j)}$ multiplets are respectively denoted $e_{\ell+2}$ and $d_{\ell+2}$ in equation (8.36) in \cite{Dolan:2001tt}.}
\end{table}

In pursuing a numerical bootstrap approach we should make no further assumptions beyond unitarity, crossing symmetry and (if possible) the absence of higher-spin currents. This leads to the following strategies.

$\mathbf{c \geqslant 3/4}$: We assume that there are no higher-spin currents. Correspondingly we set all the $\alpha_\ell$ to zero. We could in theory also demand absence of all the $\AA^{2+2j}_{[0,0,0](j,j)}$ multiplets in the numerical setup. This is however infeasible in practice, because we cannot exclude long multiplets just above the unitarity bound (\ie, multiplets of the form $\AA^{2+2j + \epsilon}_{[0,0,0](j,j)}$ with arbitrarily small $\epsilon$), and the numerics cannot reliably distinguish between the two.

$\mathbf{c < 3/4}$: In this case unitarity dictates that $\alpha_0$ has to be nonzero. We therefore set it to its lowest possible value as dictated by unitarity, namely $\alpha_0 = \frac{2}{c} - \frac{8}{3}$. The coefficient of the $\BB^{\qt,\qt}_{[2,0,2](0,0)}$ block is then precisely zero. We set to zero all the other $\alpha_\ell$ with $\ell > 0$. In this way any further contributions of higher-spin currents to the four-point functions can be written as $\AA^{2+2j}_{[0,0,0](j,j)}$ multiplets, and the $a_{\Delta,\ell} \geqslant 0$ for every unitary solution of the crossing symmetry constraints.

Of course the theory with higher-spin currents is expected to be free and thus solvable with analytic methods. We have nevertheless performed a numerical analysis also for $c < 3/4$, both for completeness and as a nontrivial consistency check of our methods.

From Table~\ref{tab:superblocks} it is clear that $\GG(u,v)$ receives contributions from both long and (semi-)short multiplets. We can therefore split split it into two parts,
\begin{equation}
\GG(u,v) = \GG^{\text{long}}(u,v) + \GG^{\text{short}}(u,v;c)~,
\end{equation}
with
\begin{equation}
\begin{split}
\GG^{\text{long}} (u,v) &= \sum_{\Delta,\ell}a_{\Delta,\ell} \GG^{(at)}_{\Delta,\ell}(z,\bar z)\,,\\
\GG^{\text{short}}(z,\bar z;c) &= \lambda^2[\BB^{\hf,\hf}_{[0,4,0](0,0)}] \GG^{(at)}_{0,0}(z,\bar z)\\
&+ \lambda^2[\BB^{\qt,\qt}_{[2,0,2](0,0)}] \GG^{(at)}_{2,0}(z,\bar z)\\
&+ \sum_{j=0}^\infty \lambda^2[\CC^{\hf,\hf}_{[0,2,0](j,j)}] \GG^{(at)}_{\ell+2,\ell+2}(z,\bar z)\\
&+ \sum_{j=0}^\infty \lambda^2[\CC^{\qt,\qt}_{[1,0,1](j,j)}] \GG^{(at)}_{\ell+4,\ell+2}(z,\bar z)\,.
\end{split}
\label{eq:Gshortblocks}
\end{equation}
The coefficients $\lambda^2[\cdot]$ are those given in Table~\ref{tab:superopecoeffs} with the $\alpha_\ell$ fixed in accordance with the above discussion. This expression highlights the utility of the chiral algebra in determining the contributions of short multiplets to the four-point function, including OPE coefficients.

Happily, Eqn.~\eqref{eq:Gshortblocks} can be explicitly summed. For $c \geqslant 3/4$ we find
\begin{equation}
\begin{split}
&\GG^{\text{short}}(z,\bar z) = \\
&\quad\frac{6}{z\zb}\left(\frac{4-2z\zb(z^2+z\zb+\zb^2-4)+(z+\zb)(z^2+z^2\zb^2+\zb^2-6)}{(1-z)^2(1-\zb)^2}+\frac{3}{c}\left(1+\frac{1}{(1-z)(1-\zb)}\right)\right)\\
&+\frac{2\log(1-z)}{(1-\zb)}\left(\frac{-3(\zb^4-6\zb^2+4\zb)+z(2\zb^4-\zb^3+4\zb^2-18\zb+12)}{(z-\zb)z^2\zb(1-\zb)}-\frac{1}{c}\left(\frac{9\zb-18}{z^2\zb}-\frac{4}{z}+\frac{4}{(z-\zb)}\right)\right)\\
&+\frac{2\log(1-\zb)}{(1-z)}\left(\frac{3(z^4-6z^2+4z)-\zb(2z^4-z^3+4z^2-18z+12)}{(z-\zb)z\zb^2(1-z)}-\frac{1}{c}\left(\frac{9z-18}{z\zb^2}-\frac{4}{\zb}-\frac{4}{(z-\zb)}\right)\right)\\
&+\frac{12\log(1-z)\log(1-\zb)}{z^2\zb^2}\left(2+\frac{3}{c}\right)\,.
\end{split}
\end{equation}
For $c < 3/4$ the nonzero $\alpha_0$ just adds a single block. An explicit expression of this type is convenient but not necessary for the numerical analysis below. The expression given in \eqref{eq:Gshortblocks} as an infinite sum can be used to numerically approximate $\GG^{\text{short}}(z,\bar z;c)$ and its derivatives to any desired precision by summing a large but finite number of terms.

\subsection{Crossing symmetry equations}
\label{subsec:crossing_equation}

We are now ready to revisit the crossing symmetry equations \eqref{eq:crossing12} and \eqref{eq:crossing13}. The first of these has the simple consequence that the spins $\ell$ in the conformal block decomposition are all even integers. The second equation is nontrivial and will be the subject of our numerical analysis. Using \eqref{suconfwardidsdo} and \eqref{eq:GGdefn} we find that the equations \eqref{eq:crossing13} are equivalent to crossing symmetry for the chiral algebra four-point function plus an additional constraint on the two-variable function $\GG(u,v)$ given by
\begin{equation}
v^2 \GG(u,v) - u^2 \GG(v,u) + (u^2 - v^2) + \frac{(u-v)}{c} = 0~.
\end{equation}
In this expression we have already substituted the solutions for the $f_i(z)$ given in \eqref{eq:fi_solutions}.

With $\GG(u,v)$ split into short and long parts, the fundamental crossing symmetry equation can be written as
\begin{equation}
\label{eq:fundamentalcrossing}
\sum_{\Delta,\ell} a_{\Delta,\ell} F^{(at)}_{\Delta,\ell} (u,v) - F^{\text{short}}(u,v;c) = 0~,
\end{equation}
with\footnote{The factors of $(z-\zb)$ are included for convenience when computing derivatives of these functions at the point $z=\zb=\frac12$. They don't add any physical content to the crossing relation given here.)}
\begin{equation}
\begin{split}
F^{(at)}_{\Delta,\ell}(u,v) &\colonequals (z-\zb)\left(v^2 \GG^{(at)}_{\Delta,\ell}(u,v) - u^2 \GG^{(at)}_{\Delta,\ell}(v,u)\right)~,\\
F^{\text{short}}(u,v;c) &\colonequals (z-\zb)\left(- v^2 \GG^{\text{short}}(u,v) + u^2 \GG^{\text{short}}(v,u) - (u^2 - v^2) - \frac{(u-v)}{c}\right)~.
\end{split}
\end{equation}
Unitarity requires that $\Delta \geqslant \ell + 2$ and $a_{\Delta,\ell} \geqslant 0$. The remainder of this paper is dedicated to the numerical analysis of Eqn.~\eqref{eq:fundamentalcrossing}. We will extract nonperturbative information about the spectrum and OPE coefficients of the long multiplets as a function of the central charge $c$ of the theory.

\section{Bootstrap methods}
\label{sec:methods}

The undetermined data in the crossing equation \eqref{eq:fundamentalcrossing} is the spectrum of long multiplets $\{\D_i,\ell_i\}$ and their (squared) OPE coefficients $a_{\Delta_i,\ell_i}=\lambda^2_i$; the central charge $c$ is a fixed input parameter. The key idea of \cite{Rattazzi:2008pe} and subsequent work is to make restrictive assumptions about this data and then proceed to derive contradictions with \eqref{eq:fundamentalcrossing}. This is generally achieved using numerical methods. 

As a standard example, we might assume that all operators with $\ell = 0$ appearing in the conformal block decomposition have scaling dimension greater than a certain fixed value $\Delta_0^\star$. (This is a nontrivial assumption only if $\Delta_0^\star > 2$.) This assumption will lead to a contradiction if one can find a real-valued linear functional $\phi$ such that
\begin{align}
&\phi \cdot 0=0~,\nonumber\\
&\phi \cdot F^{\text{short}}(u,v) = -1~, \nonumber\\
&\phi \cdot F^{(at)}_{\Delta,0}(u,v)  \geqslant 0~,\qquad \forall \Delta \geqslant \Delta_0^\star~, \label{eq:alphaconstraints}\\
&\phi \cdot F^{(at)}_{\Delta,\ell}(u,v)  \geqslant 0~,\qquad\forall \Delta \geqslant \ell + 2~,\qquad\forall\ell > 0~. \nonumber
\end{align}
By linearity, such a functional will give a strictly positive result when acting on the left-hand side of \eqref{eq:fundamentalcrossing} but will return zero when acting on the right-hand side. The existence of such a functional would therefore imply that any unitary solution of the crossing symmetry equation will have at least one scalar operator whose dimension is less than $\Delta_0^\star$.\footnote{Correspondingly, the existence of any consistent, unitary solution of \eqref{eq:fundamentalcrossing} implies that for $\phi$ satisfying \eqref{eq:alphaconstraints}, we will find that $\phi \cdot F^{(at)}_{\Delta,0}(u,v)$ is less than zero somewhere in the region $2 \leqslant \Delta \leqslant \Delta_0^\star$.} One can easily extend this example by making additional assumptions about the spectrum of operators with spin, \eg, that there exist no operators of spins zero or two with dimensions less than some values $\Delta_0^\star$ and $\Delta_2^\star$, respectively, or more generally that there are gaps up to some dimensions $\Delta_\ell^\star$ for each $\ell$.

This strategy has proven fruitful even when one restricts to linear functionals of the particular form
\begin{equation}
\label{eq:functional}
\phi\cdot f(z,\zb) = \restr{\left(\sum_{m,n=0}^{\Lambda} \alpha_{m,n} \frac{\del^m}{\del z^m} \frac{\del^n}{\del \bar z^n}f(z,\zb)\right)}{z=\bar z=\frac12}~,
\end{equation}
for some finite cutoff $\Lambda\in\Zb_{+}$ and with $\alpha_{m,n}\in\Rb$. Because of the symmetry properties of $F^{\rm short}(u,v)$ and $F^{(at)}_{\Delta,\ell}(u,v)$, we may restrict to $m<n$ and $m+n\in 2\Zb_{+}$ without loss of generality. A functional in this class is completely determined by the parameters $\{\alpha_{m,n}\}$, so the task at hand is to exhaustively search the finite-dimensional space of these parameters at fixed $\Lambda$ for a functional that satisfies the correct positivity properties. For future reference, we define the dimensionality of the space of functionals at fixed $\Lambda$ as
\begin{equation}
\label{eq:search_dimension}
N(\Lambda)\colonequals \sum_{n=0}^{\Lambda}\,\,\sum_{m=n\,({\rm mod}\,2)}^{n-1}1 = \frac{2\Lambda^2+(-1)^\Lambda-1}{8}~.
\end{equation}

In the case of the standard strategy outlined above, it is of interest to determine the minimum excludable value for $\Delta_0^\star$ with a given space of functionals determined by $\Lambda$. The simplest way to do this is to implement a binary search: starting with some trial $\Delta_{0}^{\star}$, we search for $\{\alpha_{m,n}\}$ such that \eqref{eq:alphaconstraints} is satisfied. If a feasible set of values is found, then we lower $\Delta_0^\star$ and repeat the procedure. If not, then we raise $\Delta_0^\star$ instead. We can iterate until we have converged to within an arbitrarily small distance of the optimal value of $\Delta_{0}^{\star}$.

A variant of the above strategy can be used to extract bounds for OPE coefficients. For example, let us assume that the conformal block decomposition contains a scalar operator with dimension $\hat \Delta_0$. We then look for a functional that solves the following optimization problem:
\begin{align}\label{eq:opebound}
&{\rm minimize}~(\phi \cdot F^{\text{short}}(u,v))~{\rm while}:\nonumber\\
&\quad \phi \cdot 0 = 0~,\nonumber\\
&\quad\phi \cdot F^{(at)}_{\hat \Delta_0,0}(u,v) = 1~, \nonumber\\
&\quad \phi \cdot F^{(at)}_{\Delta,\ell}(u,v)  \geqslant 0~,\qquad\forall \Delta \geqslant \ell + 2~,\qquad\forall\ell \geqslant 0~.
\end{align}
Compatibility with \eqref{eq:fundamentalcrossing} then implies that 
\begin{equation}
\lambda^2_{\hat \Delta_0,0} \leqslant \phi \cdot F^{\text{short}}(u,v)~.\\
\end{equation}
In this case no binary search is necessary: the optimal value of the bound is found directly by the minimization procedure.

As a final variant, we can bound squared OPE coefficients in the presence of gaps in the spectrum. In this scenario we assume, \eg, that the scalar spectrum contains one operator of dimension $\hat \Delta_0$ and all other scalar operators have a dimension greater than some value $\Delta_0^\star \geqslant \hat \Delta_0$. The two relevant optimization problems are then given by
\begin{align}\label{eq:opegapbound}
&{\rm minimize}~(\phi \cdot F^{\text{short}}(u,v))~{\rm while}\nonumber\\
&\quad\phi \cdot F^{(at)}_{\hat \Delta_0,0}(u,v) = \pm 1~,\\
&\quad\phi \cdot F^{(at)}_{\Delta,0}(u,v) \geqslant 0~,\qquad\forall \Delta \geqslant \Delta_0^\star~, \nonumber \\
&\quad\phi \cdot F^{(at)}_{\Delta,\ell}(u,v) \geqslant 0~,\qquad \forall \Delta \geqslant \ell + 2,~\ell > 0~. \nonumber
\end{align}
Depending on the choice of sign on the second line, the result of this optimization produces an upper or lower bound for the squared OPE coefficient of the operator of dimension $\hat\Delta_0$. More precisely,
\begin{equation} \label{eq:lambdabound}
\lambda^2_{\hat \Delta_0,0} \lessgtr \pm \left(\phi \cdot F^{\text{short}}(u,v)\right)~.
\end{equation}
If we choose the positive sign, it may happen that $\phi \cdot F^{\text{short}}(u,v) < 0$ in which case \eqref{eq:lambdabound} is in contradiction with the reality of $\lambda_{\hat\Delta_0,0}$. In that case our assumption about the spectrum cannot be reconciled with crossing symmetry and we have to modify it, \emph{e.g.}, by lowering $\Delta_0^\star$. On the other hand, if we pick the negative sign then it may happen that $\phi \cdot F^{\text{short}}(u,v) > 0$ and the bound is trivial. This occurs, for example, if we set $\hat \Delta_0\geqslant\Delta_0^\star$. 

Two different computational strategies have emerged for performing the types of functional searches described here. They proceed by formulating this problem as, respectively, a \emph{linear program} or a \emph{semi-definite program}. These are both examples of convex optimization problems for which there exist decisive computer algorithms. We have employed both of these approaches in our analysis. Below we review the setup relevant to each and give some of the technical details relevant to our implementation.

\subsection{Linear programming}
\label{subsec:lin_programming}

The inequalities \eqref{eq:alphaconstraints}, \eqref{eq:opebound}, and \eqref{eq:opegapbound} should be imposed for all $\ell \geqslant 0$, and in addition for all values of the continuous parameter $\Delta$ compatible with unitarity. They are therefore infinite in number with cardinality that of the continuum. In the linear programming approach to this problem, we render this set of inequalities finite by truncating the set of spins $\ell\leqslant \ell_{\rm max}$ and the dimensions $\Delta_{\ell}\leqslant \Delta_{\ell,{\rm max}}$ for which the inequalities are considered, and further discretizing the set of scaling dimensions. This is a reasonable approach because the building blocks of the functionals --- namely the derivatives of the functions $F^{(at)}_{\Delta,\ell}$ evaluated at $z=\zb=\hf$ --- depend continuously on $\Delta$ and have well-defined asymptotic behavior for $\Delta \gg 1$ and for $\ell \gg 1$. By taking the grid of scaling dimensions fine enough and taking large enough cutoffs, one may expect that the functionals that solve the corresponding inequalities will solve the general infinite set of inequalities. Indeed, one may use asymptotic formulae for the $F^{(at)}_{\Delta,\ell}$ to verify that the constraints are also valid beyond $\Delta_{\ell,\max}$ and $\ell_{\max}$, see \cite{Rattazzi:2008pe}. Additionally, discretization effects can be investigated experimentally by checking that the results are unchanged upon refinement of the grid.

Once the set of linear inequalities is rendered finite, the optimization problems described above become standard examples of \emph{linear programs}. In notation adapted to our purposes a general linear program is an optimization problem of the form
\begin{align}\label{eq:poly_optimization}
&{\rm minimize~over~}\a_{m,n}  & &\sum_{m,n = 0}^{\Lambda} c_{m,n} \a_{m,n}~,\nonumber\\
&{\rm subject~to} & &\sum_{m,n = 0}^{\Lambda} F_{m,n} \a_{m,n} = -1~, & &\\
& & &\sum_{m,n = 0}^{\Lambda} {\bf f}_{m,n} \a_{m,n} > {\bf 0},& & {\bf f}_{m,n}\in {\Rb}^{q}~, \nonumber
\end{align}
where the inequality on the last line is meant to hold for each component separately. The precise mapping of the parameters is self-evident for all the feasibility and optimization problems listed above. Notice that the first problem \eqref{eq:alphaconstraints} is a pure feasibility study and in that case we can simply set $c_{m,n} = 0$. The number $q$ of inequality constraints on the last line is the total number of $(\Delta,\ell)$ pairs in our discretization, which is generally very large. For most of the results in this paper we use $\ell_{\max} = 20$ and for each $\ell$ the scaling dimensions were discretized with a grid spacing ranging from $\delta\Delta = 2 \cdot 10^{-3}$ at small values of $\Delta$ up to $\delta\Delta=1$ for $\Delta\sim75$, along with a sparse set of very large dimension values up to $\Delta_{\ell,\max} = 500$, leading to a total of $9659$ constraints. We use the \texttt{IBM ILOG CPLEX} optimizer to perform the optimization.

\subsection{Semidefinite programming}
\label{subsec:SDP}

An elegant alternative to the discretization used in the linear programming approach to these problems is the semidefinite programming method introduced in \cite{Poland:2011ey} and further developed in \cite{Kos:2014bka,Simmons-Duffin:2015qma}. This method does not require discretization nor the introduction of a $\Delta_{\max}$, though it is still necessary to truncate the set of spins considered to $\ell\leqslant\ell_{\max}$ with sufficiently high cutoff that the various derivatives used in the functional are well-approximated by their large $\ell$ asymptotics for $\ell\geqslant\ell_{\max}$.

The basis for the semidefinite program approach is the existence of precise approximations for the derivatives of the atomic conformal blocks evaluated at $z=\zb=\hf$ as polynomials in the conformal dimension $\Delta$, multiplied by a universal (that is, independent of the derivatives being taken) prefactor that is positive for all values of $\Delta$ that are compatible with unitarity,
\begin{equation}
\partial_z^m\partial_{\zb}^nF^{at}_{\Delta,\ell}(\tfrac12,\tfrac12)\approx\chi_{\ell}(\Delta)P^{\ell}_{mn}(\Delta)~,\\
\end{equation}
with $P^{\ell}_{mn}(\Delta)$ polynomial and where
\begin{equation*}
\chi_{\ell}(\Delta)>0~,\qquad \Delta\geqslant 2+\ell~.
\end{equation*}
These approximations can be motivated quite generally using recursion relations for conformal blocks \cite{Kos:2014bka}, but in our case they can be seen from the explicit form of the atomic conformal blocks. In particular, we adopt the radial variables of \cite{Hogervorst:2013sma},
\begin{equation}\label{eq:rhodef}
\rho \colonequals \frac{z}{(1+\sqrt{1-z})^2}~,\qquad z=\frac{4\rho}{(1+\rho)^2}~,
\end{equation}
and similarly for $\bar\rho(\zb)$. In terms of these radial variables, the hypergeometric function \eqref{eq:hypergeometric_atomic} appearing in the atomic superconformal blocks is given by
\begin{equation}
z^{\frac{\alpha}{2}}F[\alpha,z]=(4\rho)^{\frac{\alpha}{2}}{}_2 F_1 \left(\frac12,\frac{\alpha}{2},\frac{\alpha+1}{2},\rho^2\right)~.
\end{equation}
The Taylor series expansions of the above function and its derivatives about $\rho=0$ converge uniformly (and quickly) for all $\alpha\geqslant0$ at $\rho(z=\frac12)\approx0.1716$. Keeping terms up to a fixed order in the Taylor series gives a polynomial in $\Delta$ for any $\ell$ multiplied by a positive prefactor.

With polynomial approximations of this type in place, the problem of finding a functional $\phi$ as described above becomes a simple example of a ``Polynomial Matrix Program'' \cite{Simmons-Duffin:2015qma}. Using again a notation adapted to our purposes such a program takes the general form
\begin{align}\label{eq:poly_optimization_SDP}
&{\rm minimize~over~}\a_{m,n}  & &\sum_{m,n = 0}^{\Lambda} c_{m,n} \a_{m,n}~,\nonumber\\
&{\rm subject~to} & &\sum_{m,n = 0}^{\Lambda} F_{m,n} \a_{m,n} = -1~, & &\\
& & & \forall y > 0~:~~\sum_{m,n = 0}^{\Lambda} M_{m,n}(y) \a_{m,n} \succeq 0,& & M_{m,n}\in {\Rb}[y]^{s \times s}~. \nonumber
\end{align}
where $M \succeq 0$ means that the matrix $M$ is positive semidefinite. For our purposes it suffices to consider only diagonal matrices $M_{m,n}$, more precisely
\begin{equation}
\label{eq:polynomial_matrices}
M_{mn}(y) = \begin{pmatrix}
P^0_{mn}(y + \Delta_0^\star) & 0 & \cdots & 0\\
 0 & P^2_{mn}(y + \Delta_2^\star) & \cdots & 0\\
 \vdots & \vdots & \ddots & \vdots\\
 0 & 0 & \cdots & P^{\ell_{\max}}_{mn}(y + \Delta_{\ell_{\max}}^\star)
\end{pmatrix}~.
\end{equation}
This form causes the positive semidefiniteness constraint to degenerate into a set of simple inequalities, one for each of the diagonal elements. The requirement that these inequalities hold for all $y > 0$ implies they hold for all $\Delta$ greater than the imposed gaps $\Delta^\star_\ell$. The precise mapping of the other parameters is then again straightforward for each of the problems listed above.

The precision PMP solver \texttt{SDPB} was developed to solve precisely this type of feasibility/optimization problem by transforming it to a semi-definite program and implementing an arbitrary precision primal-dual interior point method \cite{Simmons-Duffin:2015qma}. We have used \texttt{SDPB} and this approach to derive stronger bounds than were achievable using the linear programming method with a machine-precision solver. In particular, the large $\Lambda$ results for $c=3/4$ and large $c$ reported below required these methods, with which we were able to raise the cutoff as high as $\Lambda=38$.\footnote{Further details on the choice of parameters used to derive the numerical results in this paper are available from the authors upon request.}

\section{Bounds on leading-twist operator dimensions}
\label{sec:firstresults}

In this section we report bounds for the scaling dimensions of the lightest long multiplets $\AA^{\Delta}_{(\frac{\ell}{2},\frac{\ell}{2})[0,0,0]}$ for fixed spin $\ell$ that appear in the superconformal block decomposition of the ${\bf 20}^\prime$ four-point function. In other words, we are exploring the spectrum of leading-twist unprotected singlet operators (LTUSOs). Recall that Bose symmetry restricts $j$ to be integer, so $\ell$ must be even. We focus on spins $\ell\in \{0,2,4\}$, leaving the extension to higher spins for future investigation.

\subsection{Single channel bounds}
\label{subsubsec:single_channel_bounds}

\begin{figure}
\begin{center}
\begin{tabular}{ll}
\includegraphics[width=7cm]{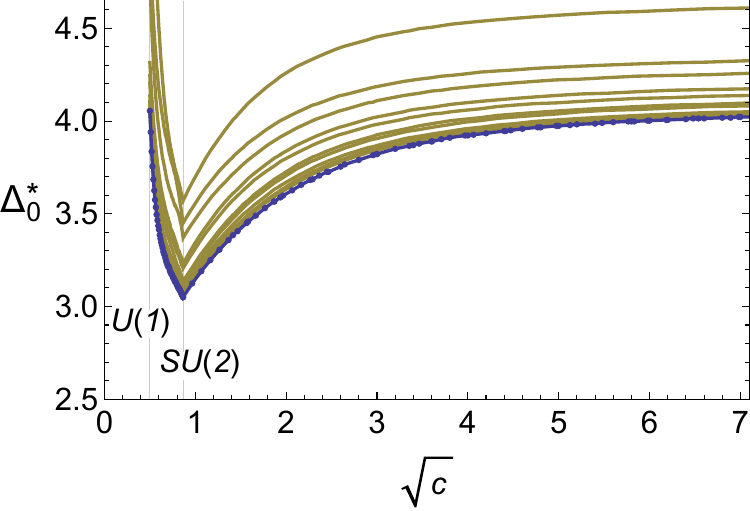} &
\includegraphics[width=7cm]{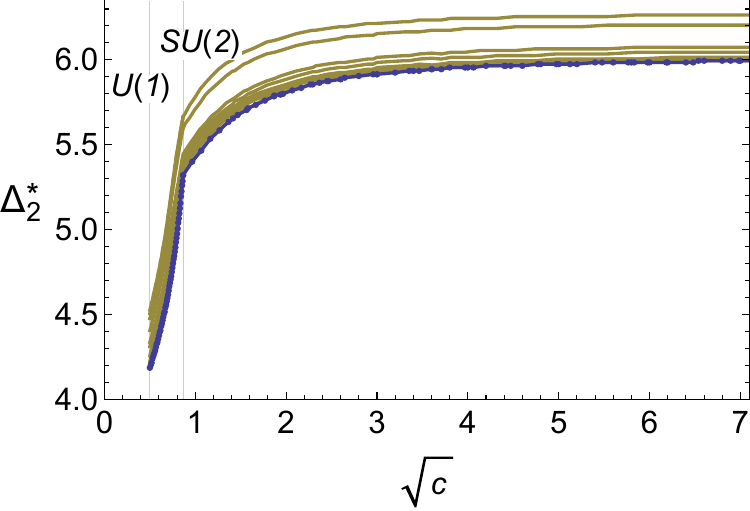} \\
\includegraphics[width=7cm]{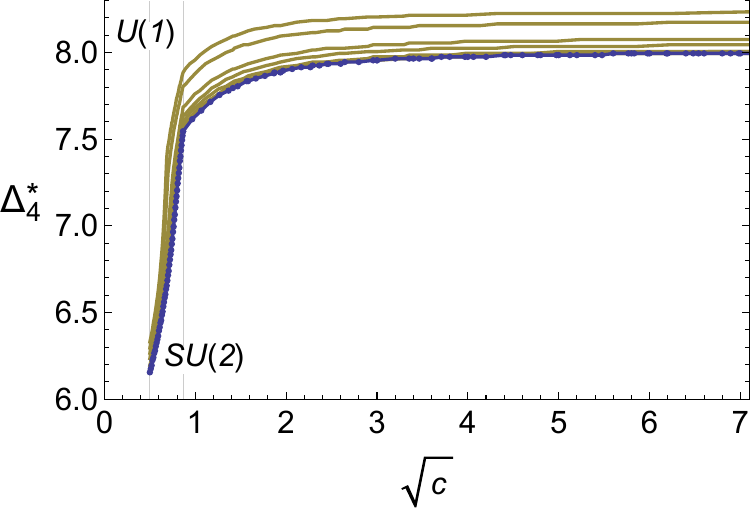} &
\end{tabular}
\end{center}
\caption{\label{fig:singlechannelbounds}Upper bounds for the dimension of the LTUSO of spin $\ell=0$, $2$, and $4$, respectively, as a function of the (square root of the) central charge.}
\end{figure}

The most straightforward operator bounds are upper bounds for the dimension of the LTUSO of a fixed spin with no restrictions imposed on the spectrum of operators of other spins. These bounds are derived by searching for functionals of the type given by \eqref{eq:alphaconstraints} and maximizing $\Delta_{\ell}^{\star}$. In Fig.~\ref{fig:singlechannelbounds} we display upper bounds of this type for spins $\ell=0,2,4$. The blue curves, which show the best bounds in the plots, were obtained using the linear programming method with $\Lambda = 17$, corresponding to $N(17)=72$ independent coefficients $\alpha_{m,n}$. The yellow curves represent bounds derived using smaller values for $\Lambda$. The curves themselves are interpolations through a finite number of data points, which are displayed explicitly only for the blue curves. We have marked with vertical lines the values $c = 3/4$ and $c = 1/4$, which are the relevant values for $\NN=4$ SYM theories with gauge algebras $\suf(2)$ and $\uf(1)$, respectively. The kink at $c = 3/4$ is related to the non-smoothness of $F^{\text{short}}(u,v)$ at this value of $c$. As was outlined above, this is due to the inclusion of higher-spin currents for $c < 3/4$.

\begin{figure}
\begin{center}
\begin{tabular}{ll}
\includegraphics[width=8cm]{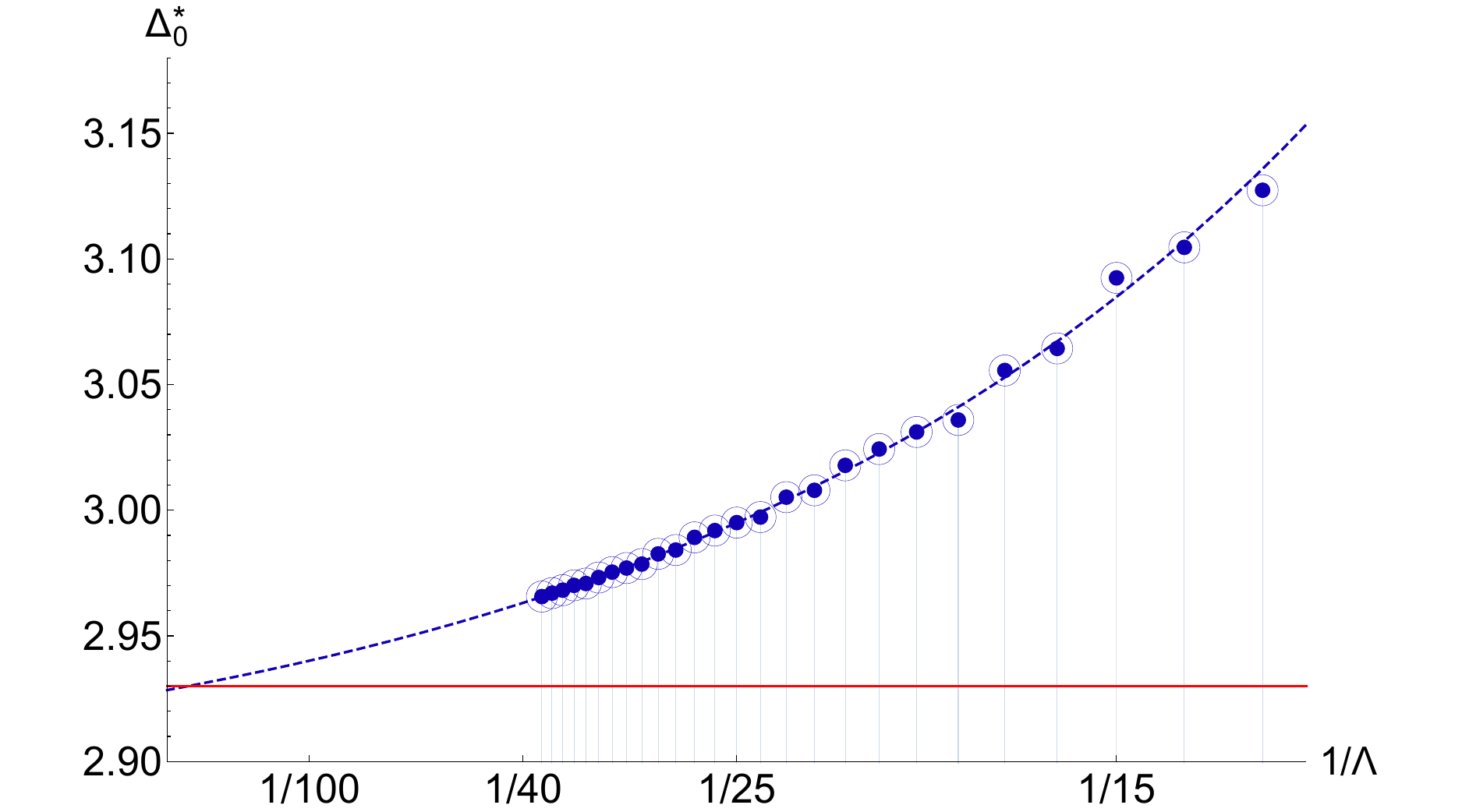} &
\!\!\!\!\!\!\!\!\!\!\!\!\!\!\!\!\!\!\includegraphics[width=8cm]{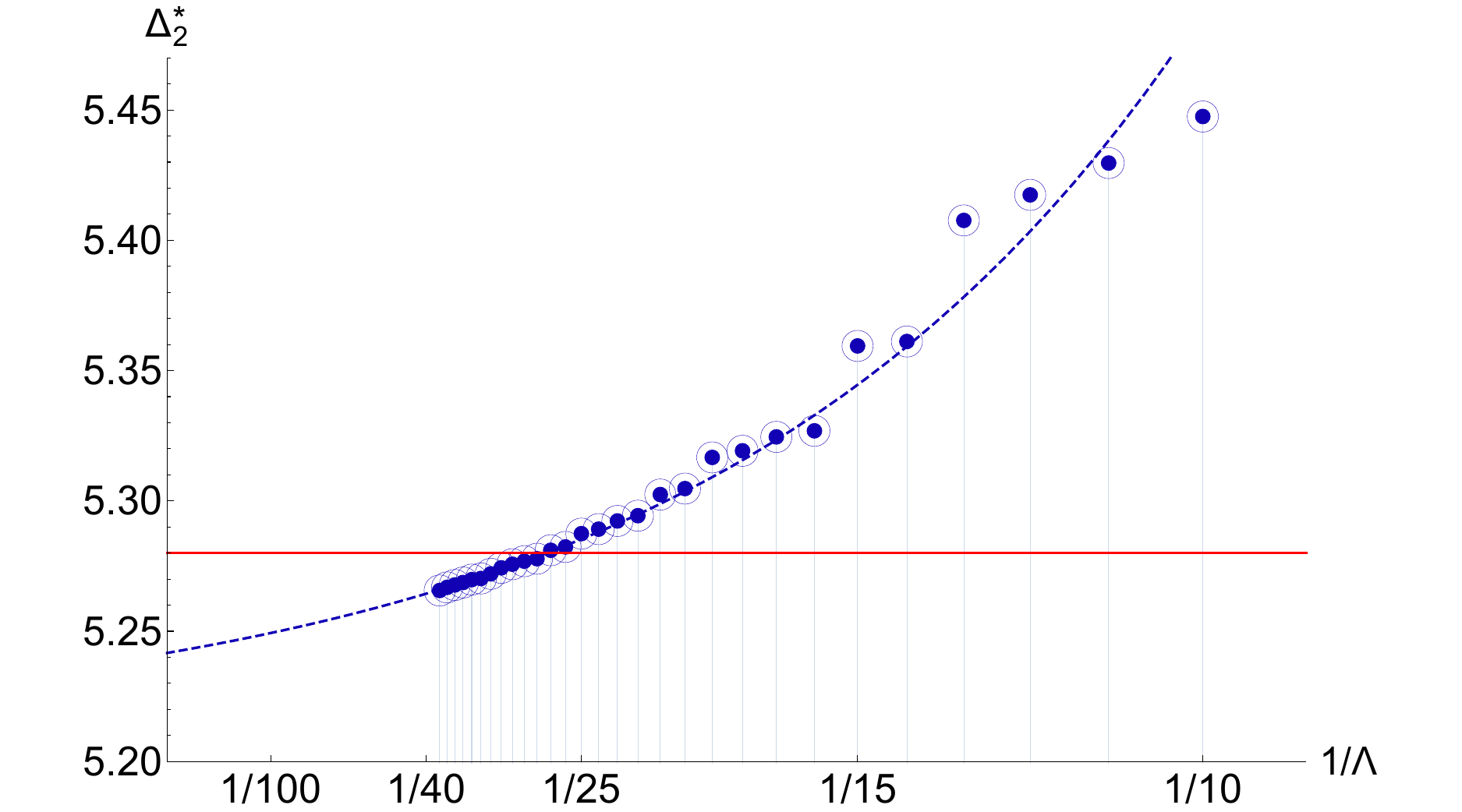} \\
\includegraphics[width=8cm]{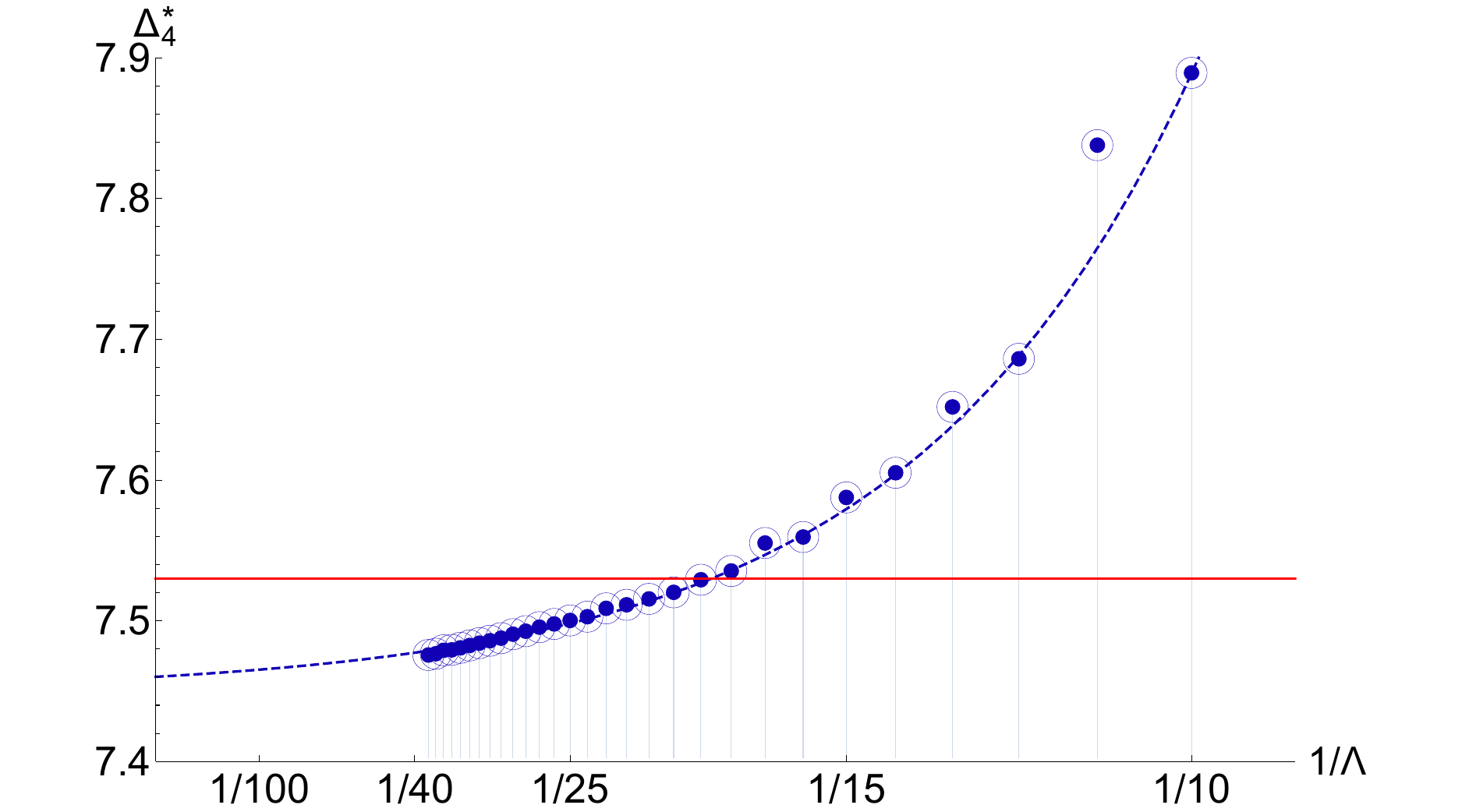} &
\end{tabular}
\end{center}
\caption{\label{fig:su2_extrap} Upper bounds for the dimensions of the LTUSOs of spins $\ell=0,2,4$ for $c = 3/4$ as a function of $1/\Lambda$ for $\Lambda=14,15,\ldots,38$. These bounds were derived using the semi-definite programming method with {\texttt sdpb} \cite{Simmons-Duffin:2015qma}. The red line denotes the ``corner estimate'' of the extremal value for these operator dimensions from \cite{Beem:2013qxa,Beem:2013hha} -- see also Section \ref{subsec:combining_bounds} below.}
\end{figure}

The plots in Fig.~\ref{fig:singlechannelbounds} suggest that the bounds are close to converging to a definite best value representing the limit $\Lambda\to\infty$. We can explore this apparent convergence by studying the bounds for many values of $\Lambda$ and considering the extrapolation to infinite $\Lambda$. It has been observed in a number of previously studied cases that these extrapolations are often quite regular, and sometimes make connections with known physics \cite{Beem:2015aoa}. We have performed such an investigation for $c=3/4$ using the semi-definite programming approach, with which we have been able to test up to $\Lambda=38$ (corresponding to a search space of dimension $N(38)=361$). Extrapolations of the single-channel bounds for $\ell=0,2,4$ are shown in Fig.~\ref{fig:su2_extrap}.

\subsubsection{Consistency at small and large central charge}
\label{subsubsec:small_large_c}

In order to interpret our bounds in the context of SYM theories, it is sensible to pay particular attention to the cases $c = 1/4$ and $c = \infty$. For these values of the central charge, there are known solutions to crossing symmetry that have a reasonable chance to realize the maximum dimensions for LTUSOs. Consequently we can use these cases to investigate whether these numerical methods are making contact with actual SCFTs.

We first consider the case of $c = 1/4$. This is the value of $c$ in the $\uf(1)$ SYM theory, which is free. The conformal block decomposition of the free-field-theory four-point function has been analyzed in \cite{Dolan:2001tt} for any gauge group. It has nonnegative OPE coefficients for all $c \geqslant 1/4$, and for $c > 1/4$ the first unprotected operator of spin $\ell$ sits at the unitarity bound $\Delta_\ell = 2 + \ell$. On the other hand, for $c = 1/4$ the coefficient of the unprotected scalar operator at the unitarity bound vanishes, and the lowest-dimension unprotected singlet operator appearing with nonzero coefficient has dimension four. (More precisely, for $c = 1/4$ the entire contribution of the dimension-two Konishi operator is accounted for by the higher-spin conserved-current block, and therefore $a_{2,0} = 0$.) This physical result is beautifully reproduced by the numerical bounds in Fig. \ref{fig:singlechannelbounds}: the spin-two and spin-four bounds are approaching the unitarity bounds of 4 and 6, respectively, at $c = 1/4$. The scalar bound, however, approaches the value 4 instead. (For $c = 1/4$ the best numerical bounds shown are given by $\Delta_0^\star=4.055$, $\Delta_2^\star=4.185$, and $\Delta_4^\star=6.155$.)

\begin{figure}[t]
\begin{center}
\begin{tabular}{ll}
\includegraphics[width=8cm]{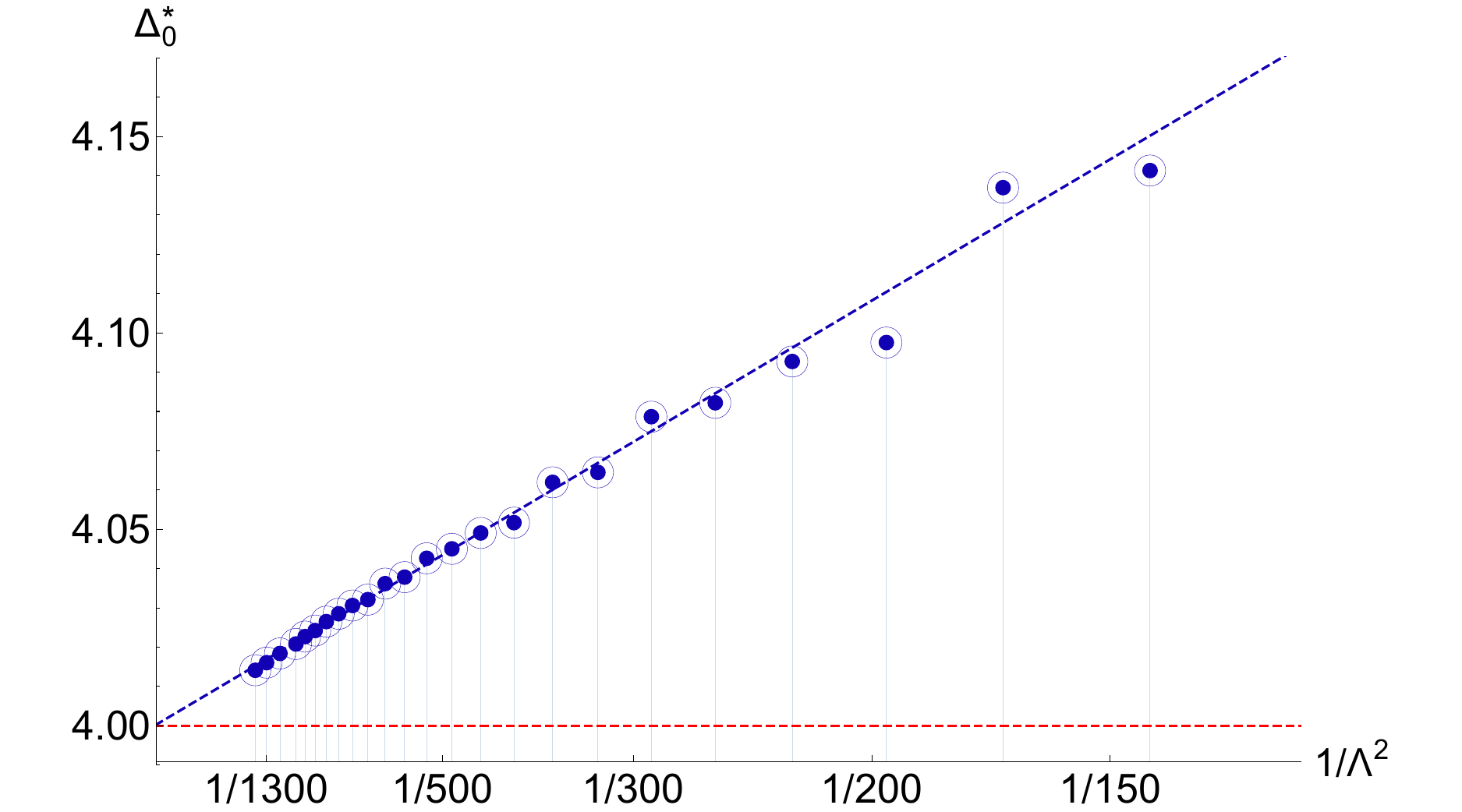} &
\!\!\!\!\!\!\!\!\!\!\!\!\!\!\!\!\!\!\includegraphics[width=8cm]{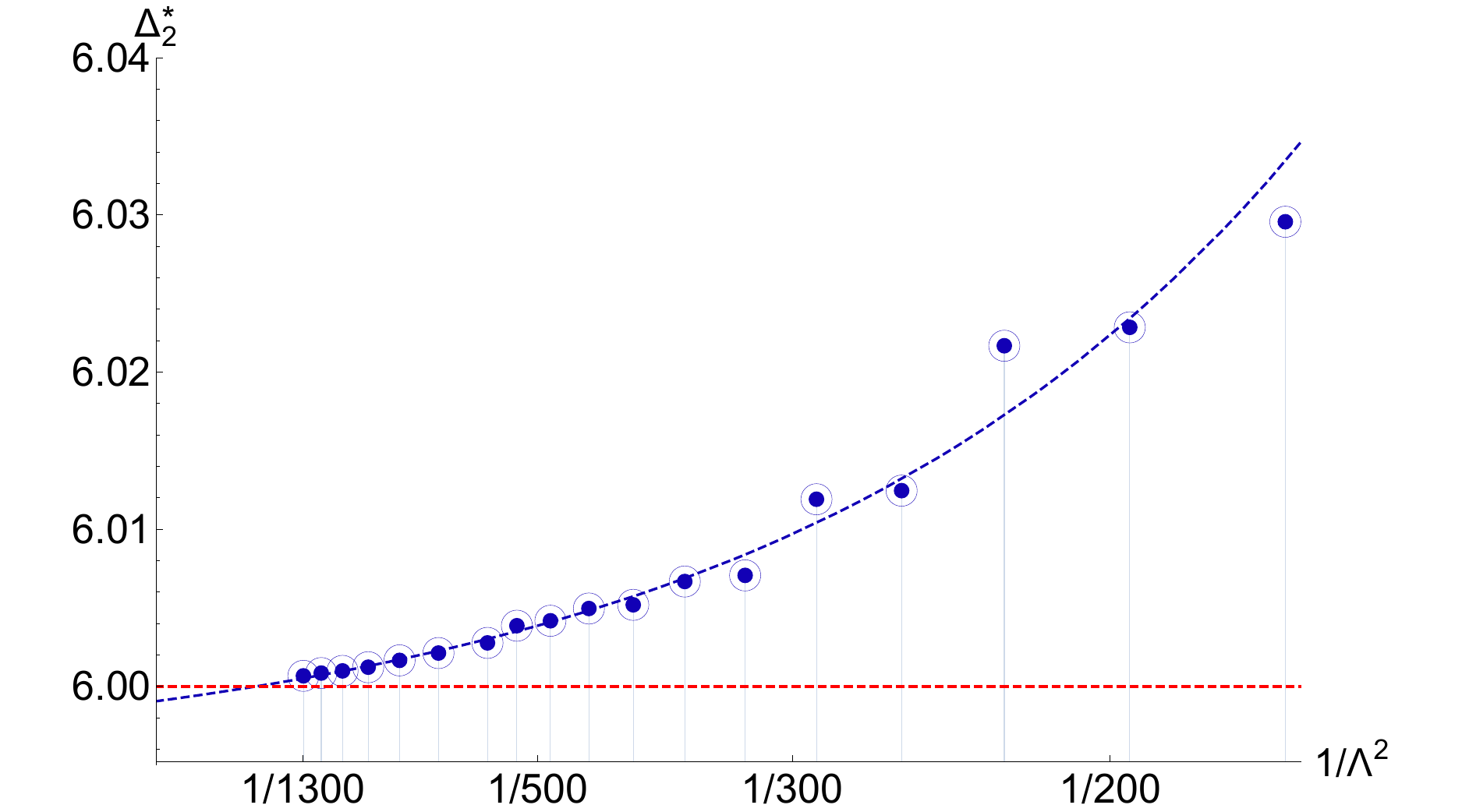}
\end{tabular}
\end{center}
\caption{\label{fig:infinite_cc}Numerical upper bounds for spin-zero and spin-two LTUSOs at infinite central charge. The curves in the two cases are a two-parameter linear fit and a three-parameter exponential fit. The mean field theory values of $\Delta_0=4$ and $\Delta_2=6$ are shown in red.}
\end{figure}

In the other extreme, we consider the case of infinite central charge. The $c\to\infty$ limit of $\NN=4$ SYM is usually considered in the context of the 't Hooft double-scaling limit in terms of $\lambda=g_{\rm YM}^2N$. At strictly infinite central charge, these theories enjoy large $N$ factorization for all values of $\lambda$, and the ${\bf 20}^\prime$ four-point function is given by the ``mean field solution'' computed as a sum of disconnected contributions,
\begin{equation} \label{eq:meanfield}
\begin{split}
\langle \OO(w_1,x_1) &\OO(w_2,x_2) \OO(w_3,x_3) \OO(w_4,x_4) \rangle = \frac{1}{x_{12}^4 x_{34}^4} \FF^{\text{m.f.}}(w_{ij};u,v) \colonequals \\
&\langle \OO(w_1,x_1) \OO(w_2,x_2)\rangle \langle \OO(w_3,x_3) \OO(w_4,x_4) \rangle +{}\\
&\langle \OO(w_1,x_1) \OO(w_3,x_3)\rangle \langle \OO(w_2,x_2) \OO(w_4,x_4) \rangle +{}\\
&\langle \OO(w_1,x_1) \OO(w_4,x_4)\rangle \langle \OO(w_2,x_2) \OO(w_3,x_3) \rangle~.
\end{split}
\end{equation}
The LTUSO for any spin in this solution is the double-trace operator of the form $\OO^{IJ}_{\bf{20'}} \del_{\mu_1}\cdots\del_{\mu_\ell} \OO^{IJ}_{\bf{20'}}$, which has dimension $4 + \ell$. This double trace value is the maximum gap for LTUSOs in the known solutions to crossing at large $c$. Happily, these operator dimensions appear to be reproduced by our numerical analysis for very large $c$. In Fig.~\ref{fig:infinite_cc} we show the upper bounds for spins $\ell=0,2$ as $\Lambda$ is increased up to a maximum value of 38 for spin zero and 36 for spin two. The best bounds derived in these two cases are $\Delta_0^\star=4.0141\pm0.0001$ and $\Delta_2^\star=6.00068\pm.00003$, and one sees by examination (and curve fitting) that the bounds are very likely approaching the mean field values of $4$ and $6$. We will investigate subleading behavior in the $1/c$ expansion below.

\subsubsection{Extremal solutions}
\label{subsubsec:extraml_sols}

It is encouraging that for large and small values of $c$, the numerical functional method produces bounds that are not only consistent with known solutions to crossing symmetry, but appear to be converging towards optimal bounds that are saturated by precisely the values for those known solutions. In other words, the numerical bounds are plausibly converging towards the best possible bounds.

For intermediate values of $c$ the numerical bounds will similarly converge towards some optimal value. We expect that there is a solution of crossing symmetry that prevents the bounds from decreasing further -- such solutions are also known as \emph{extremal solutions} because they maximize the gap in a given channel. Notice that the extremal solutions cannot correspond to weakly coupled Yang-Mills theories: weak coupling and $c > 1/4$ translates to very small gaps for all spins. We will discuss possible physical interpretations of the extremal solutions in detail below.

An approximation to the extremal solution for a given optimization problem can be recovered directly by appealing to a key result concerning the numerical strategy we have employed. For each of the linear programs formulated in Section \ref{sec:methods} there exists a \emph{dual} linear program, which amounts precisely to the search for a solution to the \emph{truncated crossing symmetry equations} obtained by considering only the constraints imposed by demanding that the $(m,n)$'th derivatives of \eqref{eq:fundamentalcrossing} vanish at $z=\zb=\frac12$ for $m,n\leqslant\Lambda$.\footnote{In the case of the semi-definite program method, we are not aware of an equally clear argument relating the dual optimization problem to the solution of the truncated crossing equation. See \cite{Beem:2014zpa} for additional discussion of this point.} This duality is actually exploited in modern numerical methods (see for example \cite{ElShowk:2012ht}): rather than finding a functional one may instead try to find a solution to the truncated equations \cite{Poland:2010wg,ElShowk:2012hu}, and in this formulation we allow a trial spectrum if such a solution can be found.\footnote{It is a theorem that in the cases under consideration the two methods are completely equivalent \cite{boyd2004convex}.} By increasing the cutoff $\Lambda$ we increase the number of equations that must be satisfied, and in this way we expect to obtain a solution to the \emph{full} crossing symmetry equation at the boundary of the allowed region as $\Lambda \to \infty$. This would be precisely the extremal solution of the previous paragraph. We can summarize the above discussion in the following conjecture.
\begin{con}
\label{conj:extremalsols}
The observed convergence of the numerical bootstrap bounds is a consequence of the existence of exact extremal solutions to the crossing equation \eqref{eq:fundamentalcrossing}. The dimensions of the first operators in the scalar, spin-two, and spin-four channels of these extremal solutions lie close to the numerical bounds shown in Fig.~\ref{fig:singlechannelbounds}.
\end{con}

There are two reasons for formulating this result as a conjecture rather than a theorem. First of all, though there exists ample numerical evidence for the convergence of the truncated solution to a full solution, there is currently no rigorous theorem concerning the convergence of the numerical results. Secondly, we cannot claim with absolute certainty that our results represent an asymptotic region, \ie, we cannot rigorously claim that the bounds obtained are very close to their optimal values. However, we hasten to add that these objections should not discourage the reader: in our view the numerical evidence in very solid, and the convergence very good.

\subsubsection{General consistency conditions}
\label{subsubsec:general_consistency}

The existence of the mean field solution \eqref{eq:meanfield} has the following interesting consequence.\footnote{The argument presented here is rather general and applies to non-supersymmetric theories as well.} Consider a solution to crossing symmetry $\FF(w_{ij};u,v)$ with positive OPE coefficients for finite $c > 3/4$ and with gap $\Delta_\ell^\star$ in the channel with spin $\ell$. The linear combination
\begin{equation}
\alpha \FF(w_{ij};u,v) + (1-\alpha) \FF^{\text{m.f.}}(w_{ij};u,v)~,
\end{equation}
interpolates between our original solution and the mean-field solution, has positive OPE coefficients for all $0 \leqslant \alpha \leqslant 1$, and from its conformal block decomposition we find an effective central charge given by $c_\text{eff} = c/\alpha \geqslant c$. It therefore provides a valid solution to crossing symmetry with gaps given by $\min(\Delta_\ell^\star,\ell+2)$ for each $\ell$. In other words, as long as $\Delta_\ell^\star \leqslant \ell + 2$ the existence of this one solution for a given value of $c$ implies the existence of a solution with the same gaps for all other values of $c_\text{eff} > c$. It follows that the bound for the first operator of spin $\ell$ is necessarily a \emph{non-decreasing} function of $c$, at least as long as $c > 3/4$ and the bound is smaller than $\ell + 2$. (For $c < 3/4$ the higher-spin currents invalidate the above argument.) This monotonicity is manifest in the numerical results of Fig.~\ref{fig:singlechannelbounds}.

In fact, this is just a simple example of a more general subtlety that arises from the possibility of taking linear combinations of solutions of crossing symmetry. This issue has particular impact in the case of the superconformal bootstrap, as it is often the case in supersymmetric theories that there will be a number of inequivalent theories that will all have some BPS operator in their spectrum with fixed quantum numbers. A linear combination of four-point functions for those theories can then affect the numerical bounds that can conceivably be derived using single correlator techniques. Linear combination effects of this type have been observed previously in the study of three-dimensional $\NN=8$ theories \cite{Chester:2014mea}.

Our numerical bounds are also consistent with the general ideas presented in \cite{Komargodski:2012ek,Fitzpatrick:2012yx} concerning the spectrum at large spin and convexity of the twist $\tau = \Delta - \ell$ of the lowest-lying operators as a function of spin. Taken together, the results of those papers imply that the OPE decomposition of our correlator must contain an operator with twist less than four for any spin $\ell \geqslant 4$. A quick inspection of Table~\ref{tab:superblocks} shows that for $c \geqslant 3/4$ this operator is necessarily the primary of a long multiplet, resulting in an analytic bound that is for example $\Delta \leqslant 8$ for the spin-four multiplets. Our numerical bounds improve on this result for low values of $c$. For very large $c$ the analytic bound is slightly stronger, which we ascribe to the finite value of the numerical cutoff $\Lambda$.

\subsection{Combining the bounds}
\label{subsec:combining_bounds}

\begin{figure}
\begin{center}
\begin{tabular}{ll}
\includegraphics[width=7cm]{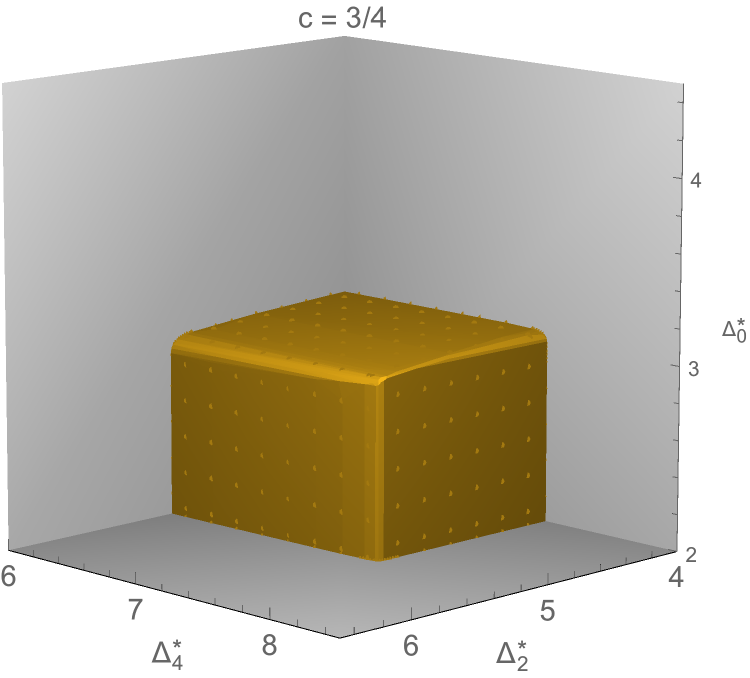} &
\includegraphics[width=7cm]{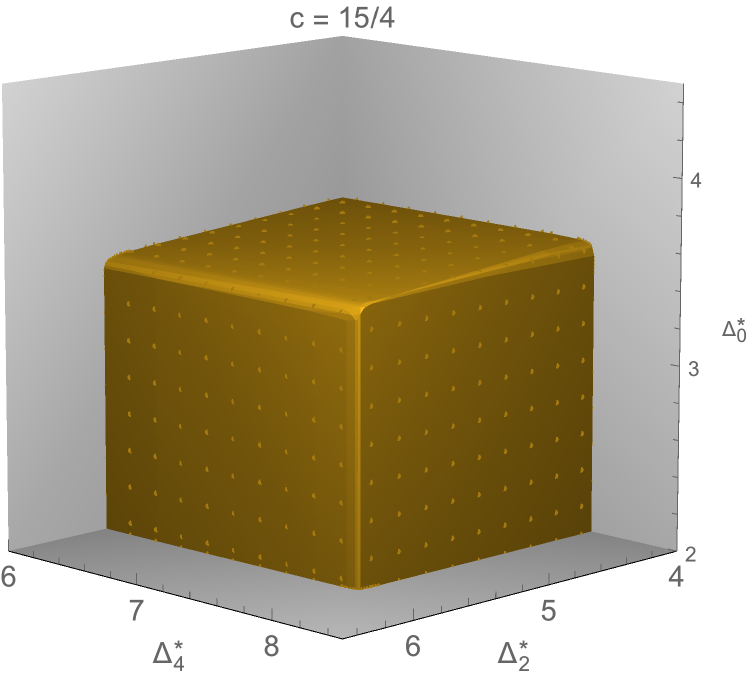} \\
\includegraphics[width=7cm]{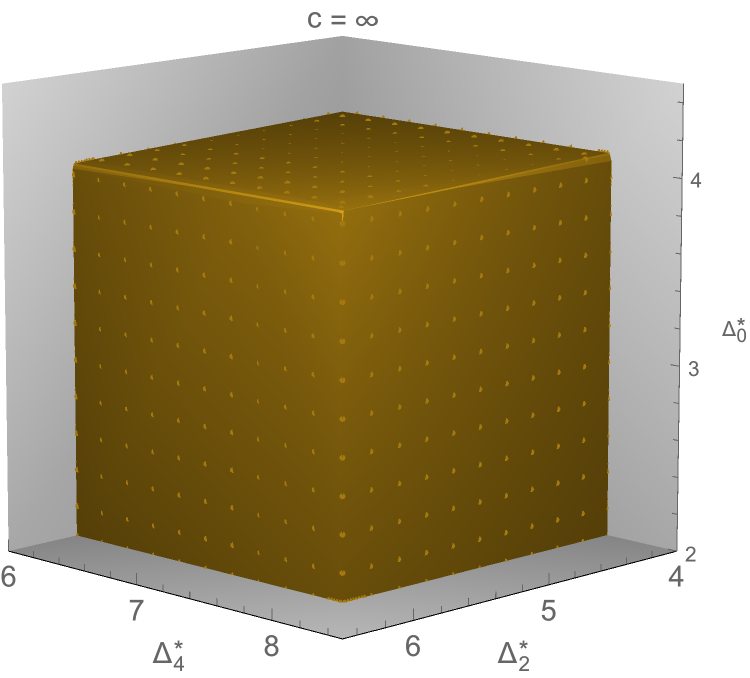} &
\end{tabular}
\end{center}
\caption{\label{fig:cubes} Combined-channel exclusion plots for the LTUSOs of spin $\ell=0,2,4$. The plots are for central charges $c=\frac{3}{4}$, $c=\frac{15}{4}$, and $c=\infty$, as labelled.}
\end{figure}

In practice, we do not expect the solutions of crossing symmetry that maximize the dimension of the LTUSO of a given spin to have operators at the unitarity bound for other spins. We can then hope to get extra information about the collection of extremal functionals by bounding the possible simultaneous gaps for several spins at once. In Fig.~\ref{fig:cubes} we plot the exclusion surface in the three-dimensional space spanned by the three gaps $(\Delta_0^\star,\Delta_2^\star,\Delta_4^\star)$. The strategy to obtain this surface is the same as the single channel bounds of the previous subsection: we perform a binary search for functionals of the type \eqref{eq:alphaconstraints}, but now with gaps in all three channels simultaneously. Consistency with crossing symmetry requires that all consistent, unitary $\NN=4$ theories with the given central charges must have a triplet of operators of spins 0, 2 and 4 corresponding to a point somwehere below the yellow surface shown in these figures.\footnote{The unitarity bound on the operator dimensions $\Delta \geqslant \ell + 2$ restricts us to the octant shown in the figures.} To orient the reader we note that the bounds presented in Fig.~\ref{fig:singlechannelbounds} (for the given values of $c$) correspond to the intersections of the surfaces of Fig.~\ref{fig:cubes} with the axes.

The exclusion surfaces in Fig.~\ref{fig:cubes} have a rather striking shape which is nearly cubic for each value of the central charge.\footnote{We have performed similar analyses for many additional values of $c$ between $3/4$ and infinity and the qualitative shape stays more or less the same.} We should consider whether this shape is consistent with theoretical expectations and if it provides insight concerning the extremal solutions of Conjecture \ref{conj:extremalsols}. First, it is important to realize that the surfaces are again necessarily monotonic in the sense that, if we move along the surface, then each $\Delta^\star_\ell$ has to be a non-increasing function of the other two scaling dimensions. This is because one cannot pass from an excluded point to an allowed point by increasing the gap in one channel (increasing one of the $\Delta^\star_\ell$).

Consider now the set $\SS$ of points $(\hat \Delta^0, \hat \Delta^2, \hat \Delta^4)$ that correspond to valid solutions to the crossing symmetry equations. From the aforementioned monotonicity argument we deduce that for each point $p \in \SS$ the numerical bound will never be able to exclude any points in the interior of the cuboid with horizontal and vertical faces and with corner point $p$. The best possible bound is therefore the union of all the cuboids spanned by all points $p \in \SS$. This is the bound that we would expect to find in the limit $\Lambda \to \infty$.

In Fig.~\ref{fig:cubes} we find what can best be described as a single approximate cuboid. Qualitatively this is the same thing we find for all other values of the central charge. This teaches us two important lessons. Firstly, the approximately cubic shape suggests that our bounds have converged relatively well, which provides us with further evidence for Conjecture \ref{conj:extremalsols}. Secondly, the expected union of cuboids should reduce to a single cuboid if there exists a \emph{single} extremal point $p_{\text{ext}}$ where all three of $\hat \Delta^0$, $\hat \Delta^2$, and $\hat \Delta^4$ are \emph{simultaneously} maximized within $\SS$. Then all other cuboids lie within the cuboid corresponding to $p_{\text{ext}}$. It is therefore natural to conjecture that this is what is happening with our bounds:
\begin{con}
For a given value of $c$, the three single-channel bounds in Fig.~\ref{fig:singlechannelbounds} all converge towards a single extremal solution which is the same for the three different spins. This solution simultaneously maximizes the dimension of the first unprotected operator for all three spins. The corresponding point $p_{\text{ext}}$ lies close to the vertex of the cuboids that we obtained numerically.
\end{con}

\begin{figure}
\begin{center}
\includegraphics[width=7cm]{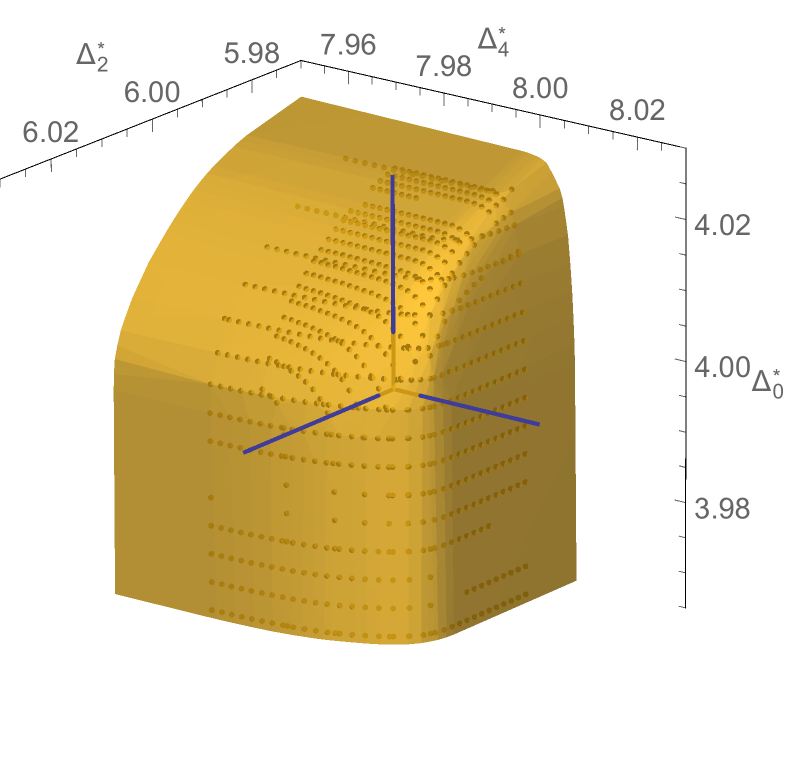}
\end{center}
\caption{\label{fig:cornerzoom}Detailed view of the corner of the cube corresponding to $c = \infty$. The superimposed axes intersect at the point $(\hat\Delta_0,\hat\Delta_2,\hat\Delta_4)=(4,6,8)$, which corresponds to the mean-field solution.}
\end{figure}

For infinite $c$ this extremal solution should be the mean-field solution \eqref{eq:meanfield}. This is borne out by Fig.~\ref{fig:cornerzoom} which provides a close up view near the corner of the cube with $c = \infty$. We see that the exclusion surface just wraps around the point $(4,6,8)$ corresponding to the mean-field solution. This detailed view of the exclusion surface also illustrates that the vertex of the cuboid is rounded off rather than perfectly sharp. Also, the faces are not perfectly horizontal or vertical: for example the top face intersects the vertical axis at the value 4.074. We suspect these are all finite $\Lambda$ effects and that the bounds approach a sharp cuboid as $\Lambda \to \infty$.

\subsection{Interpretation of the extremal solution}
\label{subsec:extremal_interpretation}

In the previous subsections we interpreted our bounds in terms of a hypothetical single extremal solution, which is completely determined in terms of the central charge $c$ of the theory. However, the existence of a solution for a certain value of $c$ does not automatically imply the existence of a corresponding full-fledged $\NN=4$ SCFT. Counterexamples are known in similar contexts: there exists a one-parameter family of solutions to crossing symmetry (with positive coefficients) in two dimensions \cite{Liendo:2012hy}, but only for given values of the central charge does this solution belong to a unitary (non-supersymmetric) CFT. Similarly, one may analytically continue the four-point functions of the $O(N)$ models to fractional $N$ and obtain consistent results for a single correlator, but nevertheless find conflicts with unitary in other correlators.

In this subsection we will consider the possibility of giving a physical interpretation of the extremal solutions away from the points $c = 1/4$ and $c = \infty$. We will see that for very large (but finite) central charge we can actually understand them relatively precisely. For specific smaller values of $c$ we will conjecture that the extremal solutions should be interpreted as physical solutions in strongly coupled $\NN=4$ SYM theories.

\begin{figure}[t]
\begin{center}
\begin{tabular}{ll}
\includegraphics[width=7cm]{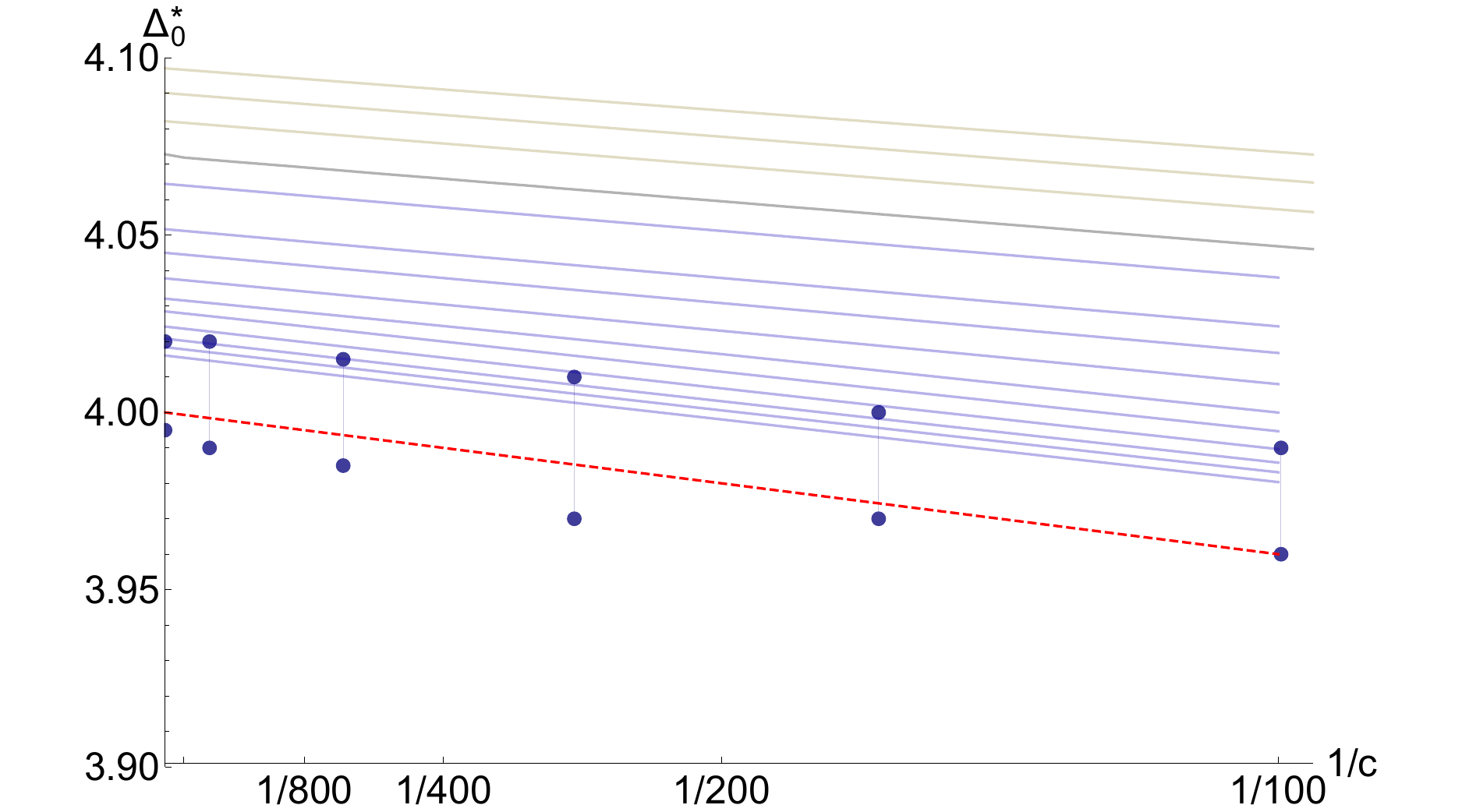} &
\includegraphics[width=7cm]{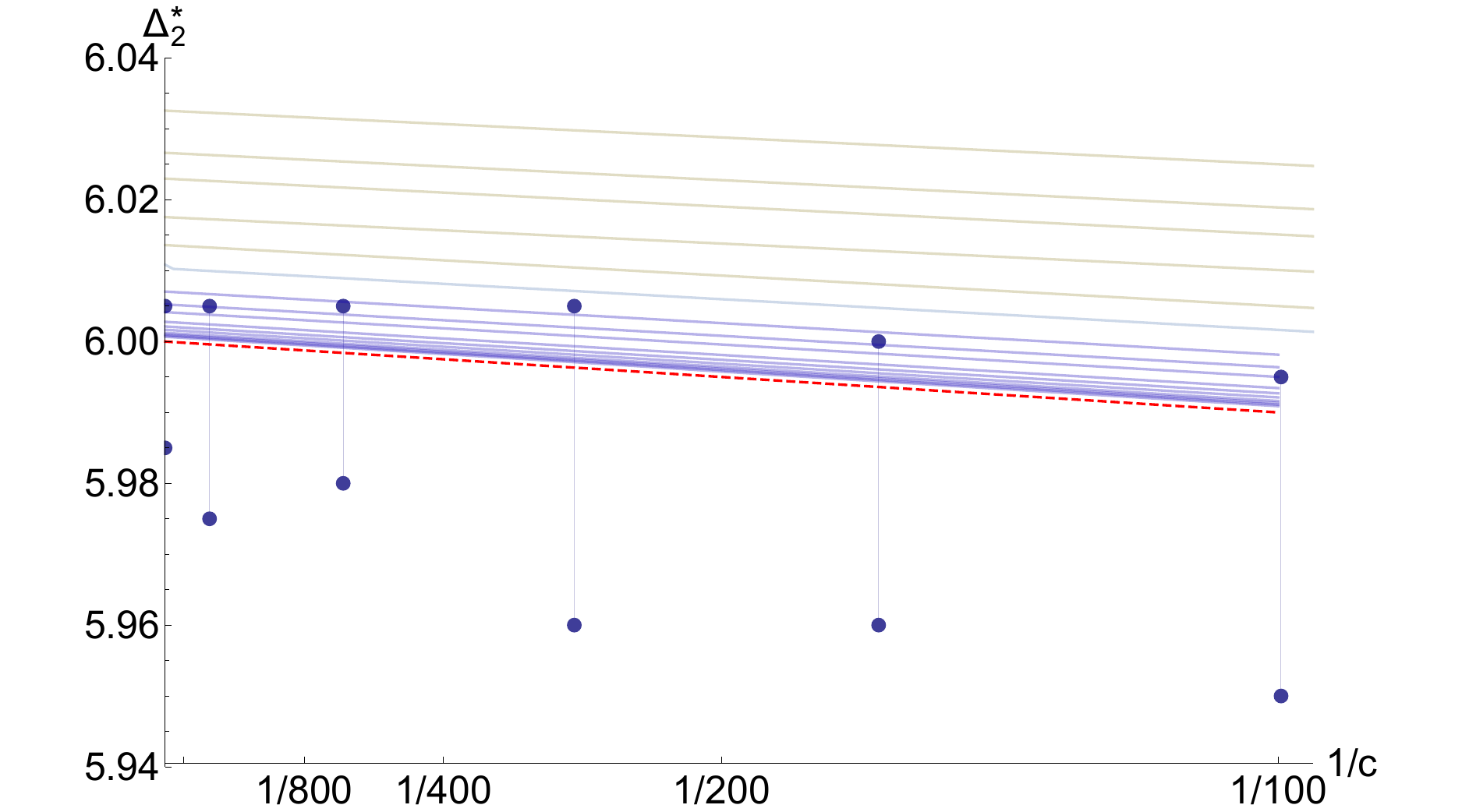}
\end{tabular}
\end{center}
\caption{\label{fig:sugra}Numerical and analytic results for LTUSO dimensions at large central charge. Gray lines correspond to the bounds of Fig.~\ref{fig:singlechannelbounds}, and blue lines were obtained using the SDP method with $\Lambda\leqslant38$. The vertical intervals represent manual estimates of the location of the `corner' of cuboids like those in Fig.~\ref{fig:cubes} for the given values of the central charge. Finally, the red line segments correspond to the leading $1/N^2$ correction to the mean-field solution obtained from supergravity and given in \eqref{sugraandim}.}
\end{figure}

\subsubsection{Interpretation for large central charge}
\label{subsubsec:large_c_interpretation}

In the 't Hooft limit of the $\NN=4$ SYM theories, the mean-field solution \eqref{eq:meanfield} is the leading-order approximation for the ${\bf 20}^\prime$ four-point function in the $1/N^2$ expansion, for all values of the coupling $\lambda$. The first subleading term depends on the 't Hooft coupling and can be computed at weak coupling using perturbation theory or at strong coupling using supergravity.

At weak coupling the first subleading correction introduces the Konishi operator to the conformal block expansion, and similarly in other spin channels one finds new operators with dimensions at the unitarity bounds. This behavior is consistent with the numerical bounds, but comes nowhere close to saturating them. The extremal solution does not correspond to weakly coupled SYM theory.

At large 't Hooft coupling we are in better shape. The first subleading term in the $1/N^2$ expansion, as computed with tree-level Witten diagrams in ${\rm AdS}_5$, introduces no new unprotected operators below the double-trace operators, but the double-trace operators get small negative anomalous dimensions. For the operators of lowest spin this anomalous dimension has been computed in \cite{Dolan:2001tt,D'Hoker:1999pj,Arutyunov:2000ku}, with the following result:
\begin{equation}
\label{sugraandim}
\begin{split}
\Delta_0 &= 4 - \tfrac{16}{N^2} + \ldots = 4 - \tfrac{4}{c} + \ldots~,\\
\Delta_2 &= 6 - \tfrac{4}{N^2} + \ldots = 6 - \tfrac{1}{c} + \ldots~,\\
\Delta_4 &= 8 - \tfrac{48}{25 N^2} + \ldots = 8 - \tfrac{12}{25 c} + \ldots~.
\end{split}
\end{equation}

In Fig.~\ref{fig:sugra} we display the large $c$ behavior of the bounds shown in Fig.~\ref{fig:singlechannelbounds}, together with a plot of \eqref{sugraandim}. We also added an admittedly crude approximation of the operator dimension obtained by manually estimating the location of the corner of a cuboid like those shown in Fig.~\ref{fig:cubes}. As shown in Fig.~\ref{fig:cornerzoom} the corner is not sharp, so our estimate is  an interval rather than a point. 

\begin{figure}[t]
\begin{center}
\begin{tabular}{ll}
\!\!\!\!\!\!\!\!\includegraphics[width=8.2cm]{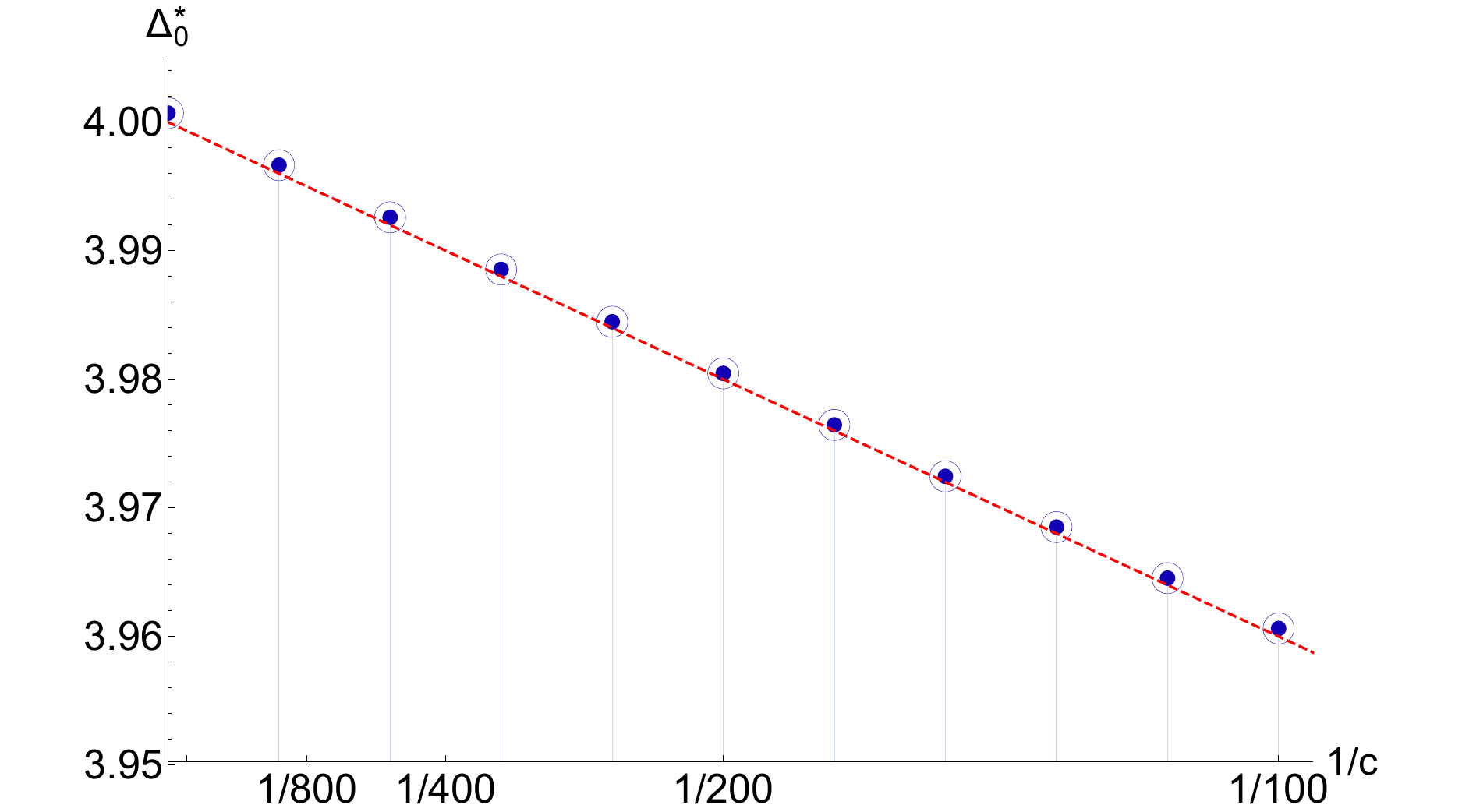} &
\!\!\!\!\!\!\!\!\!\!\!\!\includegraphics[width=8.2cm]{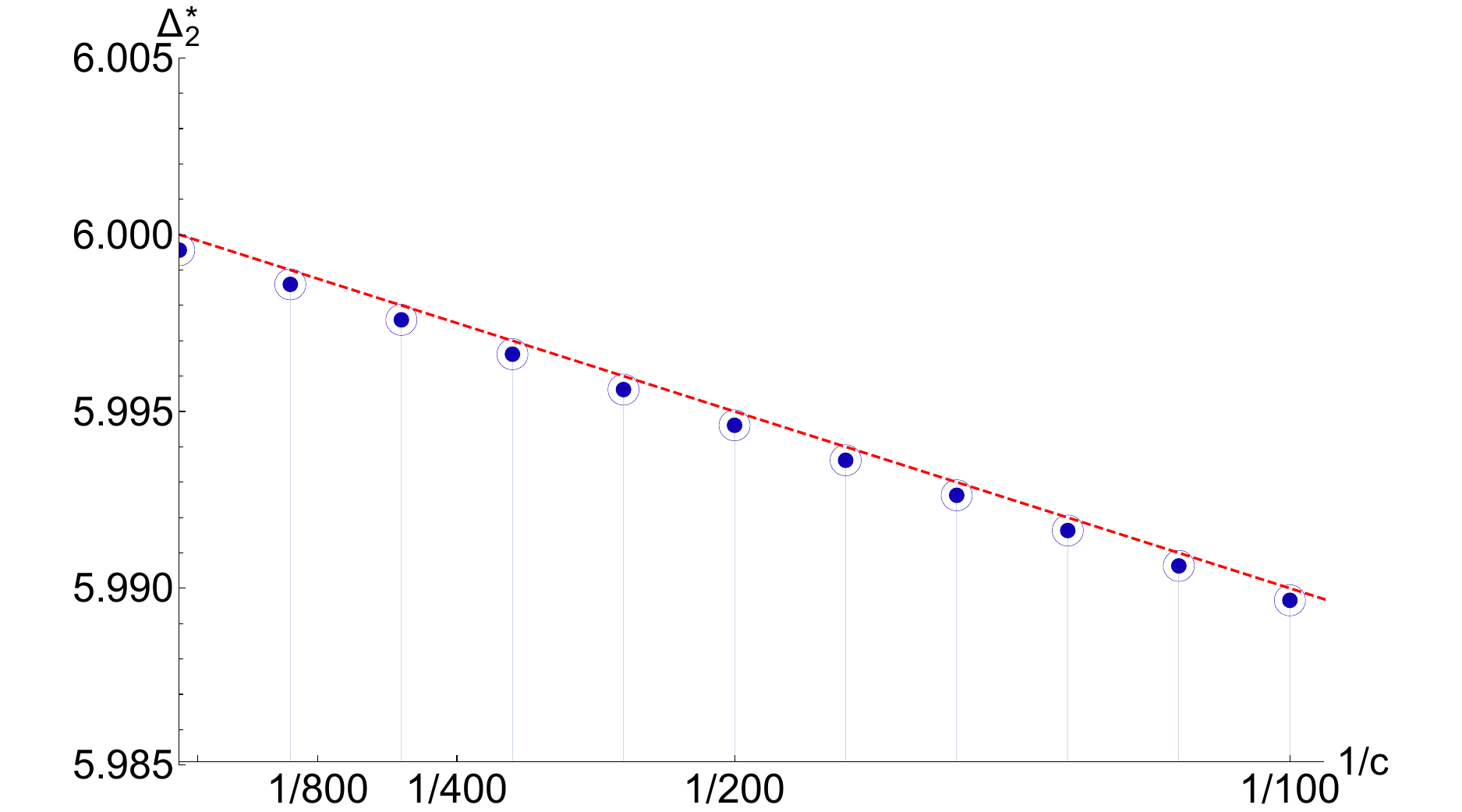}
\end{tabular}
\end{center}
\caption{\label{fig:slope_Extrap}Extrapolated upper bounds for the LTUSO in the scalar and spin-two channels at large central charge. For each data point we have extrapolated from bounds for twelve values of $\Lambda=14,\ldots,36$ to $\Lambda=\infty$. The red dashed lines are plots of the supergravity expression for the double-trace dimensions given in \eqref{sugraandim}. Since these are extrapolations, they are not strictly bounds, so there is no contradiction in the fact that they lie below the supergravity curve in the plot on the right.}
\end{figure}

We can improve our estimates of the operator dimensions in the extremal solution by extrapolating the upper bounds as a function of the cutoff $\Lambda$. In Fig.~\ref{fig:slope_Extrap} we show the result of extrapolating to infinite $\Lambda$ the sequence of upper bounds in the scalar channel for eleven large values of $c$. This strongly reinforces the conjecture that the large $c$ extremal solution is precisely the supergravity solution in the neighborhood of infinite central charge, and we can plausibly make the following conjecture:
\begin{con} For parametrically large central charge the extremal solution is the mean-field solution, and the first $1/c$ correction to the extremal solution is equal to the leading $1/N^2$ correction as computed from type IIB supergravity on ${\rm AdS}_5 \times S^5$.
\end{con}

We should emphasize that the success of the mean-field solution and its supergravity correction at large central charge was not obviously guaranteed. Naively one can take the large $c$ limit in a manner that is not controlled by planar SYM. For example we can fix the coupling constant $g_{YM}^2$ to be $O(1)$ and then take $N \gg 1$. The behavior of the theory in such limits is less tractable, and a priori we could not have excluded a solution to the crossing symmetry equations with a larger gap than observed in the mean-field solution. Our numerical bounds however teach us that this is, in fact, not possible. Analytic work supporting the same conclusion can be found in \cite{Alday:2014tsa,Alday:2016htq}, and subleading corrections in $\lambda$ were obtained in \cite{Goncalves:2014ffa}.

\subsubsection{Interpretation for finite central charge}
\label{subsubsec:su2_interpretation}

Let us now consider finite values of $c > 3/4$.%
\footnote{In principle we wish to only consider those values of $c$ corresponding to physical theories. Although our current results do not show this, we expect that a more careful analysis involving multiple correlators will restrict the allowed values of $c$ to a discrete set.}
The extremal solution can not correspond to weakly coupled $\NN=4$ SYM theory because the gaps are far larger than those that appear in the free theory. If we assume that the extremal solution is physical and does not correspond to an `exotic' theory with $\NN=4$ supersymmetry, then it must correspond to a strongly coupled point on the conformal manifold of $\NN=4$ SYM theories for the given value of the central charge. However, an important consequence of our conjecture that there is a single extremal solution for fixed $c$ is that such a point must be very special: the scaling dimensions of the LTUSO's of spin zero, two, and four should be \emph{simultaneously extremized}.

The conformal manifolds of $\NN=4$ SYM theories have a single complex dimension parametrized by the marginal coupling $\tau$. We will first consider only the theories with a simply-laced gauge algebra, for which this conformal manifold is simply the upper half plane $H$ modulo the action of $PSL(2,\mathbb Z)$.\footnote{Technically this yields an orbifold but referring to this space as a `manifold' is a common abuse of language. For the non-simply-laced theories the action of S-duality is more complicated and also changes the gauge algebra, see for example \cite{Dorey:1996hx,Argyres:2006qr}.} This conformal manifold has three distinguished points: the weakly-coupled point at $i \infty$, and the two self-dual points at $\tau = i$ and at $\tau = \exp(i \pi /3)$. At the latter two points the the theory enjoys enhanced $\mathbb Z_2$ and $\mathbb Z_3$ symmetry, respectively. It is natural to expect that the dimensions of operators should at the very least have local extrema at these fixed points.\footnote{To wit, modular invariant functions that are smooth on $H$ must be stationary at these two points, although this could theoretically be merely a saddle point. (A function that is at least $C^2$ on $H$ must however have a local extremum at the $\mathbb Z_3$ invariant point.)} The self-dual points on the conformal manifolds of the simply laced $\NN=4$ SYM theories are therefore our candidates to correspond to the extremal solutions. At present, we do not have a view on whether the $A_n$ series or the $D_n$ series will control the extremal solutions, so we make the following noncommittal conjecture:
\begin{con}
\label{conj:extremalisselfdual}
For the values of the central charge corresponding to simply-laced $\NN=4$ SYM theories of either $A$ type or $D$ type, the extremal solution is the four-point function at one of the self-dual points on the conformal manifold. At other values of the central charge the extremal solution is an unphysical continuation of these.
\end{con}

\begin{figure}
\begin{center}
\begin{tabular}{ccc}
\includegraphics[width=4cm]{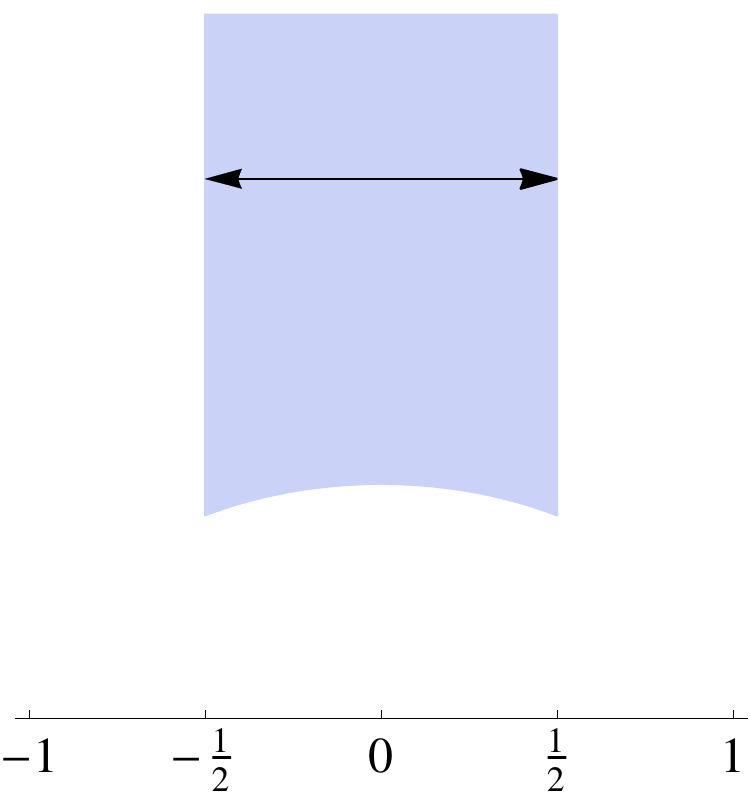} &
\raisebox{3cm}{\includegraphics[width=1cm]{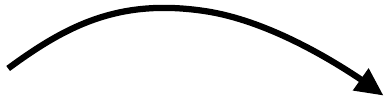}} &
\includegraphics[width=6cm]{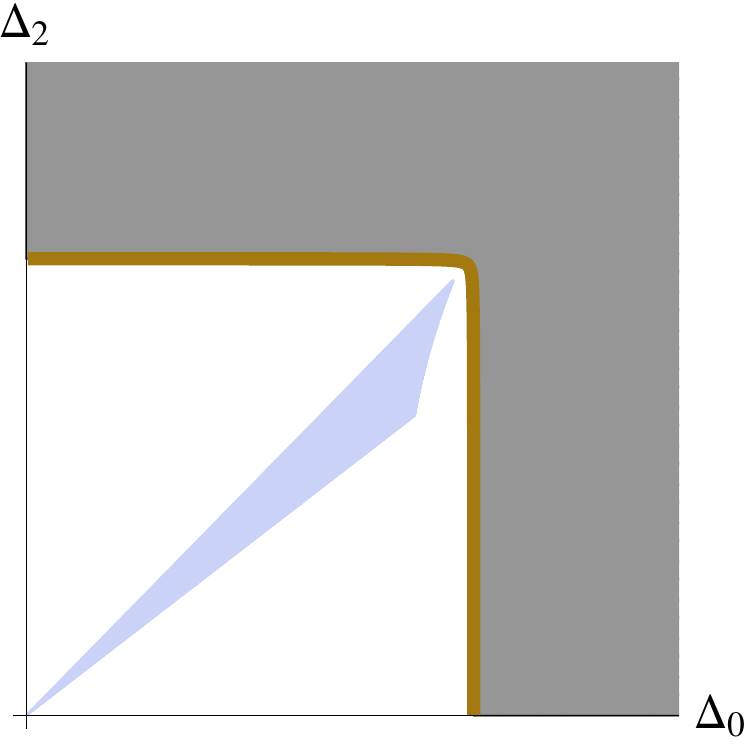}
\end{tabular}
\end{center}
\caption{\label{fig:embedding}The scaling dimensions $\Delta_0$, $\Delta_2$ are functions on the conformal manifold $\MM = H/ PSL(2,\mathbb Z)$ of $\NN=4$ SYM theories and as such define a map from $\MM$ to the allowed region shown on the right. The weakly coupled point at $i \infty$ is known to map to the lower left corner. We do not know what happens at strong coupling but have sketched one possible embedding that is consistent with the bounds. The content of conjecture \ref{conj:extremalisselfdual} is that the corner point of the allowed rectangle (or cuboid if we include $\Delta_4$) actually approximates the image of one of the two self-dual points.}
\end{figure}

Pictorially this conjecture is explained in Fig.~\ref{fig:embedding}. For a central charge $c$ that is equal to one of the SYM central charges, the set $\SS$ discussed above includes in particular the entire conformal manifold of this SYM theory, which for the simply-laced theories is $H/PSL(2,\mathbb Z)$ as shown on the left in Fig.~\ref{fig:embedding}. This conformal manifold needs to be mapped to the inside of our cuboid. A hypothetical two-dimensional version of this map is shown on the right in Fig.~\ref{fig:embedding}. If the embedding extends all the way to the corner of the cuboid, then we may naturally expect that one of the self-dual points gets mapped there. Given our limited knowledge of the strong-coupling behavior of $\NN=4$ SYM theories, we will refrain from speculating which of the two self-dual points actually gets mapped to the corner of the cuboid.

\subsubsection{Speculations on the spectrum of \texorpdfstring{$\NN=4$}{N=4} SYM at strong coupling}
\label{subsubsec:speculations}

What does the spectrum of operator dimensions of $\NN=4$ SYM look like at finite coupling? Besides our numerical results, there exist at least three sources of information to inform our opinion on this question: perturbation theory at weak coupling, S-duality, and the spectrum of the $\suf(n)$ theories in the 't Hooft limit as obtained from integrability. For definiteness, let us restrict our attention to low-lying, unprotected, $R$-symmetry-singlet, superconformal primary scalar operators. At weak coupling the lightest operator in this sector is the Konishi operator of dimension $2 + O(g^2)$, and the next operators have dimension $4 - O(g^2)$.%
\footnote{There are no $\suf(4)_R$-singlet scalars with $\Delta = 3$.}
In the remainder of this section we entertain some speculations concerning the strong-coupling behavior of this part of the spectrum. For simplicity our plots below will not show the dependence of the operator dimensions on the $\theta$ angle.

\begin{figure}
\begin{center}
\begin{tabular}{c@{\hskip 1cm}c}
\includegraphics[width=4cm]{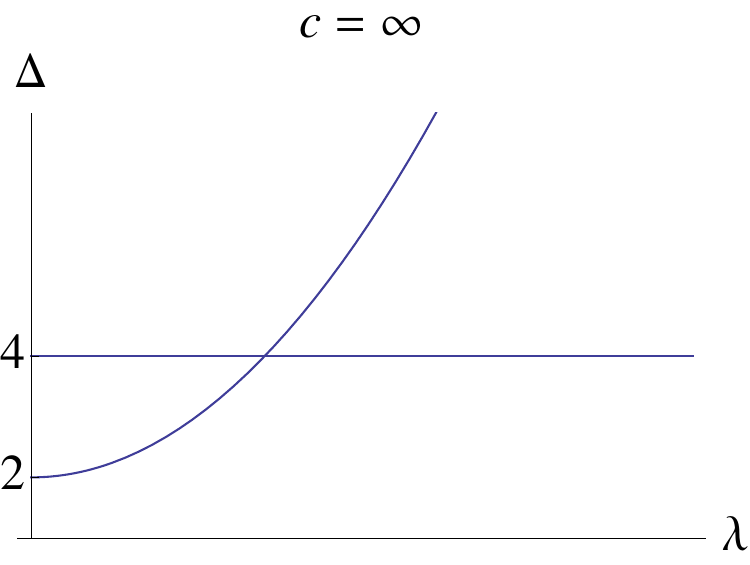}  &
\includegraphics[width=4cm]{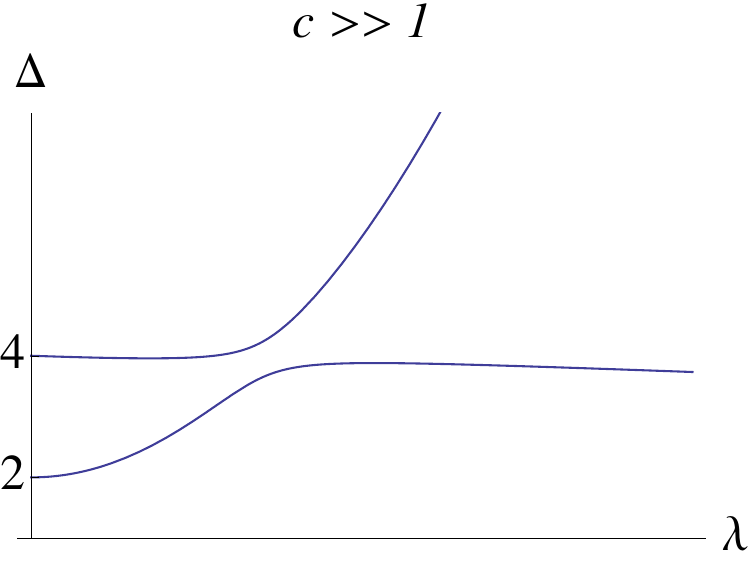}
\end{tabular}
\end{center}
\caption{\label{fig:largeNbehavior}On the left, the behavior of scaling dimensions in the strict planar limit. On the right, an approximate behavior for very large but finite $N$.}
\end{figure}

To make contact with known results let us first consider the planar limit, that is we hold $\lambda = g^2 N$ fixed and send $N \to \infty$. The dimension of the Konishi operator then grows without bound as $\lambda^{1/4}$ for very large $\lambda$. In the same limit the double-trace operator $\OO^{IJ}_{\textbf{20'}} \OO^{KL}_{\textbf{20'}}$ has $\Delta = 4$ for all $\lambda$ due to large $N$ factorization. This behavior is sketched in the left plot of Fig.~\ref{fig:largeNbehavior}, where for simplicity we omitted other operators with $\Delta = 4$ at weak coupling.\footnote{We refer to \cite{Arutyunov:2002rs} for a comprehensive one-loop analysis in the $SU(N)$ theories.} If we take $N$ large but finite then we expect this behavior to change smoothly. Furthermore, because there is no strict distinction between single- and double-trace operators anymore, we expect that levels begin to repel each other. This leads to a picture as sketched on the right of Fig.~\ref{fig:largeNbehavior}.\footnote{This picture has been studied recently in more detail in \cite{Korchemsky:2015cyx}.} Notice that in this figure $\lambda$ is plotted on the horizontal axis, so $g^2$ is actually zero in the left plot, and very small in the right plot.

\begin{figure}
\begin{center}
\begin{tabular}{c@{\hskip 1cm}c@{\hskip 1cm}c}
\includegraphics[width=4cm]{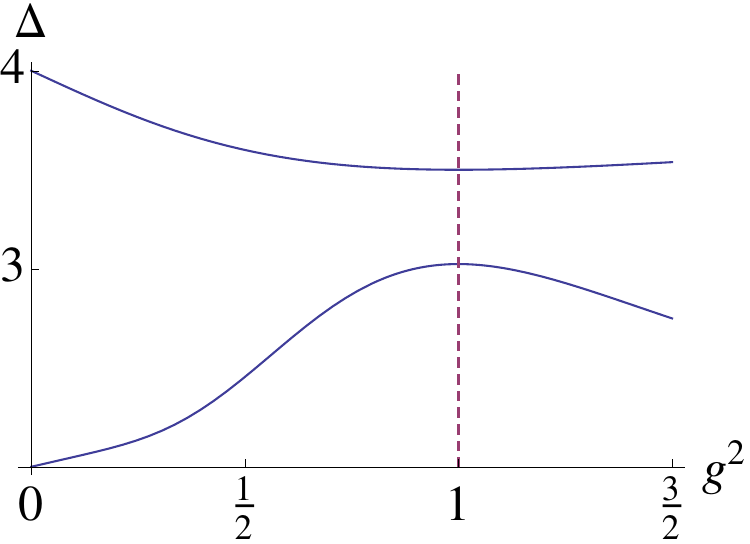}  &
\includegraphics[width=4cm]{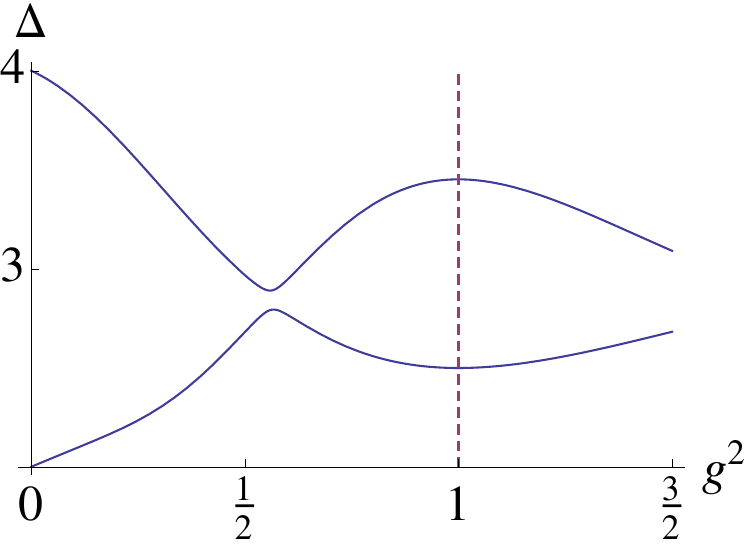}  &
\includegraphics[width=4cm]{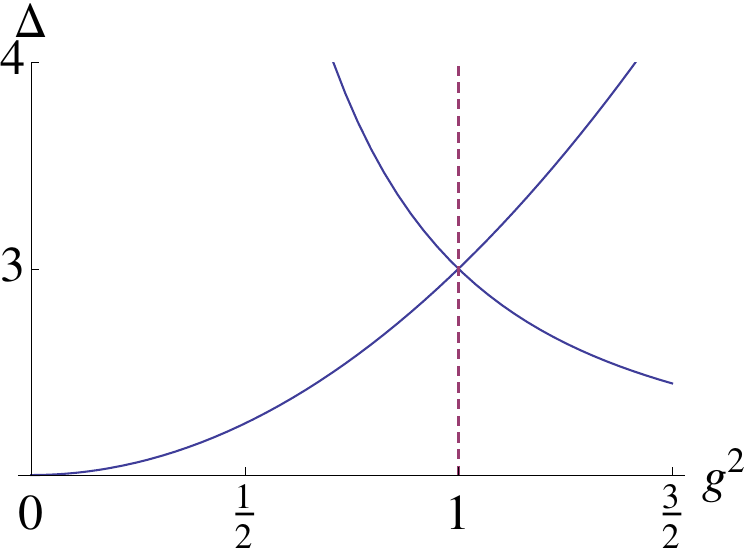}
\end{tabular}
\end{center}
\caption{\label{fig:generalbehavior}Three possible scenarios for the behavior of scalar operator dimensions as a function of the coupling. The red dotted line is the self-dual point $g=1$, and S-duality relates the spectrum for $1 < g < \infty$ to the region $0 < g < 1$. The point that maximizes the dimension of the first scalar operator presumably corresponds to the extremal solution.}
\end{figure}

Let us now consider finite values of the central charge, where we can use perturbation theory at weak coupling. In the $\suf(N)$ theories the dimension of the Konishi operator is known to four loops and given by \cite{Bajnok:2008qj,Fiamberti:2008sh,Velizhanin:2008jd,Velizhanin:2009gv}%
\footnote{For recent results on instanton corrections to unprotected operators --- including the Konishi operator --- see \cite{Alday:2016tll,Alday:2016jeo,Alday:2016bkq} and references therein.}
\begin{equation}
\Delta_{\text{kon}} = 2 + \tfrac{3Ng}{\pi}-\tfrac{3N^2g^2}{\pi^2}+\tfrac{21N^3g^3}{4\pi^3}+\left(-39 + 9\,\zeta(3) - 45\,\zeta(5)\left(\tfrac{1}{2}+\tfrac{6}{N^2}\right)\right)\tfrac{N^4g^4}{4\pi^4}+\cdots~,
\end{equation}
with $g=g_{{\rm YM}}^2/4\pi$. For the dimension four operators there is mixing already at leading order. The results of \cite{Arutyunov:2002rs} indicate that for finite $N$ at one loop there is always one operator with a negative anomalous dimension, resulting in:
\begin{equation}
\Delta_{\text{4}} = 4 - \tfrac{|f(N)| Ng}{\pi} + \cdots
\end{equation}
with $f(N)$ determined through diagonalization of a $4 \times 4$ matrix. For example, we have
\begin{equation}
f(2) = \frac{3}{2}, \qquad f(3) = 0.804\ldots, \qquad f(N) = \frac{10}{N^2} + O\left( \frac{1}{N^3} \right)\,.
\end{equation}
Our main question is now how to extend these results to strong coupling. We have sketched three possible scenarios in Fig.~\ref{fig:generalbehavior}. The simplest possible scenario is sketched on the left in Fig.~\ref{fig:generalbehavior}. The anomalous dimension of Konishi grows to a maximum at the self-dual point $g = 1$, and the dimension $4$ operators have some other unspecified behavior. For $g > 1$ the dimensions simply mirror those of $g < 1$ because of S-duality. Alternatively, the dimension of the next-to-lowest operator could approach that of the Konishi operator at some non-self dual coupling $g < 1$. Because of level repulsion this would lead to a qualitative picture as sketched in the middle of Fig.~\ref{fig:generalbehavior}. Notice that this second scenario would be in contradiction with conjecture \ref{conj:extremalisselfdual} because the extremal point now occurs where the levels approximately cross, not at the self-dual point $g = 1$. This scenario nevertheless seems attractive at larger values of $N$, as it naturally connects to the large $N$ behavior as sketched on the right in Fig.~\ref{fig:largeNbehavior}. This tension with our conjecture motivates a more detailed analysis of the corners of the cuboid exclusion surfaces at high precision (large $\Lambda$). On the right of Fig.~\ref{fig:generalbehavior} we have sketched a scenario where the Konishi operator is not a singlet under $SL(2,\mathbb Z)$, so at the self-dual point it has a dual partner with exactly the same anomalous dimension.

Of course, very little is known about the behavior of scaling dimensions in $\NN=4$ SYM theories when $g \sim 1$. One natural attempt to improve our understanding is the resummation of perturbative results. S-duality of $\NN=4$ SYM implies that this is the asymptotic behavior both around $g = 0$ and $g = \infty$. Following \cite{Sen:2013oza}, several manifestly S-dual resummation procedures were proposed in \cite{Beem:2013hha} (see also \cite{Honda:2014bza} for additional developments). As discussed in more detail in that paper, for $\suf(2)$ gauge algebra these results agree remarkably well with the results presented above, providing some additional evidence for the realization of the leftmost scenario of Fig.~\ref{fig:generalbehavior} for $c = 3/4$.

\begin{figure}
\begin{subfigure}[c]{\textwidth}
\centering
\includegraphics[width=5.5cm]{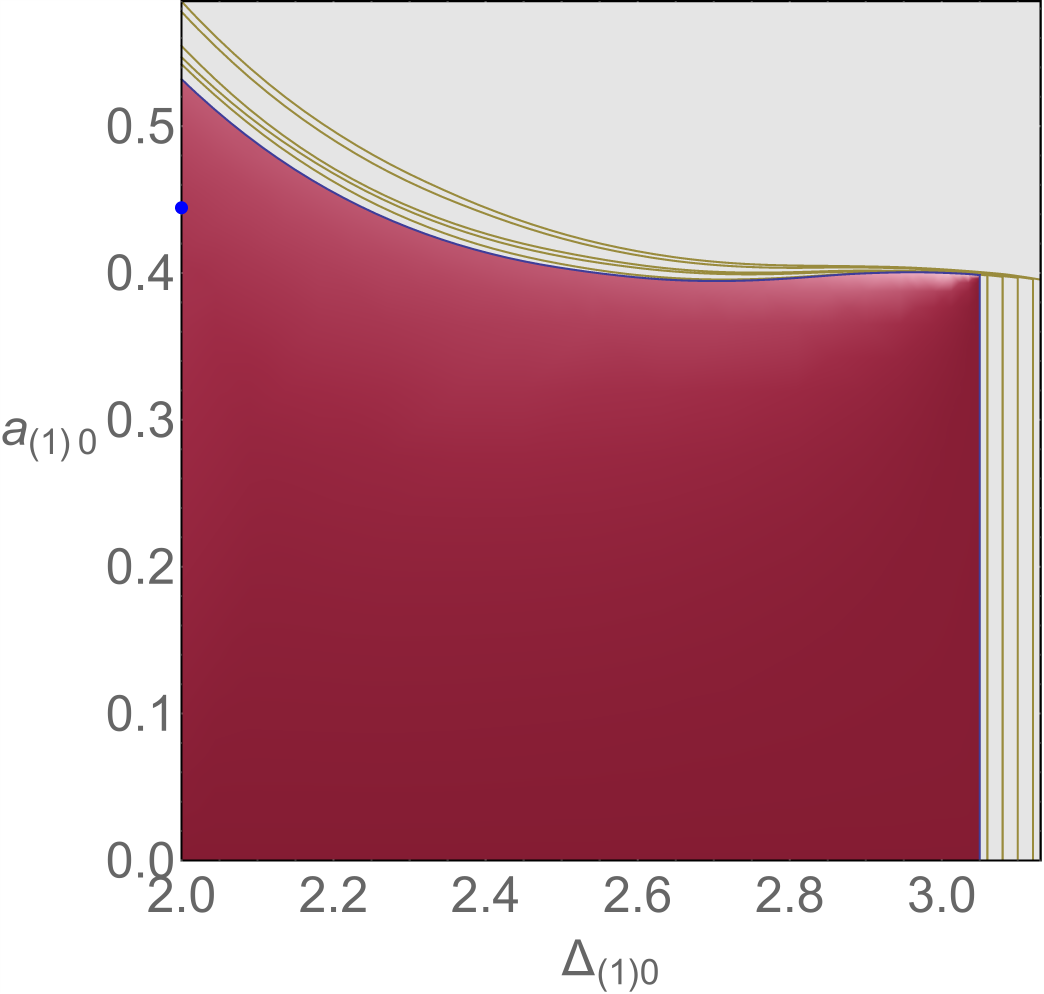}$\quad$
\includegraphics[width=9cm]{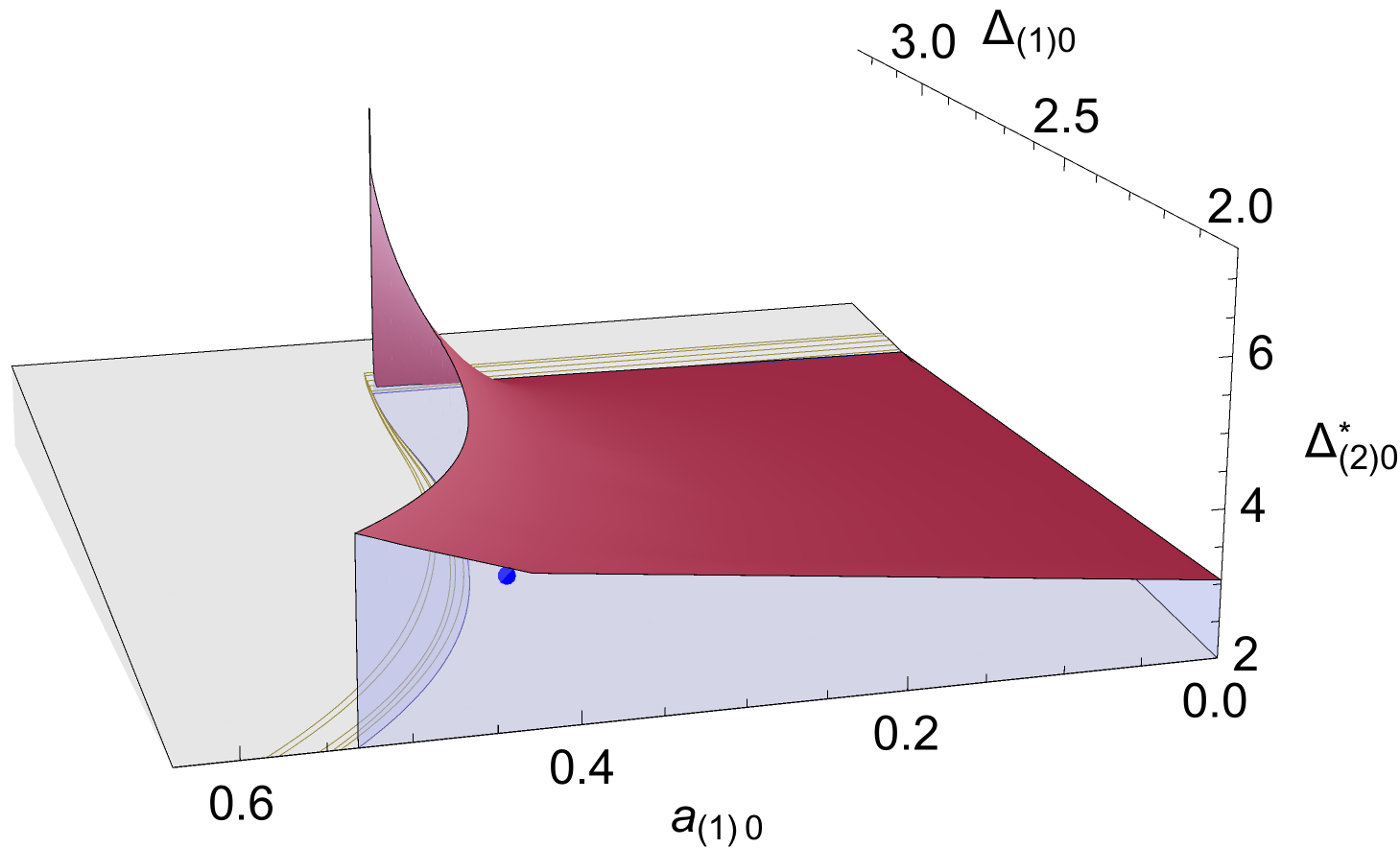}
\caption{Spin 0}
\end{subfigure}

\begin{subfigure}[b]{\textwidth}
\centering
\includegraphics[width=5.5cm]{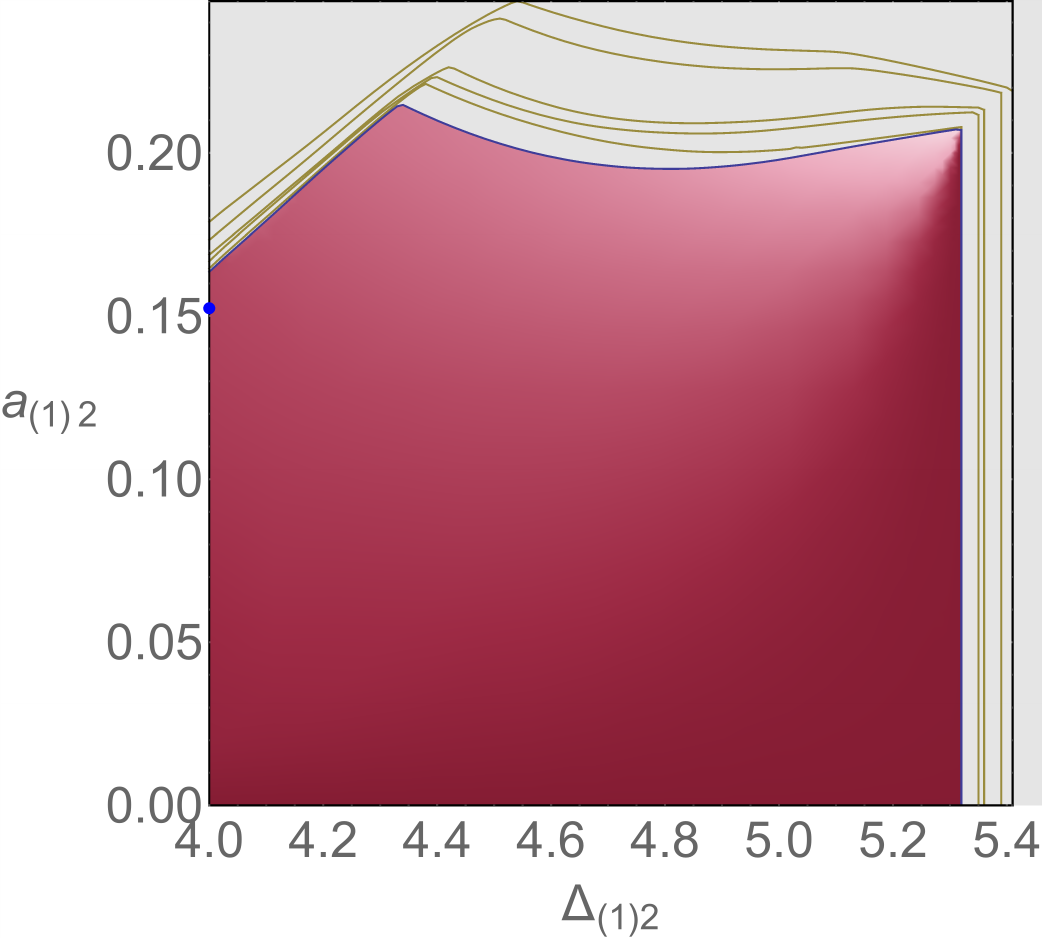}$\quad$
\includegraphics[width=9cm]{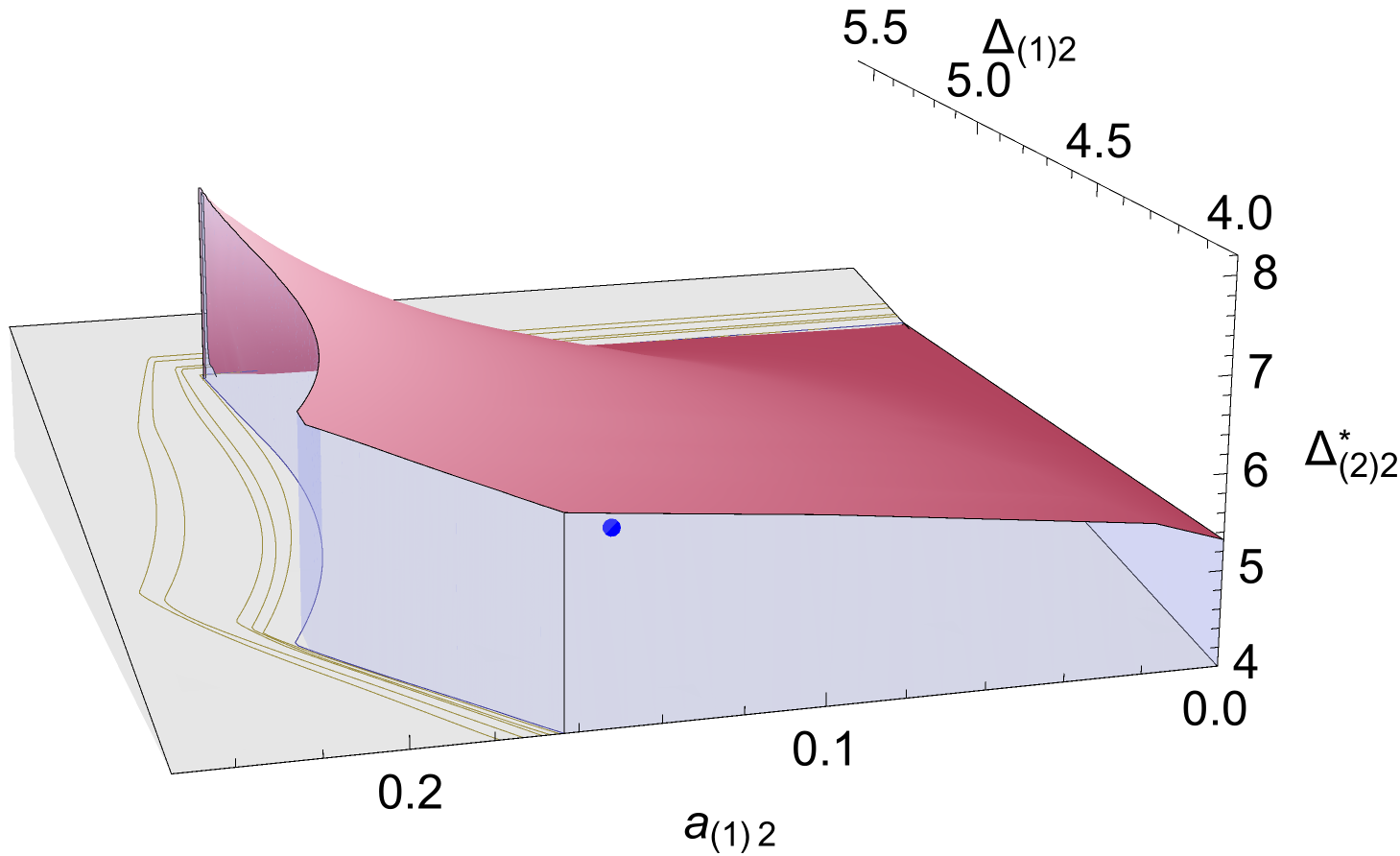}
\caption{Spin 2}
\end{subfigure}

\begin{subfigure}[b]{\textwidth}
\centering
\includegraphics[width=5.5cm]{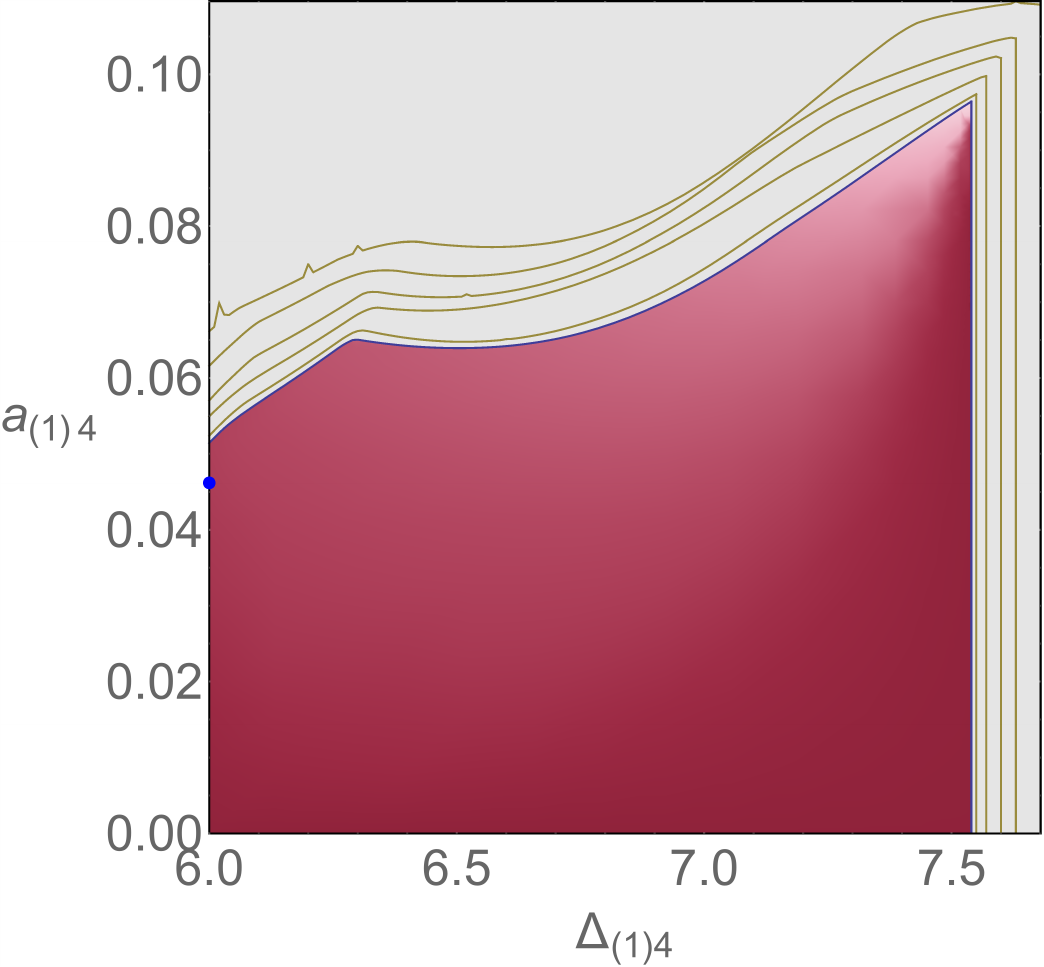}$\quad$
\includegraphics[width=9cm]{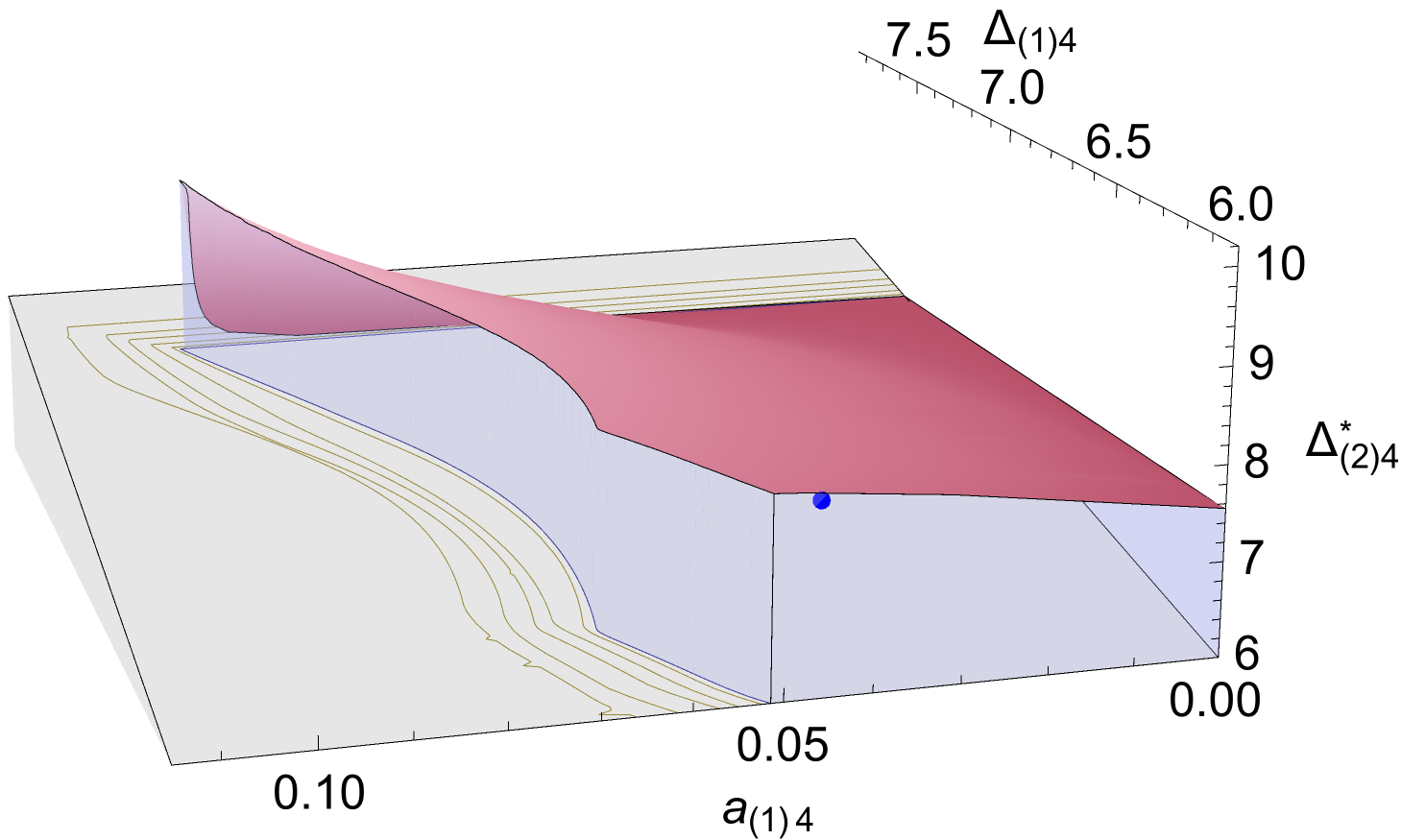}
\caption{Spin 4}
\end{subfigure}
\caption{\label{fig:nextopc34}OPE bounds for $\NN=4$ SCFTs with $c=3/4$.}
\end{figure}

\section{Bounds involving OPE coefficients}
\label{sec:secondresults}

Up to this point we have been concerned with the LTUSO of a given spin $\ell$. We will henceforth denote its dimension by $\Delta_{(1)\ell}$, and the positive coefficient of the corresponding conformal block by $a_{(1)\ell}$. In the previous section we presented a central-charge-dependent upper bound on $\Delta_{(1)\ell}$, which we denoted by $\Delta_\ell^{\star}$. In this section we go a step further and consider the \emph{next} operator of the same spin. We will call the dimension of this operator $\Delta_{(2)\ell}$ and denote its upper bound as $\Delta_{(2)\ell}^\star$. This upper bound will depend on the spin $\ell$, the central charge $c$, and the unknown values $\Delta_{(1)\ell}$ and $a_{(1)\ell}$. We have chosen to present results for $\ell \in \{0,2,4\}$ and $c \in \{\frac{3}{4}, \frac{15}{4}, \infty\}$ as before. In each case we have produced plots for $\Delta_{(2)\ell}^\star$ as a function of $\Delta_{(1)\ell}$ and $a_{(1)\ell}$. 

As in the case of the three-dimensional cubes shown in Fig.~\ref{fig:cubes}, our principal aim is to carve out as closely as possible the image of the conformal manifold in the three-dimensional space spanned by $(\Delta_{(1)\ell},a_{(1)\ell},\Delta_{(2)\ell})$. An additional virtue of this investigation is that the fixed value of $\Delta_{(1)\ell}$ can serve as a rough proxy for the interaction strength of the theory: we can approach the free theory by dialing $\Delta_{(1)\ell}$ to the unitarity bound, and describe a strongly interacting theory by dialing $\Delta_{(1)\ell}$ to its upper limit $\Delta_\ell^{\star}$. We will explain below that this leads to significant improvements in our ability to carve out physical solutions to the crossing equations.

\begin{figure}[t]
\begin{subfigure}[c]{\textwidth}
\centering
\includegraphics[width=5.5cm]{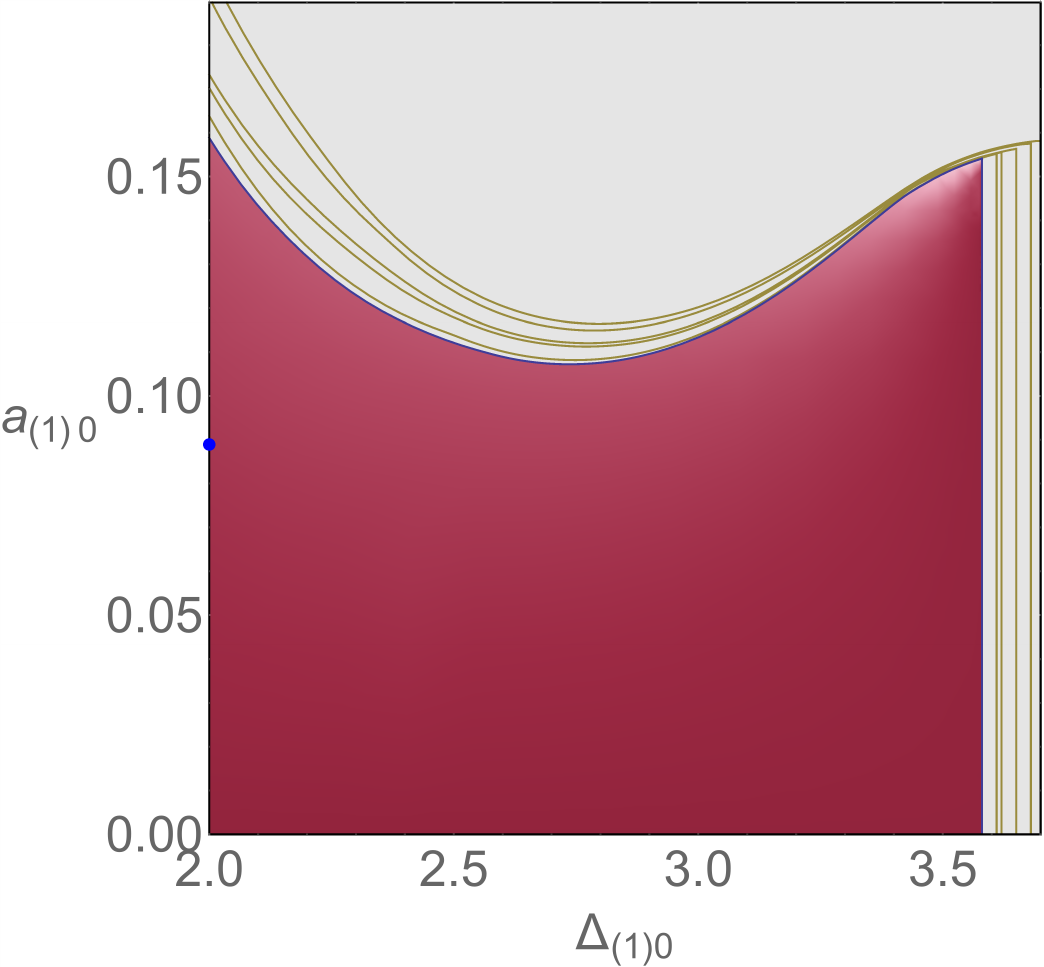}$\quad$
\includegraphics[width=9cm]{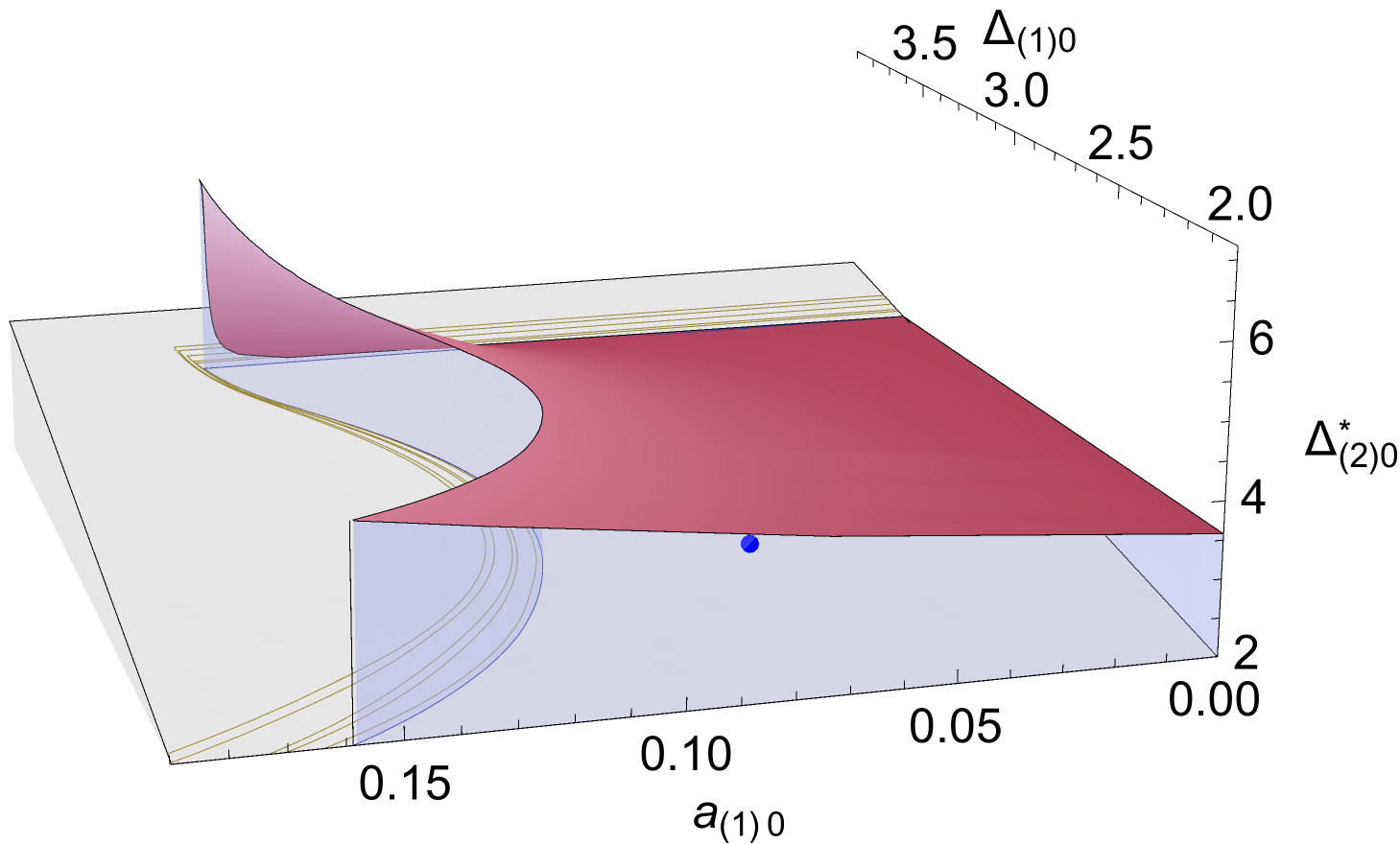}
\caption{Spin 0}
\end{subfigure}

\begin{subfigure}[b]{\textwidth}
\centering
\includegraphics[width=5.5cm]{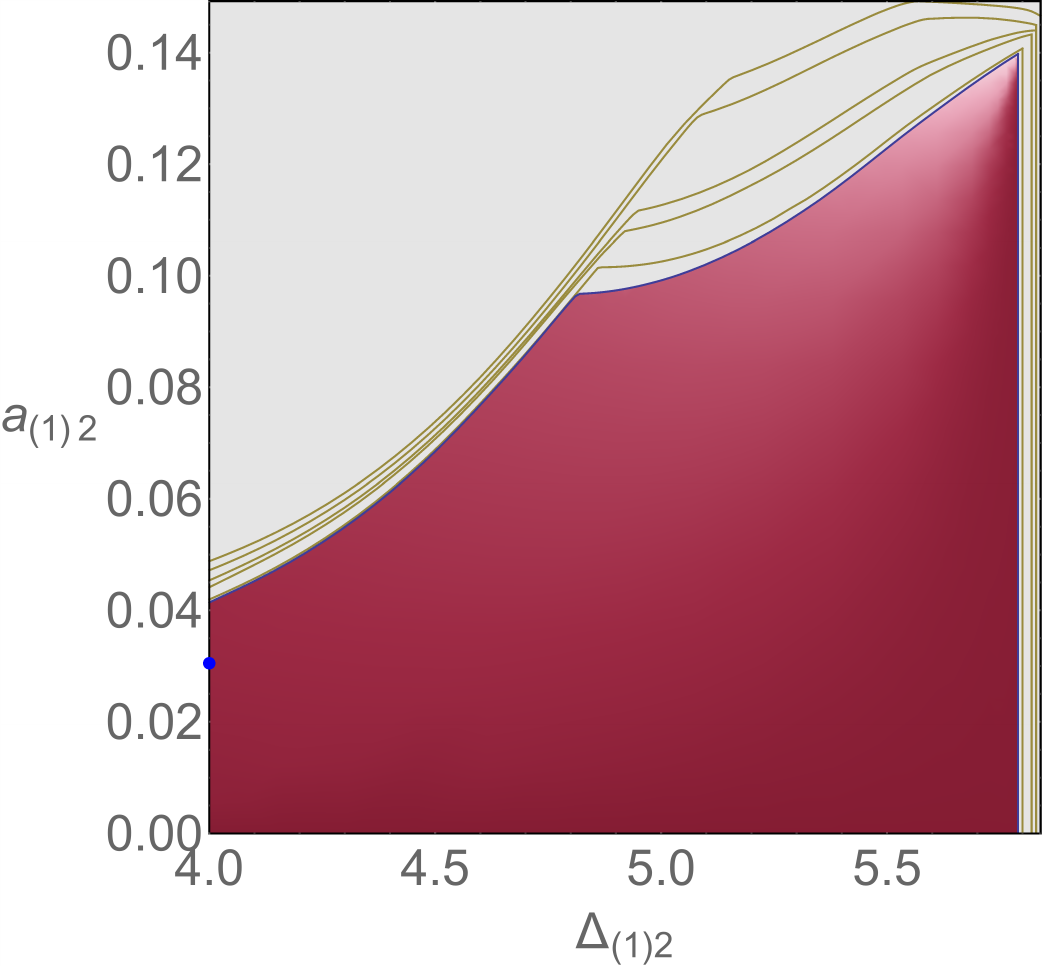}$\quad$
\includegraphics[width=9cm]{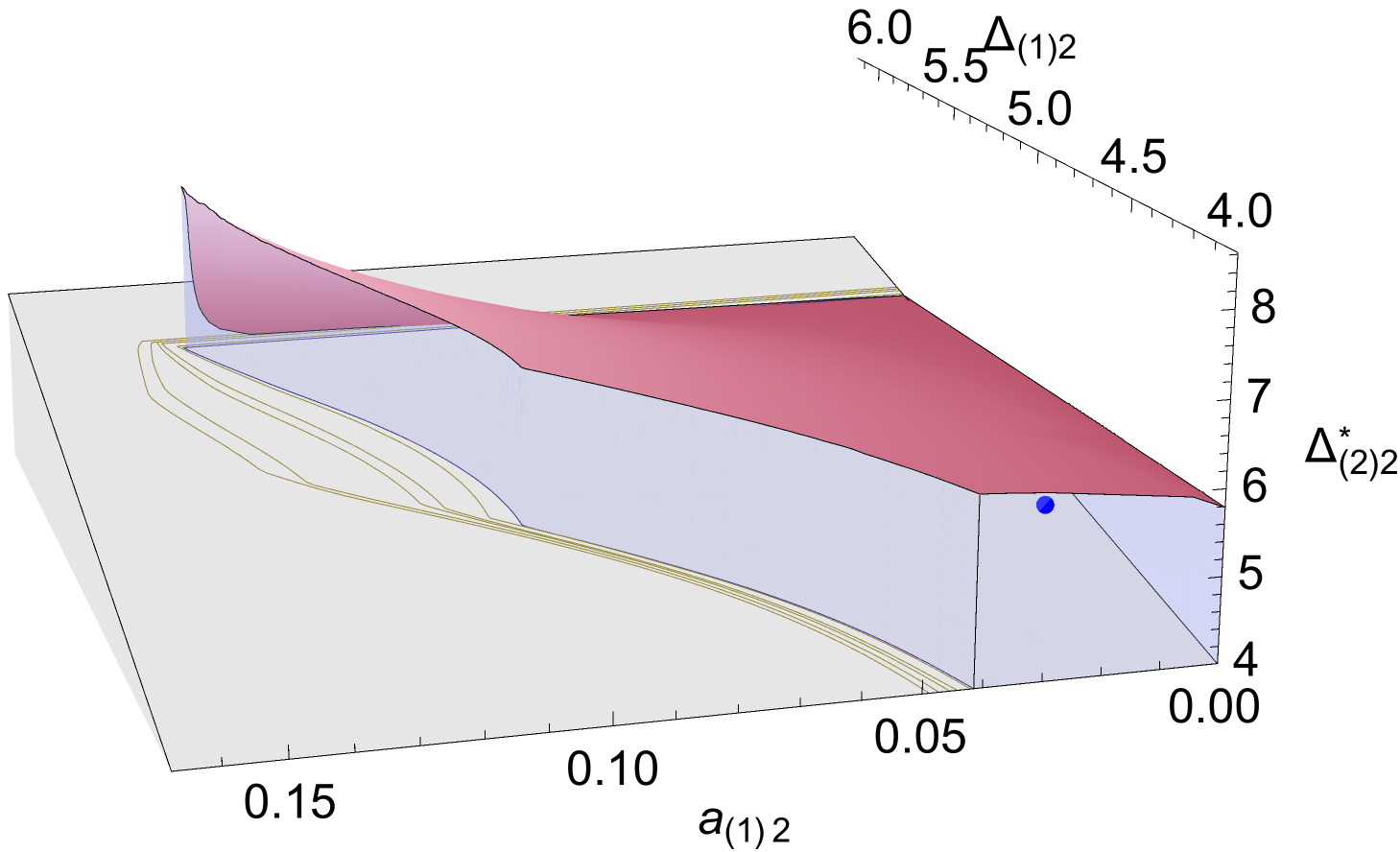}
\caption{Spin 2}
\end{subfigure}

\begin{subfigure}[b]{\textwidth}
\centering
\includegraphics[width=5.5cm]{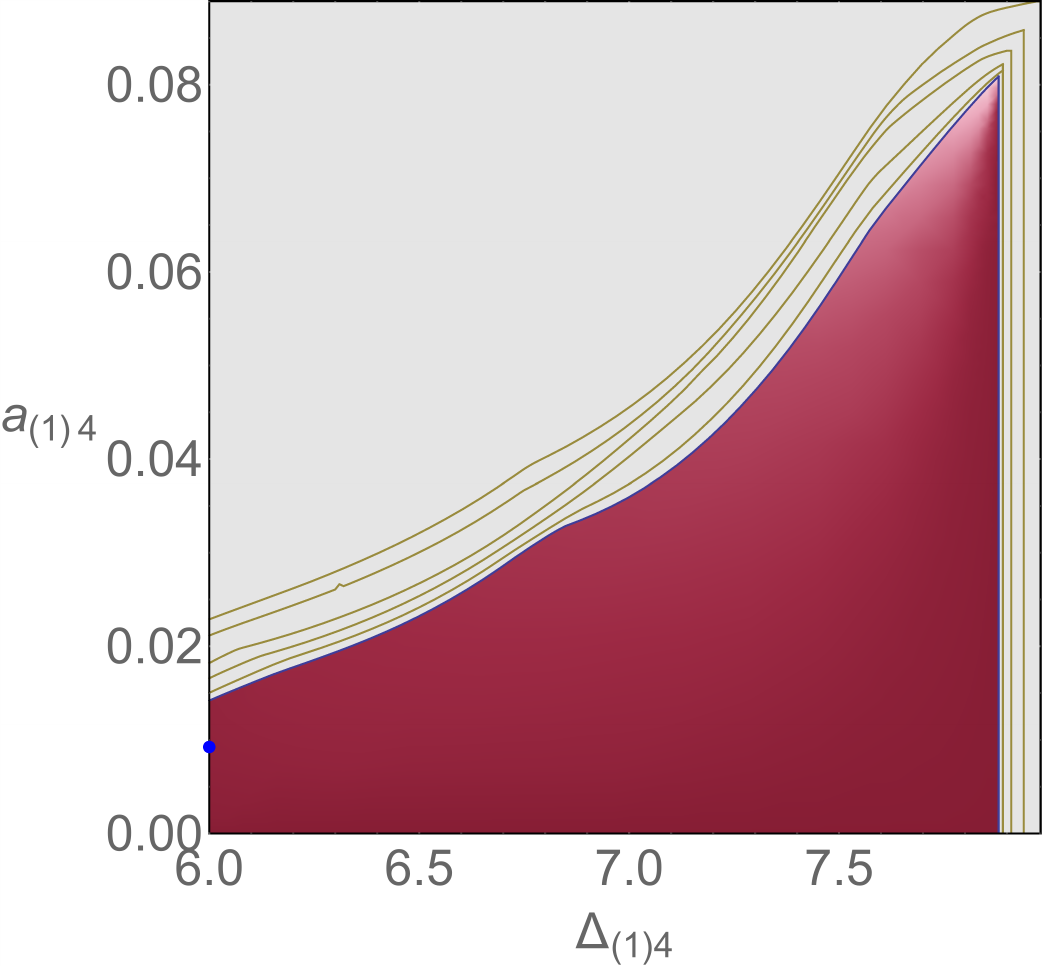}$\quad$
\includegraphics[width=9cm]{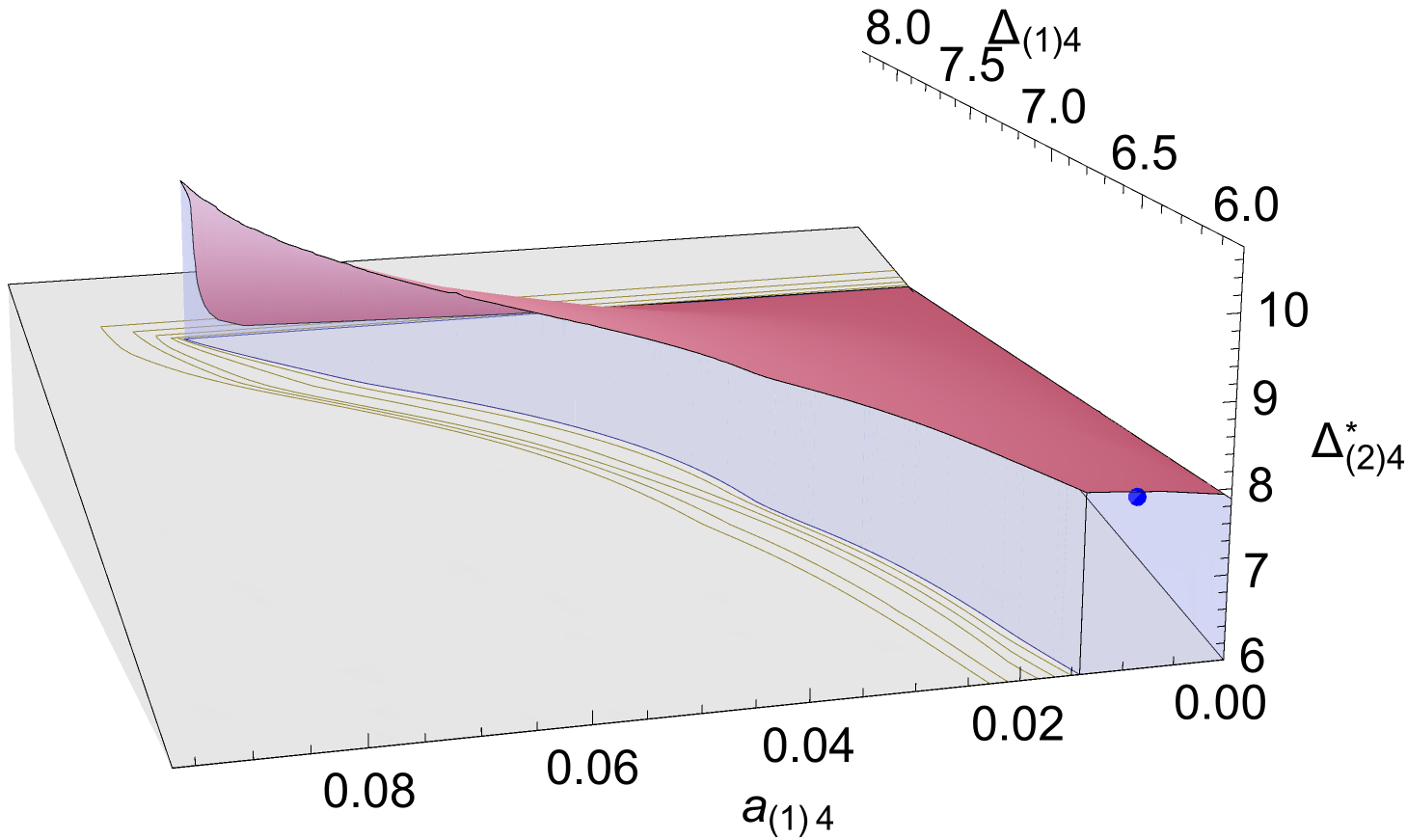}
\caption{Spin 4}
\end{subfigure}
\caption{\label{fig:nextopc154}OPE bounds for $\NN=4$ SCFTs with $c=15/4$.}
\end{figure}

\begin{figure}[t]
\begin{subfigure}[c]{\textwidth}
\centering
\includegraphics[width=5.5cm]{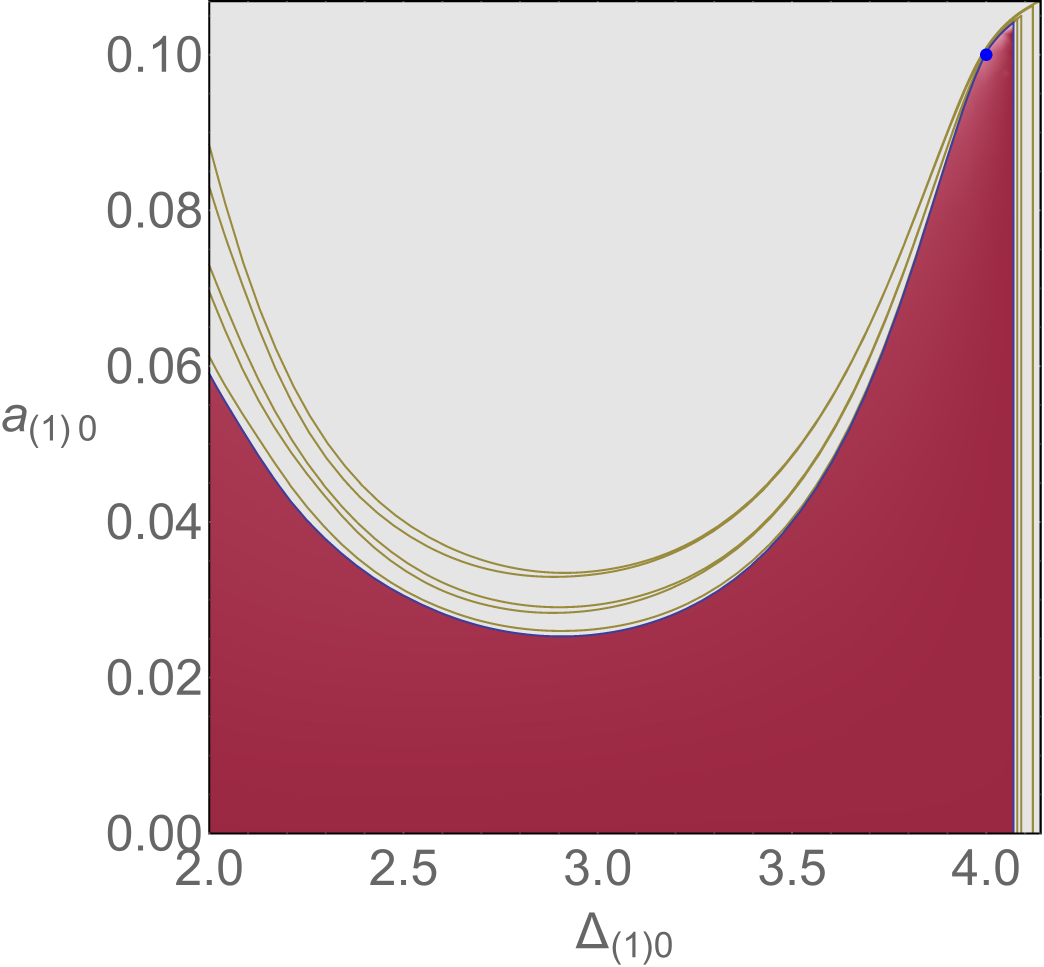}$\quad$
\includegraphics[width=9cm]{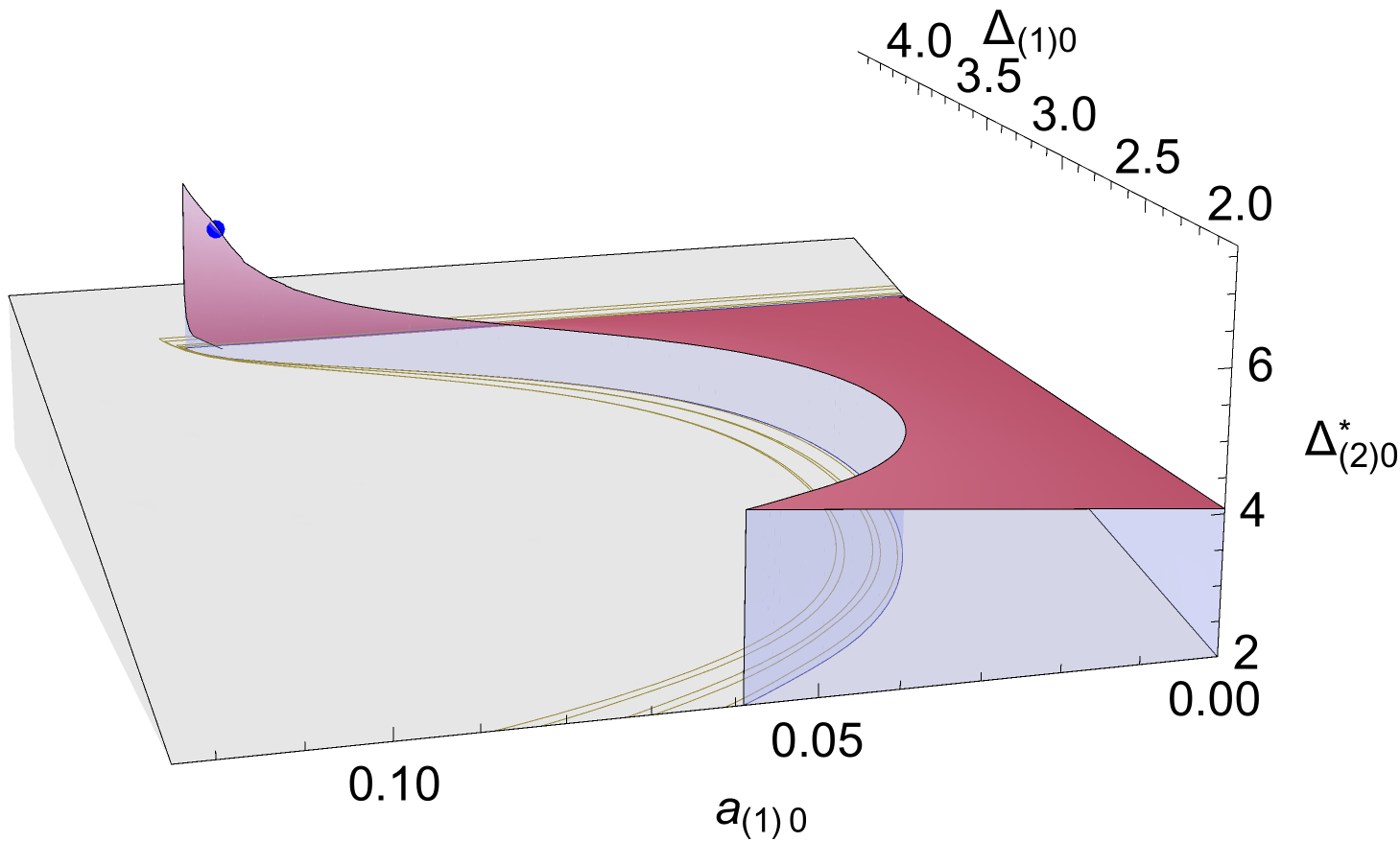}
\caption{Spin 0}
\end{subfigure}

\begin{subfigure}[b]{\textwidth}
\centering
\includegraphics[width=5.5cm]{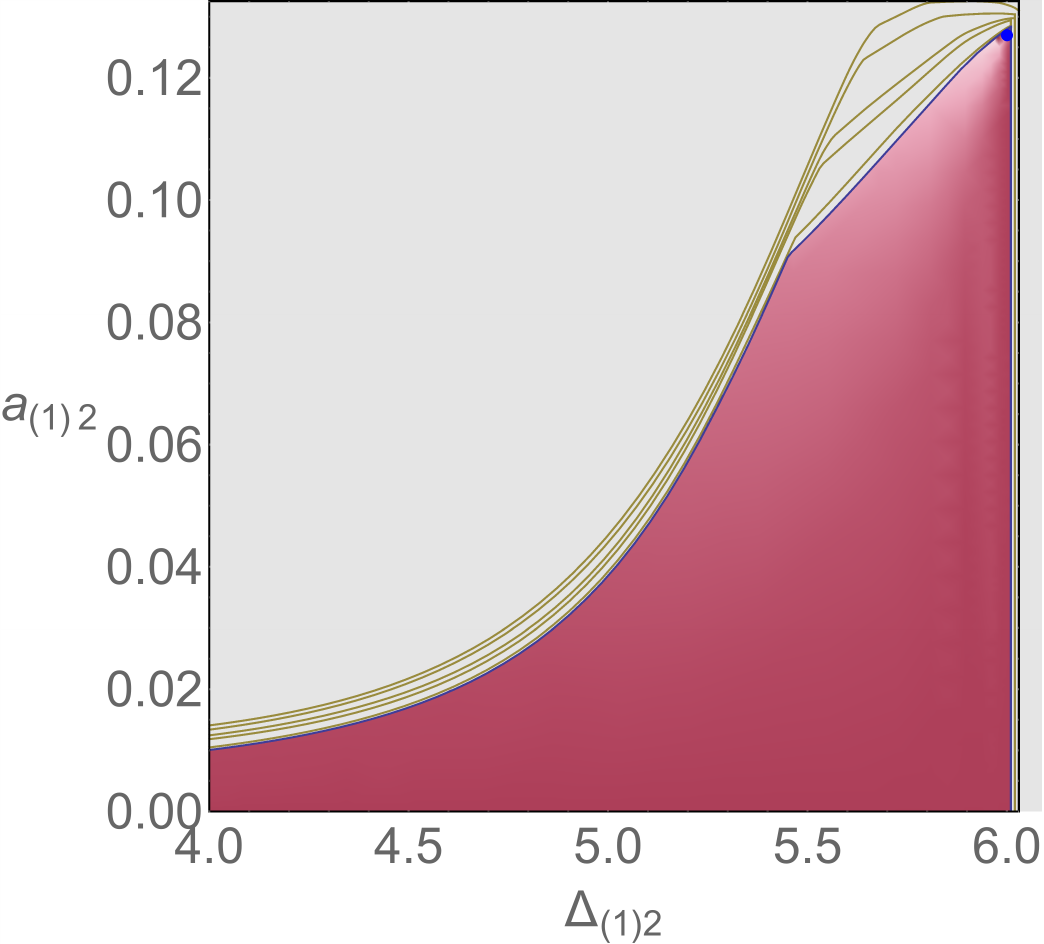}$\quad$
\includegraphics[width=9cm]{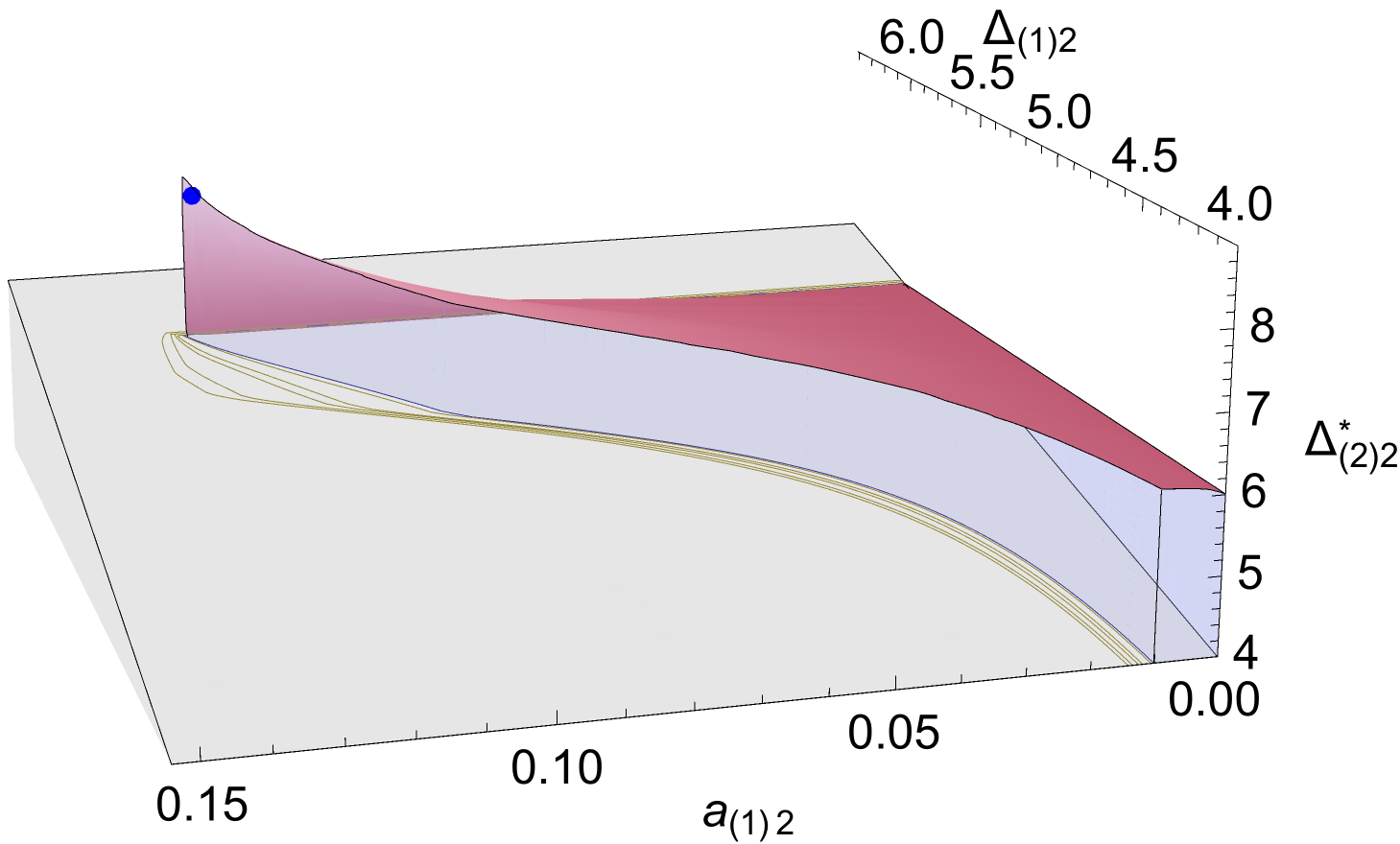}
\caption{Spin 2}
\end{subfigure}

\begin{subfigure}[b]{\textwidth}
\centering
\includegraphics[width=5.5cm]{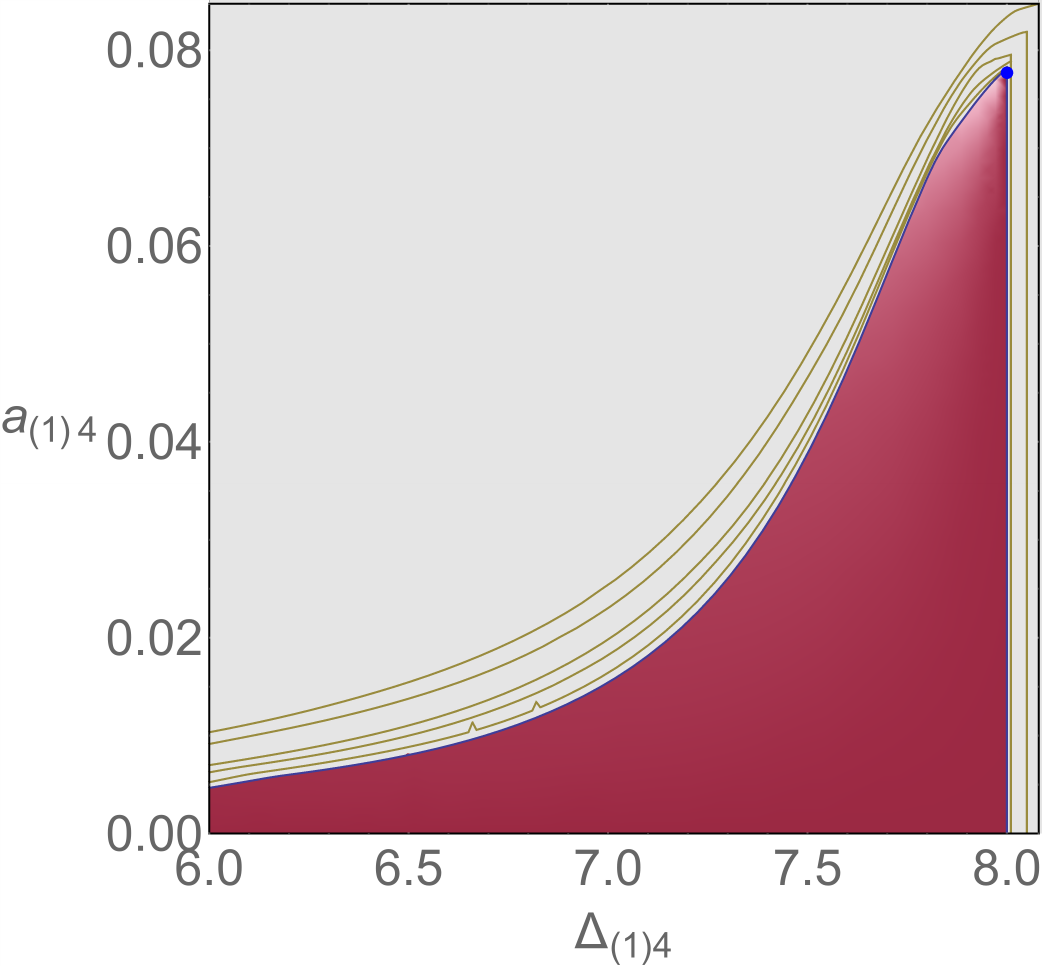}$\quad$
\includegraphics[width=9cm]{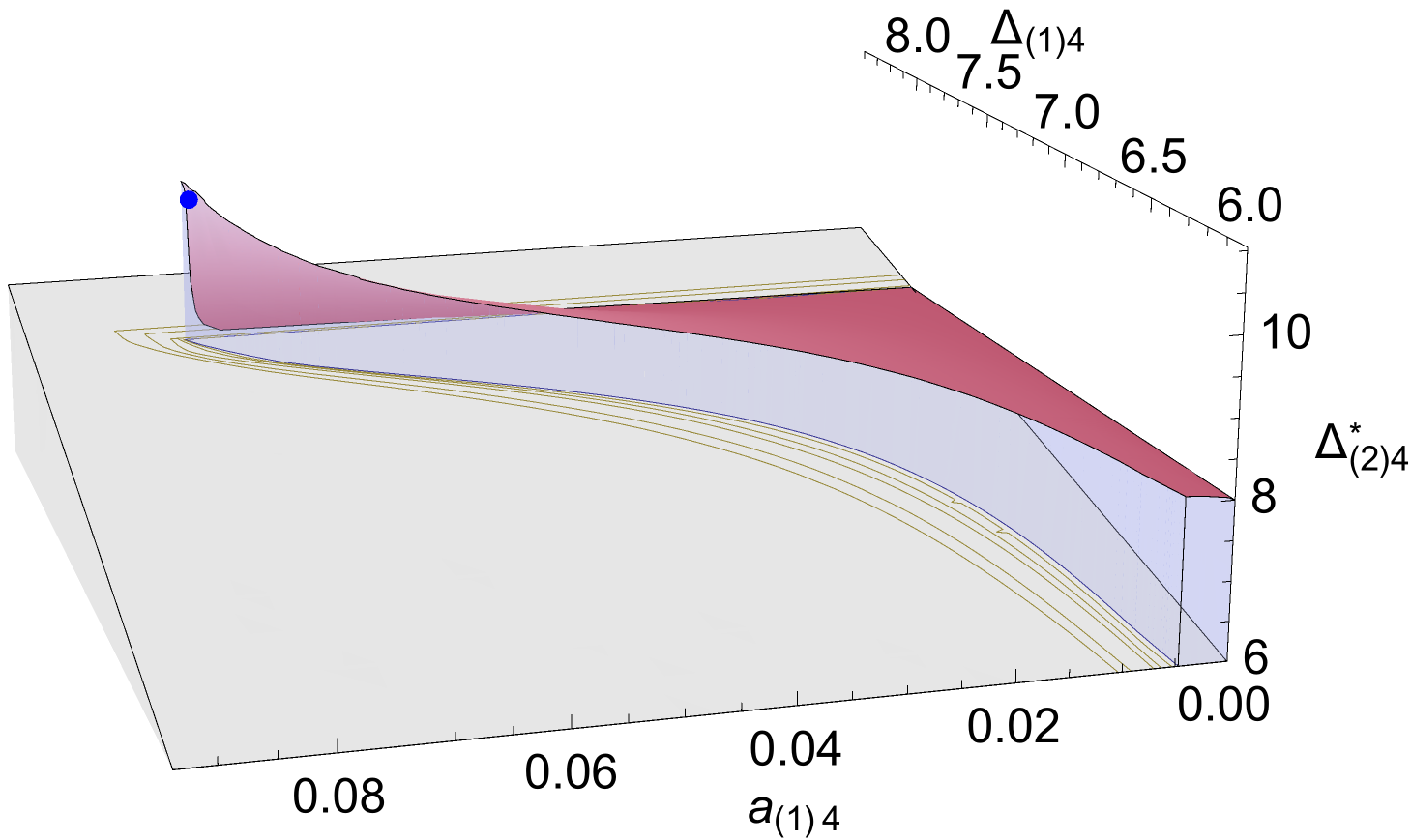}
\caption{Spin 4}
\end{subfigure}
\caption{\label{fig:nextopcinf}OPE bounds for $\NN=4$ SCFTs with $c=\infty$.}
\end{figure}

Results for $c = \frac{3}{4}$ are shown in Fig.~\ref{fig:nextopc34} and similar results for $c = \frac{15}{4}$ and $c = \infty$ can be found in Figs.~\ref{fig:nextopc154} and \ref{fig:nextopcinf}. The surfaces in the figures were obtained by fixing $\Delta_{(1)\ell}$ and $\Delta_{(2)\ell}^\star$, and deriving upper and lower bounds on $a_{(1)\ell}$ using the second optimization method described in Section~\ref{sec:methods}. The surfaces shown were obtained by interpolating through several hundreds data points per plot, the location of which were chosen carefully to obtain additional resolution near the sharper transitions. We used the linear programming method with $\Lambda = 17$, and one could probably improve the results further by adopting the semidefinite programming method and raising the cutoff. The plots in Fig.~\ref{fig:nextopc34} contain a great deal of information. We discuss the details in the next few paragraphs.

\paragraph{General constraints.} Let us first observe some general constraints in the space spanned by the triplet $(a_{(1)\ell},\Delta_{(1)\ell},\Delta_{(2)\ell}^\star)$. Unitarity restricts us to the octant with $a_{(1)\ell} \geqslant 0$ and $\Delta_{(2)\ell}^\star, \Delta_{(1)\ell} \geqslant \ell + 2$. We further have $\Delta_{(2)\ell}^\star \geqslant \Delta_{(1)\ell}$ by assumption, which is indicated by the grey wedge at the bottom of the three-dimensional plots. From the previous section we also know that $\Delta_{(1)\ell} \leqslant \Delta_{\ell}^\star$. This upper bound is the straight line that cuts off the allowed region on the right of the two-dimensional plot and at the far end in the three-dimensional plots. As in Fig.~\ref{fig:singlechannelbounds}, we have indicated in yellow the same bound obtained for lower values of the cutoff $\Lambda$.

\paragraph{Basic structure.} Suppose now that we send $a_{(1)\ell} \to 0$. In that case the first block disappears, the corresponding $\Delta_{(1)\ell}$ becomes irrelevant, and the bound obtained in the previous section then dictates that $\Delta_{(2)\ell}^\star \to \Delta_{\ell}^\star$ for all $\Delta_{(1)\ell}$. This is the constant value obtained along the rightmost boundary of the three-dimensional plots. If we now increase $a_{(1)\ell}$ away from zero, holding $\Delta_{(1)\ell}$ fixed, then $\Delta_{(2)\ell}^\star$ ought to increase (because we already satisfied the constraint of the previous section), and we can obtain a nontrivial $\Delta_{(2)\ell}^\star$ all the way until a point where $a_{(1)\ell}$ hits its own upper bound and no solution to crossing symmetry exists at all. The existence of such an upper bound is an experimental fact, and is responsible for the cutoff at the top of the two-dimensional plots and the abrupt cliff on the left side of the the three-dimensional ones. In the plots we have again used yellow lines to indicate the location of this upper bound for lower values of $\Lambda$. We find a clear dependence of the bound on $\Delta_{(1)\ell}$, with a kink for the spin-two and spin-four plots that we discuss below.

\paragraph{Behavior at $\Delta_{(1)\ell} = \Delta^\star_\ell$.} Next we consider the far side of the three-dimensional plots. Here $\Delta_{(1)\ell}$ approaches its upper bound $\Delta^\star_\ell$. As we discussed above, in that case there is a \emph{unique} approximate solution to crossing symmetry, and in this solution $a_{(1)\ell}$ is fixed to some critical value. In our plots this uniqueness is represented as follows. If our chosen value for $a_{(1)\ell}$ lies far below the critical value then we will find that $\Delta_{(2)\ell}^\star$ is approximately equal to $\Delta_\ell^\star$, because the second operator is then forced to take over the missing contribution of the first operator. On the other hand, once our assumed value for $a_{(1)\ell}$ exactly coincides with the critical value then we find that $\Delta_{(2)\ell}^\star$ will shoot up to the correct dimension of the second operator in the channel. This is precisely the peak that we observe along this edge of the plot. This peak is expected to get sharper by increasing $\Lambda$. If we dial $a_{(1)\ell}$ even further then no value for $\Delta_{(2)\ell}$ will be able to remedy the situation -- the solution to crossing symmetry ceases to exist.

\paragraph{Known solutions to crossing symmetry.} The blue dots in our plots correspond to the free theories or, for $c = \infty$, to the strict planar limit. For finite $c$ the corresponding values are given by:
\begin{equation}
\text{free field theory:} \qquad \Delta_{(1)\ell} = \ell + 2, \qquad a_{(1)\ell} = 2^{\ell + 1} \frac{((\ell+2)!)^2}{(2\ell+4)!} \frac{1}{c},\qquad \Delta_{(2)\ell}^\ell = \ell + 4\,.
\end{equation}
This is Eqn.~(8.37) of \cite{Dolan:2001tt} with the replacement of their coefficient $c$ with the inverse central charge, which is $1/c$ in our notation. For infinite central charge this coefficient goes to zero and instead we have:
\begin{equation}
\text{planar limit:} \qquad \Delta_{(1)\ell} = 4, \qquad a_{(1)\ell} = \frac{2^{\ell + 2} (\ell+2)!(\ell+3)!}{24 (2 \ell + 5)!} (\ell + 1) (\ell + 6),\qquad \Delta_{(2)\ell}^\ell = \ell + 6\,.
\end{equation}
which is obtained from Eqn.~(8.34) in \cite{Dolan:2001tt} after setting their coefficients $t = 2$ and $a = 1$. In all cases our bounds are compatible with these points, and we in fact expect the bounds to converge towards these points upon further increasing $\Lambda$. Notice that this convergence is already very clear in the large $c$ limit shown in Fig.~\ref{fig:nextopcinf} shown below.

\paragraph{Physical solutions.} So where precisely is the physics in these plots? This question is answered by envisioning a map from the conformal manifold of $\NN=4$ SYM into the octant spanned by $(a_{(1)\ell},\Delta_{(1)\ell},\Delta_{(2)\ell}^\star)$. The image must be `anchored' at the free-field theory points, and extends from these points to some surface that should lie entirely inside the domain allowed by our bounds. The conjectures of the previous sections furthermore state that it extends all the way to the peak of the three-dimensional plots where all three of $\Delta_{(1)\ell}$, $\Delta_{(2)\ell}$ and $a_{(1)\ell}$ are maximized. Consequently, the most interesting region of our plots (for finite $c$) is the curve corresponding to the upper bound on the left side of the three-dimensional plots and on the top in the two-dimensional plots. Here $a_{(1)\ell}$ and $\Delta_{(2)\ell}$ are both maximized for a given $\Delta_{(1)\ell}$. By repeating the same arguments as before, we would like to claim that the curve itself traces an `extremal path' on the conformal manifold between the free-field point and a self-dual point.\footnote{The current curve is probably a crude approximation to this path because we are still relatively far from the free-field point. This can however easily be improved by increasing $\Lambda$.} If so, then this plot presents a notable improvement over the three-dimensional plots shown in Fig.~\ref{fig:cubes}, because we are now carving out the image of an \emph{entire curve} rather than a single point on the conformal manifold.

As stated above, our eventual ambition is to improve our methods so that we can carve out the image of the conformal manifold completely. The current approach, even when pushed to very high $\Lambda$, would most likely still leave an allowed region that is too large, much like the cubes in Fig.~\ref{fig:cubes}. A more detailed analysis, with other correlation functions and/or observables, will be necessary to carve away further inconsistent points.

\paragraph{Kinks.} Along the path for the spin-two and spin-four operators we observe some rather striking kinks. Their location varies significantly with the cutoff $\Lambda$, as indicated by the yellow curves, and it is rather likely that the kinks will eventually meet the free-field point on the left of the two-dimensional plots. On the other hand, if these kinks would persist even for very large $\Lambda$ then they may represent some interesting physics on the moduli space of $\NN=4$ SYM theories. For example, the kinks may be the numerical avatar of an approximate crossing of operator dimensions as sketched in the middle plot in \ref{fig:generalbehavior}. If so then the \emph{third} operator should be relatively close to the second operator. It should be possible to investigate this possibility using current technology.

\paragraph{The free field solution.} We expect a unique solution to the crossing symmetry equations both when $\Delta_{(1)\ell}$ is equal to its upper bound $\Delta^\star_{\ell}$ and when it is equal to the unitarity bound $\ell + 2$. As discussed above, this uniqueness is clearly visible in the numerical bounds in the first case, but in the second case we do not observe the same sharp peak. The aforementioned linear combinations of solutions partially explain this behavior: in this case we can combine the free-field and the extremal solution to show that $\Delta_{(2)\ell}^\star$ will never drop below $\Delta_\ell^\star$, even when $\Delta_{(1)\ell}$ sits at the unitarity bound. This however does not explain the rather gradual ascent of our bounds towards the actual free solution, which we leave as an interesting puzzle for the future.

\acknowledgments

We are grateful to the participants of the Bootstrap 2016 workshop at the Weizmann Institute and those of the Aspen summer workshop ``From scattering amplitudes to the conformal bootstrap'' for their insightful comments on this work. We would also like to thank Fernando Alday, Agnese Bissi, Reza Doobary, Paul Heslop, Juan Maldacena, Eric Perlmutter, and David Simmons-Duffin for many helpful discussions. 

\bigskip
\noindent This work was supported in part by National Science Foundation Grant No. PHYS-1066293 and the hospitality of the Aspen Center for Physics. C.B. gratefully acknowledges support from the Frank and Peggy Taplin Fellowship at the IAS. C.B. was also supported in part by the NSF through grant PHY-1314311. This work was additionally supported by a grant from the Simons Foundation (\#494786, C.B.; \#397411, L.R.; \#488659, B.v.R.). Some computations in this paper were run on the Hyperion computing cluster supported by the School of Natural Sciences Computing Staff at the Institute for Advanced Study. The authors would also like to acknowledge the use of the University of Oxford Advanced Research Computing (ARC) facility in carrying out this work.

\bibliography{./aux/biblio}
\bibliographystyle{./aux/JHEP}
\end{document}